\documentclass[aps,twocolumn,nofootinbib]{revtex4}
\usepackage{graphicx}
\usepackage{amsmath}
\usepackage{amssymb}
\usepackage{bm}
\usepackage{mathtools}
\usepackage{hyperref}
\usepackage{xcolor}
\usepackage{xfrac}
\usepackage{bm}
\usepackage[T1]{fontenc}
\usepackage{upquote}

\DeclareMathOperator{\ch}{ch}

\DeclareMathOperator{\Th}{th}

\newcommand{\rn}{{\rm n}}

\begin{document} 

\title{\bf The information geometry of 2-field functional integrals} 

\author{Eric Smith}

\affiliation{Earth-Life Science Institute, Tokyo Institute of
Technology, 2-12-1-IE-1 Ookayama, Meguro-ku, Tokyo 152-8550, Japan}

\affiliation{Department of Biology, Georgia Institute of
Technology, 310 Ferst Drive NW, Atlanta, GA 30332, USA}

\affiliation{Santa Fe Institute, 1399 Hyde Park Road, Santa Fe, NM
87501, USA}

\affiliation{Ronin Institute, 127 Haddon Place, Montclair, NJ 07043,
USA}

\date{\today}
\begin{abstract}

2-field functional integrals (2FFI) are an important class of solution
methods for generating functions of dissipative processes, including
discrete-state stochastic processes, dissipative dynamical systems,
and decohering quantum densities.  The stationary trajectories of
these integrals describe a conserved current by Liouville's theorem,
despite the fact that there is no conserved phase space current in the
underlying stochastic process.  We develop the information geometry of
generating functions for discrete-state classical stochastic processes
in the Doi-Peliti 2FFI form, showing that the conserved current is a
Fisher information between the underlying distribution of the process
and the tilting weight of the generating function.  To give an
interpretation to the time invertibility implied by current
conservation, we use generating functions to represent importance
sampling protocols, and show that the conserved Fisher information is
the differential of a sample volume under deformations of the nominal
distribution and the likelihood ratio.  We derive a new pair of
dual Riemannian connections respecting the symplectic structure of
transport along stationary rays that gives rise to Liouville's
theorem, and show that dual flatness in the affine coordinates of the
coherent-state basis captures the special role played by coherent
states in many 2FFI theories.  The covariant convective derivative
under time translation correctly represents the geometric invariants
of generating functions under canonical transformations of the 2FFI
field variables of integration.

\end{abstract}

\maketitle

\section{Introduction: Understanding the Liouville theorems that
emerge in 2-field functional integrals for dissipative systems}

The defining feature of dissipative systems, whether classical or
quantum, is that trajectories initially distinct can merge and that
distributions or densities that differ at their initial conditions
become more similar over time as they are increasingly governed by
local generating parameters at the expense of memory.  Because such
systems are intrinsically irreversible, they obey no Liouville theorem
(see~\cite{Goldstein:ClassMech:01}) describing phase space densities
that are conserved along flow lines.

A powerful formalism for representing dissipative systems, both
classical and quantum, is that of 2-field functional integrals (2FFI)
that evolve the generating functions of distributions or densities.
2FFI methods with a shared integral form include the Doi-Peliti method
for discrete-state
processes~\cite{Doi:SecQuant:76,Doi:RDQFT:76,Peliti:PIBD:85,Peliti:AAZero:86},
which will be used in this paper, the Martin-Siggia-Rose path integral
for dynamical systems with Langevin noise~\cite{Martin:MSR:73}, and
the Schwinger-Keldysh time-loop formalism for quantum density matrices
with
decoherence~\cite{Schwinger:MBQO:61,Keldysh:noneq_diag:65}.\footnote{We
will not work with quantum 2FFI methods in this paper, and the Hilbert
space for quantum density matrices differs in some important ways from
that for classical generating functions, but the operator algebra,
causal structure, and much of the stationary-point analysis are shared
between the two~\cite{Kamenev:DP:02}.}  The modern approach traces
back much further to a nonlinear action functional due to Onsager and
Machlup~\cite{Onsager:Machlup:53} for systems with Langevin noise,
which can be derived from the 2-field formalism.

These integrals have the curious feature that their stationary-path
analysis introduces a conserved volume element and Liouville theorem
for just those systems that lack one in the state space.  Conservation
of a density along deterministic trajectories implies a form of
temporal invertibility, and since this is not reversibility in the
dynamical sense, it expresses a kind of time-\emph{duality}.  We are
interested, then, in how such conserved densities are brought
into existence by generating functions, and what is their meaning in
terms of information. 

In this paper we present the Liouville theorem and conserved volume
element of 2FFI generating functions from an Information Geometry
perspective~\cite{Amari:inf_geom:01,Ay:info_geom:17}.  To understand
the concepts and questions that are unified under a geometric
description, we give a brief synopsis of observations and claims that
will be developed in the rest of the paper. 

\subsubsection*{Synopsis of prior known facts and puzzles brought
together and explained in this paper}

Information geometry recognizes a duality with respect to the Fisher
metric, between contravariant affine coordinates representing the
intensity of a tilting weight in an exponential family such as the
family of measures that define a generating function, and covariant
coordinates which are the mean values in the resulting tilted
distributions.  Not surprisingly, those geometrically dual coordinates
will be shown to coincide with the canonically conjugate coordinates
evolving under the 2FFI Liouville theorem.

The concept of dual parallel transport that is central to information
geometry arises naturally in 2FFI generating functions, coming from
the same source as their time-duality.  These integrals are
constructed to propagate two measures through time on the same state
space, one describing dynamical underlying distributions and the other
the distribution of \textit{tilts} that define a generating function
from an exponential family over each underlying distribution.  The
independent freedom to vary these two measures leads to triples of
distributions that, when compared under the generalized Pythagorean
theorem~\cite{Nagaoka:dual_geom:82,Amari:methods_IG:00}, define an
inner product in the Fisher metric.  The inner product has the
interpretation of a sensitivity, or (more geometric) of a
direction-cosine between vector fields associated with deformations in
the base distribution and deformations in the tilt.  Saddle-point or
stationary-path conditions in the functional integral then define dual
transport laws for the two measures under the action of a symplectic
form, resulting in Liouville's theorem.

To understand the meaning of densities conserved along stationary
trajectories, we employ the interpretation of generating functions in
terms of statistical inference and more specifically of Importance
Sampling~\cite{Owen:mcbook:13}.  The tilt in a generating function
becomes the likelihood ratio relating a nominal distribution to an
importance distribution, and the conserved density of Liouville's
theorem becomes a differential volume element of sample probability.
Dual transport of a conserved inner product in the Fisher metric, in
turn, defines a transport law for the metric under coordinate
transformations respecting the symplectic structure.

The use of dual Riemannian connections that is a cornerstone of
information geometry provides a way to express dynamically meaningful
features of a theory as geometric invariants under coordinate
transformations.  For 2FFI theories we construct a particular pair of
dual connections, different from the dually flat connections
introduced by Amari and
Nagaoka~\cite{Nagaoka:dual_geom:82,Amari:methods_IG:00}, to capture
the following distinctive property of this class of theories:

Most introductions~\cite{Mattis:RDQFT:98,Kamenev:DP:02,Baez:QTSM:17}
of 2FFI methods derive them in bases of
\textit{coherent states}.  These states correspond to mixture
coordinates in the underlying distribution, and to the argument
variables $z$ that are the tilting weights in a
\emph{moment}-generating function.  They are contrasted to the
cumulant-generating function arguments $\log z$ that are coordinates
in the exponential family, and the mixture coordinate in the tilted
(importance) distribution that are dual affine coordinates under the
Amari-Nagaoka connections.  The coherent state coordinates are not
affine under the Fisher metric, and are not a basis for dualization
through the Legendre transform, yet they are often the natural affine
basis for the growth and contraction eigenvalues of the Liouville
volume element on 2FFI stationary paths.  The exponential and mixture
families on the importance distribution are related to coherent states
by a canonical transformation~\cite{Smith:LDP_SEA:11}.\footnote{This
variable transformation, which we have used
extensively~\cite{Smith:LDP_SEA:11,Smith:evo_games:15,%
Krishnamurthy:CRN_moments:17,Smith:CRN_moments:17} and will be central
in the derivation below, is sometimes adopted to compute generating
functions for counting statistics~\cite{Sinitsyn:geophase_rev:09}, but
is otherwise rarely seen.}  We will show that dual connections
describing flat parallel transport in coherent-state coordinates are
the correct concept to capture the role of this basis in relation to
the geometric role of coordinates in the exponential family.  More
generally, dual connections respecting the symplectic structure of
translation along stationary rays define a covariant procedure for
canonical transformation in 2-field integrals.

\subsubsection*{Organization of the presentation}

The derivation of the these results is organized as follows: The first
three sections review basic constructions that will be needed from
information geometry (Sec.~\ref{sec:int_lattice_CGF_IG}), importance
sampling (Sec.~\ref{sec:LDF_and_IS}), and the Doi-Peliti formalism for
2FFI generating functions (Sec.~\ref{sec:DP_gen_review}).  The aim is
to give a brief but self-contained introduction that will be
understandable to readers from each area to whom the others may be
unfamiliar, and to establish shared terms and notation.  The dualities
in information geometry and importance sampling apply to general
families of probability distributions without reference to
interpretations of time dependence, so they are presented first with
emphasis on general coordinate transformations.  The Doi-Peliti
construction then brings in the additional features needed to evolve
distributions under stochastic processes.

The main results of the paper are derived in
Sec.~\ref{sec:2_field_Liouville}.  It is shown that the conserved
density that plays the role of a phase space density in Liouville's
theorem is a Wigner function on a product space that jointly evolves
parameters associated with stochastic dynamics and with statistical
inference.  Deformations associated with underlying nominal
distributions and with likelihood ratios are shown to define dual
vector fields in the Fisher geometry on importance distributions, and
their inner product corresponds to the differential volume element
derived from the Wigner density.  The transport law for the Fisher
metric, and dual Riemannian connections respecting the symplectic
structure, are then derived.  

Sec.~\ref{sec:example_two_state} contains a simple worked example
illustrating all aspects of the 2FFI Liouville theorem and its
associated dual geometry.  Sec.~\ref{sec:conclusions} concludes,
noting where the duality between dynamics and inference developed here
extends and clarifies other studies of time-reversal duality that are
currently active topics.

\section{The dual geometry from cumulant-generating functions for
counts on integer lattices} 
\label{sec:int_lattice_CGF_IG}

Here we review the basic constructions of information geometry for
generating functions over families of probability distributions
indexed by some coordinate such as a first-moment value.  No
stochastic process or other notion of time dependence is assumed
before Sec.~\ref{sec:DP_gen_review}.  Rather than develop the
generalized Pythagorean theorem in order to arrive at the projection
theorem as originally done by Nagaoka and
Amari~\cite{Nagaoka:dual_geom:82}, we use the Pythagorean theorem to
define an inner product between vector fields associated with
independent variations in the family of underlying probability
distributions and in the weights of their generating functions, which
will be the main quantity conserved through time when time-dependence
is introduced later.  The condition for preservation of the inner
product under coordinate transformations will be existence of a
symplectic form, shown in later sections to be provided by the
generators of time evolution in stochastic processes.  The section
retains as much as possible the notation of~\cite{Amari:inf_geom:01}
Ch.~6.

\subsubsection*{Scope of systems considered}

Later results involving the symplectic geometry and Liouville theorem
associated with 2-field functional integrals apply generally across
discrete and continuous state spaces, for both classical and quantum
(Schwinger-Keldysh time-loop) methods~\cite{Kamenev:DP:02}.  However,
we will define geometric constructions only for probability
distributions on discrete state spaces, as the simplest starting point
to illustrate the idea.

For definiteness of notation, we will consider probability
distributions on integer lattices in the positive orthant in ${\mathbb
Z}^D$, appropriate to the description of population processes.
Well-developed applications include evolutionary
populations~\cite{Smith:evo_games:15} and chemical reaction
networks~\cite{Krishnamurthy:CRN_moments:17,Smith:CRN_moments:17}.
The dimensions in the lattice, indexed $i \in 1 , \ldots , D$, define
types in the population, which we will refer to as \textit{species};
population states are vectors of integer-valued coefficients $\rn
\equiv \left( {\rn}_i \right)$, with ${\rn}_i$ the count of species
$i$; probability mass functions (which we will also refer to as
probability distributions) are denoted ${\rho}_{\rn}$.

\subsection{The exponential families from generating functions for
species counts}

We will consider the simplest exponential families over a distribution
${\rho}_{\rn}$, the linear families of generating functions for the
species counts ${\rn}_i$.  Both moment-generating functions (MGF) and
cumulant-generating functions (CGF) will be used.  The ordinary
power series MGF is a function of a vector of complex variables $z
\equiv \left( z_i \right)$.  The CGF is a function of the logarithms
$\log z_i \equiv {\theta}^i$, which we introduce with a raised index
to adopt the Einstein summation convention, that pairs of raised and
lowered indices are summed.  If $\theta$ and $\rn$ are regarded as
column vectors, ${\theta}^T$ stands for transpose.  Compact notations
for the vector $z$ raised component-wise to the power $\rn$, and for
the inner product of row and column vectors, are
\begin{align}
  z^{\rn} 
& \equiv 
  \prod_{i = 1}^D
  z_i^{{\rn}_i}
\nonumber \\
  {\theta}^T \rn 
& \equiv 
  {\theta}^i {\rn}_i
\label{eq:product_conventions}
\end{align}
In terms of these, the MGF denoted $\Psi$ and the CGF denoted $\psi$,
are defined as
\begin{align}
  \Psi \! \left( z \right) 
& \equiv 
  \sum_{\rn}
  z^{\rn}
  {\rho}_{\rn}
\nonumber \\ 
\equiv 
  e^{
    \psi \left( \theta \right)
  } 
& = 
  \sum_{\rn}
  {\rho}_{\rn}
  e^{{\theta}^T \rn}
\label{eq:CGF_onetime_def}
\end{align}

$e^{{\theta}^T \rn}$ is called an \textit{exponential tilt} applied to
the distribution ${\rho}_{\rn}$.  The normalized tilted distribution
obtained by dividing by $e^{\psi \left( \theta \right)}$ is denoted 
\begin{equation}
  {\tilde{\rho}}^{\left( \theta \right) }_{\rn} \equiv 
  {\rho}_{\rn}
  e^{
    {\theta}^T \rn - 
    \psi \left( \theta \right)
  }
\label{eq:rho_tilt_theta}
\end{equation}
Normalization of ${\tilde{\rho}}^{\left( \theta \right)}_{\rn}$
implies a value for the first moment, which we denote $n \! \left(
\theta \right)$, of
\begin{equation}
  n \! \left( \theta \right) \equiv 
  {
    \left< \rn \right>
  }_{
    {\tilde{\rho}}^{\left( \theta \right)}
  } = 
  \frac{\partial \psi}{\partial \theta}
\label{eq:expect_n_from_psi}
\end{equation}

\subsection{The Fisher metric on the exponential family of tilted
distributions} 
\label{sec:Fisher_metric_intro}

\subsubsection{The variance as local metric, and the Fisher distance
element introduced}

On a single underlying distribution ${\rho}_{\rn}$, a small change of
coordinate $\delta \theta$ in the exponential family leads to a change
in the tilted distribution of 
\begin{equation}
  \delta 
  \log {\tilde{\rho}}^{\left( \theta \right)}_{\rn} = 
  \delta {\theta}^T
  \left( 
    \rn - n \! \left( \theta \right)
  \right)
\label{eq:dlogrho_dtheta}
\end{equation}
The standard geometry on the exponential family of tilted
distributions is introduced by using the variance under
${\tilde{\rho}}^{\left( \theta \right)}_{\rn}$ to define a distance
element between coordinates separated by a small increment $\delta
\theta$ of
\begin{align}
  \delta s^2 
& = 
  \delta {\theta}^i
  \delta {\theta}^j
  \frac{
    {\partial}^2 
    \psi \! \left( \theta \right)
  }{
    \partial {\theta}^i
    \partial {\theta}^j
  }
\nonumber \\ 
& \equiv 
  \delta {\theta}^i
  \delta {\theta}^j
  \sum_{\rn}
  {\tilde{\rho}}_{\rn}^{\left( \theta \right)} 
  \left( 
    {\rn}_i - n_i \! \left( \theta \right)
  \right)
  \left( 
    {\rn}_j - n_j \! \left( \theta \right)
  \right)
\nonumber \\ 
& \equiv  
  \delta {\theta}^i
  \delta {\theta}^j
  g_{ij} \! \left( \theta \right)
\label{eq:Fisher_element}
\end{align}
$g\! \left( \theta \right)$, defined here as the Hessian of $\psi \! 
\left( \theta \right)$, is the Fisher metric tensor, introduced in
this usage by Rao ~\cite{Rao:info_metric:45} (reprinted
as~\cite{Rao:info_metric:92}).  Using the differential geometry
notation in which ${\left\{ \partial / \partial {\theta}^i
\right\}}_{i = 1}^D$ is the set of basis elements in the tangent space
to the exponential family, the Fisher metric is an inner product,
which we denote
\begin{align}
  g_{ij} 
& \equiv 
  \left< 
    \frac{\partial}{\partial {\theta}^i} , 
    \frac{\partial}{\partial {\theta}^j} 
  \right> 
\label{eq:Fisher_as_innerprod}
\end{align}

\subsubsection{Coordinate dualization, Legendre transform, and the
Large-Deviation function}

Eq.~(\ref{eq:expect_n_from_psi}) implies that the Fisher metric in
Eq.~(\ref{eq:Fisher_element}) is a coordinate transformation from
contravariant to covariant coordinates: 
\begin{equation}
  g_{ij} \! \left( \theta \right) = 
  \frac{
    {\partial}^2 
    \psi \! \left( \theta \right)
  }{
    \partial {\theta}^i
    \partial {\theta}^j
  } = 
  \frac{
    \partial n_i \! \left( \theta \right) 
  }{
    \partial {\theta}^j
  }
\label{eq:g_as_coord_change}
\end{equation}
If $\psi \! \left( \theta \right)$ is convex, the
transformation~(\ref{eq:g_as_coord_change}) is invertible.  The inverse
of the coordinate transform, and with it the Fisher metric, is
obtained from the Legendre transform of $\psi \! \left( \theta \right)$, 
\begin{equation}
  {\psi}^{\ast} \! \left( n \right) \equiv 
  {
    \left[ 
      {\theta}^T n  - 
      \psi \! \left( \theta \right)
    \right] 
  }_{\theta \left( n \right)}
\label{eq:psi_Legend_to_phi}
\end{equation}
where $\theta \left( n \right)$ is the maximizer of the argument in
Eq.~(\ref{eq:psi_Legend_to_phi}) over $\theta$ values.  ${\psi}^{\ast}
\!  \left( n \right)$ is the \textit{Large-Deviation function} (LDF),
which will be used in Sec.~\ref{sec:LDF_and_IS}.

The Legendre transform is constructed to give 
\begin{equation}
  \frac{\partial {\psi}^{\ast} \! \left( n \right)}{\partial n_i} = 
  {\theta}^i \! \left( n \right)
\label{eq:Legend_grad}
\end{equation}
so 
\begin{equation}
  \frac{\partial {\theta}^i}{\partial n_j} = 
  \frac{
    {\partial}^2 {\psi}^{\ast} \! \left( n \right)
  }{
    \partial n_i
    \partial n_j
  } \equiv 
  g^{ij} \! \left( n \right)
\label{eq:g_inv_as_coord_trans}
\end{equation}
the inverse of $g \! \left( \theta \right)$ from
Eq.~(\ref{eq:g_as_coord_change}).  It follows that the distance
element~(\ref{eq:Fisher_element}) can be expressed in the dual
covariant coordinates as 
\begin{equation}
  \delta s^2 = 
  \delta n_i 
  \delta n_j 
  g^{ij} \! \left( n \right)
\label{eq:Fisher_element_ns}
\end{equation}

The Fisher metric can be obtained as the projection of the Euclidean
metric in ${\mathbb R}^D$ under a spherical embedding of the
distribution ${\rho}_{\rn}$, briefly reviewed in
App.~\ref{sec:Fisher_general_sphere}, providing a third set of
coordinates for the tilted distribution~${\tilde{\rho}}^{\left( \theta
\right) }$.  We note this embedding because it provides an interesting
perspective on families of base distributions that are also
exponential, which play several important roles in Doi-Peliti theory,
and which we review next.

\subsubsection{Exponential families on multinomial distributions}

In general, the distribution ${\rho}_{\rn}$ on which one wants to
define an information geometry could have any structure, and could
require arbitrarily much information to specify.  An important
sub-class of distributions, however, are those formed as products of
Poisson marginal distributions over the independent counts ${\rn}_i$,
or sections through such products of marginals.

The Poisson 
\begin{equation}
  {\rho}^{\left( n_i \right)}_{{\rn}_i} = 
  e^{-n_i}
  \frac{
    n_i^{{\rn}_i}
  }{
    {\rn}_i !
  }
\label{eq:Poisson_form}
\end{equation}
is a minimum-information distribution; the expectations of its
\textit{factorial moments}~\cite{Baez:QTRN_eq:14}, defined (again,
component-wise) as ${\rn}^{\underline{k}} \equiv {\rn} ! / {\left( \rn
- k \right)} !$, are $n^k$ for all $k$.  Products of Poisson marginals
or multinomial distributions
\begin{equation}
  {\rho}^{\left( n \right)}_{\rn} = 
  \frac{1}{N^N}
  \left(
    \frac{
      N !
    }{
      {\rn}_1 ! , \ldots , {\rn}_D !
    }
  \right)
  \prod_{i = 1}^D
  n_i^{{\rn}_i} 
\label{eq:multinom_form}
\end{equation}
arise as approximations to more complicated distributions in 2FFI
stationary-point methods, and are also an important class of exact
solutions for some applications such as chemical reaction network
models~\cite{Anderson:product_dist:10}.  The stationary-point
solutions in functional integrals are important whether or not they
provide close approximations, as they define families of coordinate
transformations that will be the basis to construct a dual symplectic
geometry in Sec.~\ref{sec:2_field_Liouville}.  

Distributions of the form~(\ref{eq:multinom_form}) are both mixture
families in the coordinates ${\left\{ n_i \right\}}_{i = 1}^D$, and
exponential families in a suitable coordinate $\eta \propto \log n$
(introduced later in Sec.~\ref{sec:canonical_trans}), which acts
additively with the exponential coordinate $\theta$ of the CGF.  A
consequence of the simplification of the moment hierarchies in
multinomial families is that the Fisher spherical embedding can be
reduced to only $D$ dimensions in the coordinates $\left\{ n_i / N
\right\}$, where $N \equiv \sum_j n_j$, as reviewed in 
App.~\ref{sec:Fisher_reduced_sphere}.  The reduction of functions of
possibly-complicated distributions ${\rho}_{\rn}$ such as
Eq.~(\ref{eq:Fisher_element}) to functions of the same form involving
only their coordinates $n_i$ arises repeatedly in the use of 2FFI
stationary-point methods,\footnote{An example is the reduction of a
complicated similarity transform of a potentially infinite-dimensional
transition matrix for a stochastic process, originally due to Hatano
and Sasa~\cite{Hatano:NESS_Langevin:01}, to a similarity transform of
the same form involving only first-moment values due to
Baish~\cite{Baish:DP_duality:15}, which we will use in
Sec.~\ref{sec:canonical_trans}.} so we note it in passing here.

\subsection{The base and the tilt: inner products between vector
fields describing two sources of variation}

The most basic use of information geometries takes the distance
element~(\ref{eq:Fisher_element}) as a point of departure to consider
the geometry in the Fisher metric on a single exponential family of
distributions ${\tilde{\rho}}^{\left( \theta \right)}_{\rn}$, and
develops the dual Riemannian connections on exponential coordinates
$\theta$ and mixture coordinates $n$ that define parallel transport of
${\tilde{\rho}}^{\left( \theta \right)}_{\rn}$ within that family.
Here we wish to consider families of families, in which generating
functions with coordinates $\theta$ are defined over families of
distributions indexed by independent coordinates.  Derivations of the
Fisher metric from a divergence in that $2D$-dimensional family will
define inner products between vectors in the two $D$-dimensional
subspaces, to which we attach an interpretation in terms of
statistical inference in Sec.~\ref{sec:LDF_and_IS}.

Therefore, in place of ${\rho}_{\rn}$ in
Eq.~(\ref{eq:CGF_onetime_def}), let $\left\{ {\rho}^{\left( n_0
\right)} \right\}$ denote a family of distributions that we will call
\textit{base distributions}, where ${\rho}^{\left( n_0 \right)}$  has
first moment $n_0$.  Over each base distribution define a family of
CGFs and associated tilted distributions, denoted
\begin{align}
  e^{
    \psi \left( \theta , n_0 \right)
  } 
& \equiv 
  \sum_{\rn}
  {\rho}^{\left( n_0 \right)}_{\rn} 
  e^{\theta \cdot \rn}
\nonumber \\ 
  {\tilde{\rho}}^{\left( \theta , n_0 \right)}_{\rn} 
& \equiv 
  {\rho}^{\left( n_0 \right)}_{\rn} 
  e^{
    \theta \cdot \rn - 
    \psi \left( \theta , n_0 \right)
  }
\label{eq:CGF_tilt_n0_def}
\end{align}

The mixed change in $\psi \left( \theta , n_0 \right)$ with two
coordinates $\delta \theta$ in the tilt and $\delta n_0$ in the base,
has an expression as a limit of the extended Pythagorean
theorem~\cite{Nagaoka:dual_geom:82,Amari:methods_IG:00} for
Kullback-Leibler (KL) divergences which is also an inner product in
the Fisher metric:
\begin{align}
& 
  D \! \left(
    {\tilde{\rho}}_{\rn}^{\left( \theta , n_0 + dn_0 \right)} 
  \, \right\rVert \left. 
    {\tilde{\rho}}_{\rn}^{\left( \theta , n_0 \right)}     
  \right) + 
  D \! \left(
    {\tilde{\rho}}_{\rn}^{\left( \theta , n_0 \right)} 
  \, \right\rVert \left. 
    {\tilde{\rho}}_{\rn}^{\left( \theta + d\theta , n_0 \right)}     
  \right) 
\nonumber \\ 
&  \mbox{} - 
  D \! \left(
    {\tilde{\rho}}_{\rn}^{\left( \theta , n_0 + dn_0 \right)} 
  \, \right\rVert \left. 
    {\tilde{\rho}}_{\rn}^{\left( \theta + d\theta , n_0 \right)}     
  \right) 
\nonumber \\ 
& = 
  \sum_{\rn}
  \left(
    {\tilde{\rho}}_{\rn}^{\left( \theta , n_0 + dn_0 \right)} - 
    {\tilde{\rho}}_{\rn}^{\left( \theta , n_0 \right)} 
  \right) 
  \log 
  \left(
    \frac{
      {\tilde{\rho}}_{\rn}^{\left( \theta + d\theta , n_0 \right)}     
    }{
      {\tilde{\rho}}_{\rn}^{\left( \theta , n_0 \right)} 
    }
  \right)
\nonumber \\ 
& \rightarrow  
  \delta {\theta}^i
  \delta n_{0j}
  \frac{
    {\partial}^2 \psi \! \left( \theta , n_0 \right)
  }{
    \partial {\theta}^i
    \partial n_{0j}
  } 
\nonumber \\ 
& \equiv 
  \delta {\theta}^i
  \delta n_{0j}
  \left< 
    \frac{\partial}{\partial {\theta}^i} , 
    \frac{\partial}{\partial n_{0j}}
  \right> 
\label{eq:triangle_to_Fisher}
\end{align}
In the first equality of Eq.~(\ref{eq:triangle_to_Fisher}), $\delta
\theta$ and $\delta n_0$ need not be small; this is the standard
quantity used in the projection theorem defining a notion of
orthogonality to an exponential family involving any three
distributions separated from $\left( \theta, n_0\right)$ by $\delta
\theta$ and $\delta n_0$.  The second, limiting equivalence takes
$\delta \theta \rightarrow 0$ and $\delta n_0 \rightarrow 0$, to
express differences of ${\tilde{\rho}}_{\rn}$ and $\log
{\tilde{\rho}}_{\rn}$ in terms of the mixed second partial derivative
of the KL divergence and hence the Fisher metric.

Recognizing that $\partial \psi / \partial {\theta}^i = n_i \! \left(
\theta , n_0 \right)$, the inner product in the final line of
Eq.~(\ref{eq:triangle_to_Fisher}) is just the sensitivity 
\begin{align}
  \left< 
    \frac{\partial}{\partial {\theta}^i} , 
    \frac{\partial}{\partial n_{0j}}
  \right> = 
  \frac{\partial n_i}{\partial {n_0}_j}
\label{eq:sensitivity}
\end{align}
of the mean in the tilted distribution to variations in the base.

\subsection{Preservation of the inner product in connection with
Liouville's theorem}

A coordinate change from the mean in the base distribution to the mean
in the tilted exponential distribution produces the metric in mixed
coordinates that by construction is the Kronecker $\delta$: 
\begin{align}
  \left< 
    \frac{\partial}{\partial {\theta}^i} , 
    \frac{\partial}{\partial n_j}
  \right> = 
  {\delta}_i^j 
\label{eq:metric_delta}
\end{align}
and the coordinate inner product 
\begin{align}
  \delta {\theta}^i
  \delta n_{0j}
  \left< 
    \frac{\partial}{\partial {\theta}^i} , 
    \frac{\partial}{\partial n_{0j}}
  \right> 
& \equiv 
  \delta {\theta}^i
  \delta n_i
\label{eq:innerprod_expfam}
\end{align}

We may ask, for what one-parameter families of coordinate systems
$\left( \theta \! \left( t \right) , n \! \left( t \right) \right)$ is
the coordinate inner product~(\ref{eq:metric_delta}) conserved across
the family?  A one-parameter family of coordinates generates a
one-parameter family of maps of vector fields $\delta \theta$ and
$\delta n$ by the action
\begin{align}
  {
    \left( 
      \frac{d}{dt} \delta \theta
    \right)
  }^i
& = 
  \delta {\theta}^j
  \frac{\partial}{\partial {\theta}^j}
  {\dot{\theta}}^i
\nonumber \\ 
  {
    \left( 
      \frac{d}{dt} \delta n
    \right) 
  }_i
& = 
  \delta n_j
  \frac{\partial}{\partial n_j}
  {\dot{n}}_i
\label{eq:edge_change_rates}
\end{align}
The condition 
\begin{equation}
  \frac{d}{dt}
  \left( 
    \delta {\theta}^i \, 
    \delta n_i 
  \right) = 
  0 
\label{eq:inner_prod_vanish}
\end{equation}
will be met whenever 
\begin{equation}
  \frac{ 
    \partial {\dot{\theta}}^i
  }{
    \partial {\theta}^j
  } = 
  - \frac{
    \partial {\dot{n}}_j
  }{ 
    \partial n_i
  } 
\label{eq:symplect_cond}
\end{equation}
where $\cdot$ denotes $d/dt$.  Eq.~(\ref{eq:symplect_cond}) is
satisfied if there is a symplectic form $\mathcal{L} \! \left( \theta
, n \right)$ in terms of which the velocity vectors along trajectories
can be written
\begin{align}
  {\dot{\theta}}_i
& = 
  \frac{\partial}{\partial n_i}
  \mathcal{L} \! \left( \theta , n \right)
\nonumber \\ 
  {\dot{n}}_i
& = 
  - \frac{\partial}{\partial {\theta}_i}
  \mathcal{L} \! \left( \theta , n \right)
\label{eq:Hamiltonian_var}
\end{align}
Alternatively, if Eq.~(\ref{eq:symplect_cond}) holds everywhere, the
form $\mathcal{L} \! \left( \theta , n \right)$ can be constructed by
integration. 

Eq.~(\ref{eq:symplect_cond}) relates the dual contravariant and
covariant coordinates under the Fisher metric as canonically conjugate
variables in a Hamiltonian dynamical system.  We return in
Sec.~\ref{sec:DP_gen_review} to derive symplectic forms $\mathcal{L}
\! \left( \theta , n \right)$ from the generators of stochastic
processes, but first we note a sampling interpretation of the
generating functions~(\ref{eq:CGF_onetime_def}) that will provide
intuition for the meaning of transporting an invariant inner
product~(\ref{eq:metric_delta}) along trajectories.

\section{Finite-system models as sample estimators; the
Large-Deviation function, and importance sampling}
\label{sec:LDF_and_IS}

\subsubsection*{Why develop an Importance Sampling interpretation of
generating functions? -- context from the wider applications of 2FFI
duality} 

Looking ahead to Sec.~\ref{sec:DP_gen_review}, the 2-field structure
in Doi-Peliti theory that we will describe geometrically in terms of
dual connections and the Fisher metric reflects a natural duality that
is present in the quadrature of any time-dependent stochastic process.
The integral of the generator of time translations over any finite
time interval allows expectations of random variables at a later time
to be evaluated in measures derived by time evolution of probability
distributions specified at earlier times.  The evolution kernel that
connects the two may be regarded either as a forward-time propagator
of probability distributions or a reverse-time propagator of the
random variables.  The duality between these two interpretations of
the time-evolution kernel was developed by Kolmogorov in his study of
the ``backward'' generator or its (adjoint) ``forward''
generator~\cite{Gardiner:stoch_meth:96}, and is the same as the
equivalence of the Schr{\"{o}}dinger picture (time evolution of
states) and the Heisenberg picture (time evolution of operators) in
quantum mechanics~\cite{Cohen-Tannoudji:QM:77}.

A quite large literature has sprung up over the past 20 years making
use of this forward/backward duality of time-dependent stochastic
processes, focused on how path weights in time-dependent generating
functionals may be used to exchange the roles of the generator and its
adjoint.\footnote{It is impossible to fairly represent the motivations
and scope of what has now become a significant fraction of work
spanning dynamical systems and statistical mechanics.  The study of
generating functions for reverse-time trajectories began in dynamical
systems~\cite{Evans:shear_SS:93,Gallavotti:dyn_ens_NESM:95,Gallavotti:dyn_ens_SS:95,Cohen:NESM_2_thms:99},
and was later taken up in similar form for non-equilibrium stochastic
processes~\cite{Jarzynski:eq_FE_diffs:97,Jarzynski:neq_FE_diffs:97,Kurchan:fluct_thms:98,Searles:fluct_thm:99,Crooks:NE_work_relns:99,Crooks:path_ens_aves:00,Hatano:NESS_Langevin:01,Chernyak:PI_fluct_thms:06,Kurchan:NEWRs:07,Jarzynski:fluctuations:08,Esposito:fluct_theorems:10}.
Reviews of parts of this literature from different stages in its
development and from different domain perspectives
include~\cite{Evans:fluct_thm:02,Harris:fluct_thms:07,Chetrite:fluct_diff:08,Seifert:stoch_thermo_rev:12}.}
In that literature, when ``physical
interpretations''~\cite{Seifert:stoch_thermo_rev:12} are assigned to
backward propagation, the assignment is done in terms of time-reversal
of physically traversed paths.  Such an interpretation is
inappropriate for the analysis we wish to provide of Liouville's
theorem and the symplectic structure of 2FFI constructions, on
multiple grounds: i) it is needlessly restrictive: we wish to study
stochastic processes in which paths together with their time-reverses
may or may not be defined within the dynamics; ii) dynamical reversal
is not fundamental to the Kolmogorov forward/backward duality: the
sense in which the time-reversed propagation of random variables is
``anti-causal'' is inherent in the adjoint relation
itself~\cite{Smith:LDP_SEA:11}.  Any mapping onto physical
time-reversal depends on the strong and
\emph{independent} requirement that a system's trajectory space
contain an image of its own adjoint,\footnote{The one widely-developed
interpretation of generating-function duality not based on explicit
trajectory reversal is the excess-heat theory by Hatano and
Sasa~\cite{Hatano:NESS_Langevin:01,Verley:HS_FDT:12}.  Even here,
however, the heat interpretation depends on microscopic reversibility
and local equilibrium to assign interpretations of heats associated
with maintaining non-equilibrium steady states and excess heats
associated with changes of non-equilibrium state.  Thus dynamical
reversal is still intrinsic to the interpretation that has led this
construction to be widely used.} which we do not generally wish to
impose; iii) the physical time-reversal interpretation
\emph{substitutes} time-reversed for forward dynamics; to understand
Liouville's theorem in 2FFI theories, both maps under the forward
generator and its adjoint must be co-present and meaningful.

It is in order to furnish an interpretation of the 2FFI symplectic
transport structure, comparable to the phase-space density transport
interpretation for Hamiltonian mechanics, that we appeal to
statistical inference to assign informational meanings to the tilt
weights used in generating functions.  Beyond simply using the
generating function as a mathematical device to extract moments
from probability distributions, an importance sampling application
gives the specific interpretation of a likelihood ratio to
non-uniform as well as uniform weights as a transformation of measure
for samples.  It is then easy to understand how dual forward and
backward generators can jointly propagate the images of regions in a
probability distribution and regions of concentration or dilution
under the likelihood ratio through time, and how densities of rays
in these paired coordinates can reflect conserved information about
the expectations of random variables in evolving distributions.  This
section notes the main steps in the importance-sampling
interpretation.

\subsection{States as samples; system scaling and sample aggregation} 

Statistical inference concerns the distributions and convergence
properties of sample estimators for the parameters that define some
underlying process, as a function of the scaling of samples under some
aggregation rule.  

There is no formal distinction between the ensemble output of a
stochastic process as a model of the distribution of states taken by a
finite-particle system on a discrete state space, and the output of
the same process considered as a distribution of samples from the
generating process.  If sequences of states are modeled, the same
equivalence holds for the stochastic process as a specifier of samples
of trajectories.

The state of a system with multiple components defines a notion of the
size of a sample from the generating process, and a rule for changing
the number of components (\textit{e.g.} the population size) defines
an aggregation procedure over samples.  The aggregation rule may
correspond to simple repeated sampling from subsystems without
replacement, or it may define a class of independent scaling
behaviors, as the independent variation of different conserved
quantities of the stoichiometry in a chemical reaction network does.
The Laplace transform to the generating function, and its Legendre
transform to the Large-Deviations Function, have direct
interpretations in terms of sampling procedures and exponential
scaling approximations in importance sampling.

\subsection{Legendre transform and  large deviations in the
interpretation of importance sampling}
\label{sec:indicator_LDF}

In the terminology of importance sampling~\cite{Owen:mcbook:13}, the
base distribution ${\rho}^{\left( n_0 \right)}$ corresponds to the
\textit{nominal} distribution, and the tilted (and normalized)
distribution ${\tilde{\rho}}_{\rn}^{\left( \theta , n_0 \right)}$ of
Eq.~(\ref{eq:CGF_tilt_n0_def}) plays the role of an \textit{importance
distribution}. The combined tilt and normalization $e^{{\theta}^T \rn
- \psi \left( \theta , n_0 \right)}$ is the corresponding
\textit{likelihood ratio}, also called the \textit{Radon-Nikodym
derivative} of the measure between the base and the importance
distributions.

Importance distributions ${\tilde{\rho}}^{\left( \theta , n_0
\right)}$ can be chosen to concentrate the density of samples away
from the mode of ${\rho}^{\left( n_0 \right)}$ to values of $\rn$ that
are more informative about observables of interest.  Tilts are
typically chosen to minimize some cost function, such as the variance
of samples.  The large-deviation function can be derived as a leading
exponential approximation to the tail weight of the base distribution,
in a protocol tuned to minimize sample variance, as shown in the
following construction from~\cite{Siegmund:IS_seq_tests:76}.

To illustrate with an example in one dimension, an estimate of the
probability that a particle count $\rn$ exceeds some bound $\bar{n}$
can be obtained by sampling values of the random variable 
\begin{equation}
  h^{\left( \bar{n} \right)}_{\rn} \equiv 
  I \! \left\{ \rn > \bar{n} \right\} 
\label{eq:h_interval_def}
\end{equation}
the indicator function for $\rn > \bar{n}$.  In the base
distribution ${\rho}^{\left( n_0 \right)}$, the probability for $\rn >
\bar{n}$ is
\begin{equation}
  P \! \left( \rn > \bar{n} \mid n_0 \right) = 
  {\left< h^{\left( \bar{n} \right)} \right>}_{\left( n_0 \right)}
\label{eq:esc_prob_0}
\end{equation}

An unbiased estimator for $P \! \left( \rn > \bar{n} \mid n_0 \right)$
can be obtained by using the tilted distribution
${\tilde{\rho}}_{\rn}^{\left( \theta , n_0 \right)}$ of
Eq.~(\ref{eq:CGF_tilt_n0_def}) and instead of accumulating the values
$\left\{ 0 , 1 \right\}$ of the indicator $h^{\left( \bar{n}
\right)}$, accumulating values of the tilted
observable\footnote{Normally an un-normalized tilted measure and its
compensating observable are defined only in terms of the exponential
weight $e^{\theta \rn}$, since $\psi \left( \theta , n_0 \right)$ is
not known.  Here to simplify the calculations and avoid introducing
further notations, we include the CGF and work with the normalized
distribution ${\tilde{\rho}}^{\left( \theta , n_0
\right)}$.}
\begin{align}
  {\tilde{h}}^{\left( \theta , \bar{n} \right)}_{\rn}
& \equiv 
  e^{
    \psi \left( \theta , n_0 \right) - \theta \rn 
  }
  h_{\rn}
\label{eq:IS_tilts}
\end{align}
The tilted estimator is unbiased because 
\begin{equation}
  {
    \left< {\tilde{h}}^{\left( \theta , \bar{n} \right)} \right>
  }_{
    \left( \theta , n_0 \right)
  } = 
  {\left< h^{\left( \bar{n} \right)} \right>}_{\left( n_0 \right)} 
\label{eq:IS_no_bias}
\end{equation}
A few lines of algebra, provided in App.~\ref{sec:app_IS_LDF_alg},
show that an exponential bound for the estimator at any choices of
$\theta$ and $\bar{n}$ is given by 
\begin{align}
  {
    \left< {\tilde{h}}^{\left( \theta , \bar{n} \right)} \right>
  }_{
    \left( \theta , n_0 \right)
  } 
& \le 
  e^{
    \psi \left( \theta , n_0 \right) - \theta \bar{n} 
  }
\label{eq:IS_est_bound_short}
\end{align}

The variance of the same sample estimator has a corresponding bound
(see Eq.~(\ref{eq:IS_var_def}))
\begin{align}
& \lefteqn{
  {
    \left< 
      { \left( {\tilde{h}}^{\left( \theta , \bar{n} \right)} \right) }^2
    \right>
  }_{\left( \theta , n_0 \right)} - 
  {
    \left< {\tilde{h}}^{\left( \theta , \bar{n} \right)} \right>
  }_{
    \left( \theta , n_0 \right)
  }^2 
} 
\nonumber \\ 
& \le  
  e^{
    \psi \left( \theta , n_0 \right) - \theta \bar{n} 
  }
  {\left< h^{\left( \bar{n} \right)} \right>}_{\left( n_0 \right)} - 
  {\left< h^{\left( \bar{n} \right)} \right>}_{\left( n_0 \right)}^2 
\label{eq:IS_var_def_short}
\end{align}
The parameter $\theta$ that minimizes the bound on sample
variance~(\ref{eq:IS_var_def_short}) also gives the tightest
bound~(\ref{eq:IS_est_bound_short}) on the tail weight.  It is the
minimizing argument $\theta \! \left( \bar{n} \right)$ of
Eq.~(\ref{eq:psi_Legend_to_phi}), so the bound is given in terms of
the LDF as.
\begin{align}
  {\left< h^{\left( \bar{n} \right)} \right>}_{\left( n_0 \right)} 
& \le 
  e^{- {\psi}^{\ast} \left(\bar{n} \right)}
\label{eq:IS_est_bound_LDF}
\end{align}

Without further assumptions about ${\rho}^{\left( n_0 \right)}$ it is
not possible to say more about the ratio ${\left< h^{\left( \bar{n}
\right)} \right>}_{\left( n_0 \right)} / e^{- {\psi}^{\ast} \left(\bar{n}
\right)}$.  The relevant additional property, which is also associated
with the use and tightness of saddle-point
approximations~\cite{Goutis:saddlepts:95} in Doi-Peliti theory, is the
onset of large-deviations scaling; that is, if $n_0$ and $\bar{n}$ are
increased together in proportion to some scale factor $N$ as $\bar{n}
= N \bar{\nu}$, $n_0 = N {\nu}_0$, the following two limits should
exist:
\begin{align}
  \lim_{N \rightarrow \infty}
  \frac{1}{N}
  {\psi}^{\ast} \! \left( N \bar{\nu} \right) 
& = 
  {\bar{\psi}}^{\ast} \! 
  \left( \bar{\nu} \right)
\nonumber \\ 
  \lim_{N \rightarrow \infty}
  \frac{1}{N}
  \psi \! 
  \left( 
    \theta \! \left( N \bar{\nu} \right) ; 
    N {\nu}_0
  \right) 
& = 
  \bar{\psi} \! \left( \bar{\theta} \right)
\label{eq:LD_scaling}
\end{align}
Then the variance-minimizing tilt $\theta$ likewise has a limit, the
variance ${\partial}^2 \psi / \partial {\theta}^i \partial {\theta}^j$
in Eq.~(\ref{eq:Fisher_element}) scales as $N$, and the relative
variance scales as $1/N$.  App.~\ref{sec:app_IS_LDF_alg} shows that in
this limit the log ratio $\log \left[ {\left< h^{\left( \bar{n}
\right)} \right>}_{\left( n_0 \right)} / e^{- {\psi}^{\ast} \left(\bar{n}
\right)} \right] \le \mathcal{O} \! \left( N^{1/2} \right)$, compared
to ${\psi}^{\ast} \! \left( \bar{n} \right) \sim N$.

\section{Doi-Peliti 2-field functional integrals, and dual mappings
induced by time translation}
\label{sec:DP_gen_review}

A 2FFI formalism to compute time-dependent generating functions and
functionals for stochastic processes with discrete state spaces has
been developed based on the operator linear algebra for
moment-generating functions due to
Doi~\cite{Doi:SecQuant:76,Doi:RDQFT:76}, and a coherent-state basis
expansion due to Peliti~\cite{Peliti:PIBD:85,Peliti:AAZero:86} that
converts the Doi time-evolution operator into a functional integral.
The Doi-Peliti method is now widely known~\cite{Baez:QTSM:17}, so we
will review here only definitions of essential terms and the detailed
construction of the core elements, most importantly the coherent-state
representation of the identity operator acting on generating
functions.  App.~\ref{sec:DP_review} provides some further supporting
definitions and algebra.  Self-contained introductory derivations
following the notation used here are available
in~\cite{Smith:LDP_SEA:11,Smith:evo_games:15,Krishnamurthy:CRN_moments:17,%
Smith:CRN_moments:17}.

\subsection{Time evolution of moment- and cumulant-generating
functions} 

The essential geometric constructions of
Sec.~\ref{sec:int_lattice_CGF_IG} for tilted distributions, and the
importance-sampling interpretation of Sec.~\ref{sec:LDF_and_IS}, are
defined on distributions ${\rho}_{\rn}$ without reference to any
notion of embedding in a dynamical system.  Here we add the feature of
a continuous-time master equation
\begin{align}
  \frac{d {\rho}_{\rn}}{dt} 
& = 
  \sum_{{\rn}^{\prime}}
  {\rm T}_{\rn {\rn}^{\prime}} 
  {\rho}_{{\rn}^{\prime}} .
\label{eq:ME_genform}
\end{align}
that evolves densities ${\rho}_{\rn}$ along a coordinate $t$.  In
physical applications $t$ is time, but for the purpose of this paper,
$t$ may be any one-dimensional coordinate along which we can define a
continuous mapping of probability distributions, or even just a
continuous family of coordinate transformations in which to describe
the geometry of a single distribution regarded as fixed.  ${\rm
T}_{\rn {\rn}^{\prime}}$ is known as the \textit{transition matrix},
and is the representation of the generator of the stochastic process
acting in the space of probability distributions.  It is
left-stochastic, meaning $\sum_{\rn} {\rm T}_{\rn {\rn}^{\prime}} =
1$, $\forall {\rn}^{\prime}$, ensuring conservation of probability.
The matrix elements ${\rm T}_{\rn {\rn}^{\prime}}$ can depend on the
time $t$, though in the example developed in
Sec.~\ref{sec:example_two_state} we will use a time-independent
generator for simplicity.

\subsubsection{Laplace transform converts that master equation on
distributions to a Liouville equation on moment-generating functions} 

The Doi-Peliti method works not with the distribution ${\rho}_{\rn}$
in the discrete basis, but with the moment-generating
function~(\ref{eq:CGF_onetime_def}) that is its Laplace transform.
From the form of the transition matrix ${\rm T}_{\rn {\rn}^{\prime}}$
it is possible to derive a
\emph{Liouville equation} (see any
of~\cite{Smith:LDP_SEA:11,Smith:evo_games:15,Krishnamurthy:CRN_moments:17,%
Smith:CRN_moments:17}) for the time-dependence of the MGF, 
\begin{align}
  \frac{\partial}{\partial t} 
  \Psi \! \left( z \right)
& = 
  - \mathcal{L} \! 
  \left( z , \frac{\partial}{\partial z} \right)
  \Psi \! \left( z \right) , 
\label{eq:Liouville_eq_multi_arg}
\end{align}
$\mathcal{L} \! \left( z , \partial / \partial z \right)$, called the
\textit{Liouville operator}, is the representation of the generator of
the stochastic process acting in the space of generating functions,
and is conventionally defined with the minus sign of
Eq.~(\ref{eq:Liouville_eq_multi_arg}) because its spectrum is
non-negative.  The Liouville operator will play the role of a
Hamiltonian as a generator of symplectomorphisms in the derivation
below.

\subsubsection{The Doi operator representation of the Hilbert space
of generating functions}

For many purposes, the properties of the generating function as an
analytic function of a complex variable are not needed, and the
algebra of the MFG as a formal power series is sufficient.  To capture
only this algebra, in a notation that is more convenient than that of
analytic functions for computing the quadrature of
Eq.~(\ref{eq:Liouville_eq_multi_arg}), the Doi
formalism~\cite{Doi:SecQuant:76,Doi:RDQFT:76} replaces the variable
$z$ and derivative $\partial / \partial z$ with formal raising and
lowering operators 
\begin{align}
  z_i 
& \rightarrow 
  a^{\dagger}_i ; & 
  \frac{\partial}{\partial z_i} 
& \rightarrow 
  a^i 
\label{eq:a_adag_defs}
\end{align}
as are conventionally used in quantum mechanics or quantum field
theory~\cite{Baez:QTSM:17}.  In the condensed
notation~(\ref{eq:CGF_onetime_def}) for vector inner products, we will
regard $a$ as a column vector and $a^{\dagger}$ as a row vector.

Associated with operators $a^{\dagger}_i$ and $a^i$ are bilinear
\textit{number operators} ${\hat{n}}_i \equiv a^{\dagger}_i a^i$ (no
Einstein sum), of which the basis monomials $z^{\rn}$ under the
mapping~(\ref{eq:a_adag_defs}) correspond to number eigenstates.  A
Hilbert space of generating functions, and an inner product
corresponding to the evaluation of the MGF at argument $z = 1$, are
defined and given a standard bracket notation reviewed in
App.~\ref{sec:DP_review}.  Number states are denoted $\left| \rn
\right)$, and the MGF $\Psi \! \left( z \right)$ is represented as a
state vector $\left| \Psi \right)$ defined in terms of number states
as 
\begin{equation}
  \sum_{\rn}
  {\rho}_{\rn} 
  \left| \rn \right) \equiv 
  \left| \Psi \right)
\label{eq:MGF_abstract}
\end{equation}
The Liouville equation~(\ref{eq:Liouville_eq_multi_arg}) becomes 
\begin{align}
  \frac{d}{dt} 
  \left| \Psi \right) 
& = 
  - \mathcal{L} \! 
  \left( a^{\dagger} , a \right)
  \left| \Psi \right) . 
\label{eq:Liouville_eq_aadag}
\end{align}
in which the the Liouville operator $\mathcal{L} \! \left( a^{\dagger}
, a \right)$ is the former function $\mathcal{L} \! \left( z ,
\partial / \partial z \right)$ under the
substitution~(\ref{eq:a_adag_defs}). 

The analytic form~(\ref{eq:CGF_onetime_def}) of the MGF can be
recovered using a variant of the \emph{Glauber norm} that defines the
inner product on the Hilbert space, as
\begin{align}
  \left( 0 \right|
  e^{z a}
  \left| \Psi \right) = 
  \Psi \! \left( z \right) . 
\label{eq:Glauber_to_FG}
\end{align}
In Eq.~(\ref{eq:Glauber_to_FG}), $z$, like $a^{\dagger}$, is regarded
as a row vector and $z a$ is the scalar product of $z$ and $a$.  We
will need the analytic form to relate the Doi-Peliti functional
integral to the coordinates $\theta$ in which the geometry of
Sec.~\ref{sec:int_lattice_CGF_IG} has been constructed.

\subsection{The 2FFI representation of the identity on distributions}

\subsubsection{Coherent states and the Peliti construction of the
functional integral}

The objective in introducing the Doi operator formalism is to more
conveniently compute the quadrature of the Liouville
equation~(\ref{eq:Liouville_eq_multi_arg}), formally written 
\begin{align}
  {\left| \Psi \right)}_T 
& = 
  \mathcal{T}
  e^{
    - \int_0^T dt 
    \mathcal{L} \left( a^{\dagger}, a \right)
  }
  {\left| \Psi \right)}_0
\nonumber \\ 
& \equiv 
  \lim_{\delta t \rightarrow 0}
  \mathcal{T}
  \prod_{k = 1}^{T / \delta t}
  e^{
    - \delta t 
    {
      \mathcal{L} \left( a^{\dagger}, a \right)
    }_{k \delta t}
  }
  {\left| \Psi \right)}_0
\label{eq:Liouville_quadrature}
\end{align}
${\left| \Psi \right)}_T$ is the generating function for the
distribution ${\rho}_{\rn}$ evolved to time $t = T$ from the
generating function ${\left| \Psi \right)}_0$ for an initial
distribution given at time $t = 0$.  $\mathcal{T}$ denotes
time-ordering of the exponential integral, defined operationally in
the second line of Eq.~(\ref{eq:Liouville_quadrature}), in terms of a
time-ordered product of applications of $\mathcal{L} \! \left(
a^{\dagger}, a \right)$ evaluated at the sequence of times $k \delta
t$. 

The 2FFI method of solution due to
Peliti~\cite{Peliti:PIBD:85,Peliti:AAZero:86} makes use of the
\textit{coherent states} as a basis for the expansion of arbitrary
generating functions.  Coherent states, defined in terms of a (column)
vector $\phi$ of complex coefficients by 
\begin{align}
  \left| \phi \right) 
& \equiv 
  e^{
    \left( a^{\dagger} - 1 \right) \phi
  } 
  \left| 0 \right) 
\label{eq:coh_state_def}
\end{align}
are the generating functions of products of Poisson distributions with
component-wise mean values ${\phi}^i$, and eigenstates of the lowering
operator:
\begin{equation}
  a^i 
  \left| \phi \right) = 
  {\phi}^i 
  \left| \phi \right) 
\label{eq:coh_st_eig_a}
\end{equation}

An essential feature of the of the Doi Hilbert space and inner
product, explained in App.~\ref{sec:DP_review}, is that dual to each
basis basis vector in the number basis (corresponding to the monomials
$z^{\rn}$) is a projection operator extracting the amplitude with
which that state appears in $\left| \Psi \right)$.  For the number
basis these are just the probability values ${\rho}_{\rn}$.  In the
same manner, dual to coherent states there are projection operators
defined in terms of (row) vectors ${\phi}^{\dagger}$ of complex
coefficients that are the complex conjugates of the components of
$\phi$.  Using the normalization consistent with
Eq.~(\ref{eq:coh_state_def}) for $\left| \phi \right)$ the dual
projectors are defined as 
\begin{align}
  \left( \phi \right|
& \equiv 
  e^{
    \left( 1 - {\phi}^{\dagger} \right) 
    \phi 
  }
  \left( 0 \right|
  e^{
    {\phi}^{\dagger} a 
  } . 
\label{eq:dual_phi_states}
\end{align}

Unlike the number states and their dual projectors, which are
complete, the coherent states and their dual projectors are
\emph{over-complete}.  The inner product of a state at parameter
${\phi}_1$ with a projector at parameter ${\phi}^{\dagger}_2$ is given
by 
\begin{align}
  \left( {\phi}^{\dagger}_2 \, \right| \! 
  \left. \vphantom{{\phi}^{\dagger}_2} {\phi}_1 \right)
& = 
  e^{
    - {\phi}^{\dagger}_2 
    \left( {\phi}_2 - {\phi}_1 \right) 
  }
\label{eq:inner_prod_21}
\end{align}
The inner product~(\ref{eq:inner_prod_21}) is very important because
it is the source of a ``kinetic term'' analogous to $p \dot{q}$ in
classical Hamiltonian mechanics that causes the Liouville operator to
behave as a Hamiltonian in the Doi-Peliti field theory.

Although they are overcomplete, the coherent states and their dual
projectors furnish a representation of the identity in the space of
generating functions, as shown in App.~\ref{sec:Peliti_CS_alg}, 
\begin{align}
  \int 
  \frac{
    d^D \! {\phi}^{\dagger}
    d^D \! \phi
  }{
    {\pi}^D
  }
  \left| \phi \right) 
  \left( \phi \right| 
& = 
  I 
\label{eq:field_int_ident_short}
\end{align}
When a copy of the identity~(\ref{eq:field_int_ident_short}) is
inserted into the quadrature~(\ref{eq:Liouville_quadrature}) between
each increment of evolution of length $\delta t$, and the generating
function is evaluated at argument $z$ using the inner
product~(\ref{eq:Glauber_to_FG}), the resulting MGF or CGF of
Eq.~(\ref{eq:CGF_onetime_def}) can be written as the functional
integral 
\begin{align}
  e^{{\psi}_T \left( \theta \right)} = 
  \int_0^T 
  {\mathcal{D}}^D \! {\phi}^{\dagger} 
  {\mathcal{D}}^D \! {\phi} \, 
  e^{
    \left( z - {\phi}^{\dagger}_T \right) {\phi}_T - 
    S + 
    {\psi}_0 
    \left( 
      \log {\phi}^{\dagger}_0 
    \right)
  } 
\label{eq:gen_fn_twofields_nosource}
\end{align}

$e^{{\psi}_0 \left( \log {\phi}^{\dagger}_0 \right)}$ is the
coherent-state expansion of the initial generating function ${\left|
\Phi \right)}_0$ in Eq.~(\ref{eq:Liouville_quadrature}).  $S$ in
Eq.~(\ref{eq:gen_fn_twofields_nosource}) has the form of a
Lagrange-Hamilton \textit{action} functional, 
\begin{align}
  S 
& = 
  \int_0^T dt
  \left\{ 
    - \left( d_t {\phi}^{\dagger} \right)
    \phi + 
    \mathcal{L} \! \left( {\phi}^{\dagger} , \phi \right)
  \right\} 
\label{eq:CRN_L_genform}
\end{align}
in which the kinetic term comes from the log of the inner
product~(\ref{eq:inner_prod_21}) in the continuous-time limit, and the
Liouville operator $\mathcal{L} \! \left( {\phi}^{\dagger} , \phi
\right)$, with ${\phi}^{\dagger}$ and $\phi$ replacing operators
$a^{\dagger}$ and $a$, playing the role of the Hamiltonian.  The
action~(\ref{eq:CRN_L_genform}) will be the source of the
relation~(\ref{eq:Hamiltonian_var}) leading to a conserved volume
element and Liouville theorem in the information geometry for
dissipative stochastic processes.

\subsubsection{The Peliti functional integral as a statistical model}

The Peliti basis of coherent states~(\ref{eq:coh_state_def}) defines a
\textit{statistical model} for the stochastic process.  The role of
the projection operators in the representation of
unity~(\ref{eq:field_int_ident_short}) in populating the model can be
clarified by splitting the functional
integral~(\ref{eq:CRN_L_genform}) at any intermediate time $t$, in the
same fashion as the Chapman-Kolmogorov equation splits the time
evolution of a probability distribution by a sum over intermediate
states.  This is done by integrating
Eq.~(\ref{eq:gen_fn_twofields_nosource}) up to time $t$, inserting an
explicit representation of unity in terms of a pair of fields $\left(
{\phi}_{\ddagger}^{\dagger} , {\phi}_{\ddagger} \right)$, and resuming
the functional integral on the generating function extracted by that
representation of unity:
\begin{widetext}
\begin{align}
  e^{{\psi}_T \left( \theta \right)} = 
  \int_{t + \delta t}^T 
  {\mathcal{D}}^D \! {\phi}^{\dagger}
  {\mathcal{D}}^D \! {\phi} \, 
  e^{
    \left( z - {\phi}^{\dagger}_T \right) {\phi}_T - 
    S 
  } 
  \int d^D \! {\phi}_{\ddagger}
  e^{
    \left( 
      {\phi}^{\dagger}_{t + \delta t} - 1 
    \right) 
    {\phi}_{\ddagger}
  }
  \int 
  \frac{
    d^D \! {\phi}^{\dagger}_{\ddagger}
  }{
    {\pi}^D
  }
  e^{
    - \left( {\phi}^{\dagger}_{\ddagger} - 1 \right) 
    {\phi}_{\ddagger}
  }
  e^{
    {\psi}_t 
    \left( 
      \log {\phi}^{\dagger}_{\ddagger}
    \right)
  } 
\label{eq:gen_fn_twofield_CKdecomp}
\end{align}
\end{widetext}

In Eq.~(\ref{eq:gen_fn_twofield_CKdecomp}) the argument $\log
{\phi}^{\dagger}_{\ddagger}$ is an affine tilt coordinate in an
exponential family of generating functions on whatever base
distribution the functional integral produces at time $t$.  The
(component-wise) product ${\phi}^{\dagger}_{t + dt} {\phi}_{\ddagger}$
(see also Sec.~\ref{sec:orig_action_angle} below) is a dual mixture
coordinate, corresponding to the mean of $n$ in the importance
distribution with $\left| {\phi}_{\ddagger} \right)$ as the base
distribution and ${\phi}^{\dagger}_{t + dt}$ as the likelihood
ratio.  It is the mean of this importance distribution, together
with the log-likelihood that is its dual coordinate in the exponential
family, that must transform under symplectomorphism to satisfy the
condition~(\ref{eq:Hamiltonian_var}) for preservation of inner
products.  The next section shows that the stationary-path conditions
provide the necessary mapping.

\subsection{Stationary paths of the 2-field action functional}

A deterministic approximation to the mean values through time in the
base distribution, and to the mean weights in the likelihood ratio, is
given by the saddle point of $e^{-S}$ in
Eq.~(\ref{eq:gen_fn_twofields_nosource}).  The functional has a saddle
point where the variational derivative of $S$ vanishes, leading to the
stationary-path equations of motion
\begin{align}
  {\dot{\phi}^{\dagger}}_i
& = 
  \frac{\partial}{\partial {\phi}_i}
  \mathcal{L} \! \left( {\phi}^{\dagger} , \phi \right)
\nonumber \\ 
  {\dot{\phi}}_i
& = 
  - \frac{\partial}{\partial {\phi}^{\dagger}_i}
  \mathcal{L} \! \left( {\phi}^{\dagger} , \phi \right)
\label{eq:Hamiltonian_var_CS}
\end{align}
which are of Hamiltonian form.

The final-time boundary value for the field ${\phi}^{\dagger}$ is
given by the vanishing derivative of the exponent in
Eq.~(\ref{eq:gen_fn_twofields_nosource}) with respect to ${\phi}_T$,
resulting in ${\phi}^{\dagger}_T = z$.  The initial time boundary
value for $\phi$ is given, after an integration by parts, by the
vanishing derivative with respect to ${\phi}^{\dagger}_0$, and depends
on the form of the CGF ${\psi}_0$ and the stationary-path value of its
argument ${\phi}^{\dagger}_0$.  The two conditions are solved
self-consistently through the equations~(\ref{eq:Hamiltonian_var_CS}).

As the discussion following Eq.~(\ref{eq:gen_fn_twofield_CKdecomp})
highlights, joint stationary values at intermediate times $t$, in a
generating function with argument $z$ imposed at a final time $T$,
represent the bundle of rays in the statistical model that dominate
the contribution to the importance distribution at later times.  For
this reason the stationary trajectory in the base distribution is not
generally independent of the trajectory for the tilt in systems with
non-linear equations of motion.  The ray interpretation is well
developed in Freidlin-Wentzell theory~\cite{Freidlin:RPDS:98} for
eikonal approximations to boundary-value problems for diffusion
equations, and with respect to 2FFI solutions for first-passage times
and escape 
trajectories~\cite{Maier:escape:93,Maier:scaling:96,Maier:oscill:96,%
Maier:exit_dist:97}.

Our use of the stationary-path equations~(\ref{eq:Hamiltonian_var_CS})
will be to define a 1-parameter family of coordinate transformations
in the manifold of coordinates for base distributions and likelihood
functions.  These may be viewed as dynamical maps of distributions
through time, or they may simply be used as alternative coordinate
systems in which to evaluate inner products of vector fields on a
fixed distribution.

\subsection{Canonical transformations of the field variables of
integration} 
\label{sec:canonical_trans}

An important class of changes of variable in 2-field integrals are
those corresponding to the \textit{canonical transformations} in
Hamiltonian mechanics~\cite{Goldstein:ClassMech:01}.  The canonical
transformations preserve the form of the kinetic term in the
action~(\ref{eq:CRN_L_genform}), and thus the separation into
conjugate field pairs with a preserved volume element.  Three
canonical transformations are heavily used in 2FFI generating
functions, and each of them plays a role in our construction of a dual
symplectic geometry.

\subsubsection{An action-angle transform from coherent-state variables
to number-potential variables}
\label{sec:orig_action_angle}

A transformation from coherent-state fields to what would be
``action-angle'' variables in classical mechanics is defined by 
\begin{align}
  {\phi}_i^{\dagger}
& \equiv 
  e^{{\theta}^i}
& 
  {\phi}_i
& \equiv 
  e^{-{\theta}^i}
  n_i 
\label{eq:AA_std_defs}
\end{align}
$n$, corresponding to the bilinear ${\phi}^{\dagger} \phi$ has the
interpretation of the number coordinate in the importance
distribution, and its conjugate $\theta$ has the interpretation of a
potential, such as the chemical potential in a chemical-reaction
application~\cite{Smith:LDP_SEA:11,Krishnamurthy:CRN_moments:17,%
Smith:CRN_moments:17}. $\theta$ also corresponds directly to the
coordinate $\log z$ which is the affine coordinate in the exponential
family of tilted distributions~(\ref{eq:rho_tilt_theta}).  $n$ and
$\theta$ are thus dual mixture and exponential coordinates with
respect to the Fisher
geometry~\cite{Amari:inf_geom:01,Ay:info_geom:17}. 

The action~(\ref{eq:CRN_L_genform}) in action-angle variables becomes 
\begin{align}
  S 
& = 
  \int_0^T dt 
  \left\{ 
    - \left( d_t \theta \right)
    n + 
    \mathcal{L} \! \left( \theta , n \right)
  \right\} 
\label{eq:CRN_L_genform_AA}
\end{align}
$\mathcal{L} \! \left( \theta , n \right)$ is 
$\mathcal{L} \! \left( {\phi}^{\dagger} , \phi \right)$ with
${\phi}^{\dagger}$ and $\phi$ written as functions of $n$ and $\theta$
by Eq.~(\ref{eq:AA_std_defs}). 
The stationary-path equations corresponding to
Eq.~(\ref{eq:Hamiltonian_var_CS}) in action-angle variables are
\begin{align}
  {\dot{\theta}}_i
& = 
  \frac{\partial}{\partial n_i}
  \mathcal{L} \! \left( \theta , n \right)
\nonumber \\ 
  {\dot{n}}_i
& = 
  - \frac{\partial}{\partial {\theta}_i}
  \mathcal{L} \! \left( \theta , n \right)
\label{eq:Hamiltonian_var_AA}
\end{align}

\subsubsection{Descaling with respect to the instantaneous steady-state
mean number}

A second class of canonical transformations, performed by transferring
a scale factor from $\phi$ to ${\phi}^{\dagger}$, was first used in
Doi-Peliti integrals by Baish~\cite{Baish:DP_duality:15}.  Let ${\bm
n}$ be the saddle-point value of the field $n$ in
Eq.~(\ref{eq:AA_std_defs}) in the steady state that would be
annihilated by ${\rm T}_{\rn {\rn}^{\prime}}$ at the parameters it possesses
at some time.  If ${\rm T}_{\rn {\rn}^{\prime}}$ and thus $\mathcal{L}$ is
explicitly time-dependent, then the scale factor ${\bm n}$ will
generally be different for each time.  Define coherent state fields
rescaled locally by ${\bm n}$ as
\begin{align}
  {\phi}^{\dagger}_i
  {\bm n}_i 
& \equiv 
  {\varphi}^{\dagger}_i 
& 
  \frac{
    {\phi}_i
  }{
    {\bm n}_i
  } 
& \equiv 
  {\varphi}_i
\label{eq:Baish_trans_only}
\end{align}
The form of the action in fields $\left( {\varphi}^{\dagger} , \varphi
\right)$ remains as in Eq.~(\ref{eq:CRN_L_genform}) but the Liouville
function $\mathcal{L}$ which was the ``Hamiltonian'' for the
stationary-path equations~(\ref{eq:Hamiltonian_var_CS}) is replaced by
a possibly-shifted function
\begin{equation}
  \mathcal{\tilde{L}} \! \left( {\varphi}^{\dagger} , \varphi \right) \equiv 
  \mathcal{L} \! \left( {\phi}^{\dagger} , \phi \right) + 
  \sum_i 
  \left( 
    {\varphi}^{\dagger}_i
    {\varphi}^i
  \right) 
  d_t \log {\bm n}_i
\label{eq:Baish_Kamiltonian}
\end{equation}

Canonical transformation by coherent-state rescaling is closely
related to a class of similarity transforms of the transition matrix
introduced by Hatano and Sasa~\cite{Hatano:NESS_Langevin:01}, and now
widely used as the basis for a class of \textit{integral fluctuation
theorems}~\cite{Verley:HS_FDT:12}.  The Hatano-Sasa dualization and
the Baish dualization differ in the important respect that Hatano and
Sasa rescale the entire \emph{transition matrix} ${\rm T}_{\rn
{\rn}^{\prime}}$ by values of the stationary probability density
${\rho}_{\rn}$, requiring knowledge of infinitely many scale factors,
whereas the rescaling~(\ref{eq:Baish_trans_only}) requires only $D$
scale factors (the stationary-point numbers of the $D$ species ${\bm
n}_i$).  The two coincide exactly in the case that the distribution is
a coherent state, for which all probability values are expressed in
terms of the mean.  This is just the case in which the Fisher
spherical embedding of a general probability distribution, reviewed in
App.~\ref{sec:Fisher_general_sphere}, projects to a spherical
embedding of the same form in terms of only the mean number, as shown
in App.~\ref{sec:Fisher_reduced_sphere}.

\subsubsection*{Dual action-angle transform in descaled coherent
states} 

An instance of the action-angle transformation that we have not seen
used before, but which directly gives the dual geometry for the
symplectic structure of Liouville evolution, is one applied to the
rescaled coherent-state fields $\left( {\varphi}^{\dagger} . \varphi
\right)$:
\begin{align}
  {\varphi}^{\dagger}_i 
& \equiv 
  e^{-{\eta}_i} n_i
& 
  {\varphi}_i
& \equiv 
  e^{{\eta}_i}
\label{eq:dual_AA}
\end{align}
The action~(\ref{eq:CRN_L_genform}) in the new variables becomes
\begin{equation}
  S = 
  \int_0^T dt 
  \left\{ 
    \left( d_t \eta \right) n + 
    \mathcal{\tilde{L}} \! \left( n , \eta \right)
  \right\}
\label{eq:S_dual_form}
\end{equation}
where the modified Liouville function $\mathcal{\tilde{L}}$ from
Eq.~(\ref{eq:Baish_Kamiltonian}) must be used, such that in the new
variables
\begin{equation}
  \mathcal{\tilde{L}} \! \left( n , \eta \right) \equiv 
  \mathcal{L} \! \left( \theta , n \right) + 
  \sum_{i = 1}^D
  n_i 
  d_t \log {\bm n}_i
\label{eq:Kamiltonian_genform}
\end{equation}
$\eta$ is the affine coordinate in an exponential family of
distributions produced by tilting a reference which is a product of
Poisson marginals, or an equivalent section through that product, such
as a multinomial. 

\subsubsection{Time-translation along stationary paths}
\label{sec:time_trans_canonical}

The third class of canonical transformations are the coordinate
transformations generated by time-translation along stationary paths.
For any coordinate system $\left( \theta , n \right)$, the bundle of
stationary trajectories $\left( \bar{\theta} , \bar{n} \right)$
passing through those values at a time $t$, together with a time-shift
$\Delta t$, generate new coordinates
\begin{equation}
  \left( {\theta}^{\prime} , n^{\prime} \right) \equiv 
  {
    \left. 
      \left( 
        \bar{\theta} \! \left( t + \Delta t \right) , 
        \bar{n} \! \left( t + \Delta t \right) 
      \right)
    \right| 
  }_{
    \left( 
      \bar{\theta} \left( t \right) , 
      \bar{n} \left( t \right) 
    \right) = 
    \left( \theta , n \right)
  }
\label{eq:time_shift_coords}
\end{equation}
which also obey Eq.~(\ref{eq:Hamiltonian_var_AA}), possibly with
shifted parameters in $\mathcal{L}$.

We will be interested in the class of dual Riemannian connections that
can be imposed on 2-field coordinate systems that respect the
symplectic structure of the generator $d/dt$ of canonical
transformations by time translation.

\section{The Liouville theorem connecting dynamics to inference
induced by 2-field stationary trajectories}
\label{sec:2_field_Liouville}

Liouville's theorem in classical mechanics describes conservation of a
\textit{phase space} density over position coordinates and their
conjugate momenta.  The discrete-state stochastic processes for which
Doi-Peliti methods were invented do not possess conjugate momentum
coordinates, and their measures over positions concentrate as
trajectories coalesce.  For other groups of 2FFI methods such as
Martin-Siggia-Rose~\cite{Martin:MSR:73}, although the states may be
those of dynamical systems, the dissipative context for which these
methods are applied likewise do not conserve density in the resulting
phase spaces.  Instead, the 2-field integrals themselves provide
fields that relate to the coherent-state parameters as conjugate
momenta under the Liouville function $\mathcal{L}$, which admit (among
other possibilities) the interpretation of likelihood ratios.

This section constructs the conserved density in a 2FFI-``phase
space'' in which the conjugate coordinates represent
forward-propagating nominal distributions and backward-propagating
likelihoods with the interpretation of sampling protocols for
inference.  The conserved density is only the scalar representation of
symplectic structure; the deformations in the Liouville volume element
carried on stationary paths imply further transport laws for vector
and tensor fields including the Fisher metric.  Those are derived next
by writing the inner product of vector fields from base and tilt
variations as the basis for the $2D$-dimensional differential of the
CGF.

\subsection{The Wigner function from the 2-field identity operator
plays the role of a phase-space density}

The scalar density in 2FFI that fills the role of a phase space
density in classical Hamiltonian mechanics is the Wigner
function~\cite{Wigner:quantum_density:35}, of which versions exist for
both classical and quantum systems.\footnote{Wigner introduced this
function to treat quantum density matrices, such as those arising in
the Schwinger-Keldysh time-loop
2FFI~\cite{Schwinger:MBQO:61,Keldysh:noneq_diag:65}, and it is closely
related to the Glauber-Sudarshan
``$P$-representation''~\cite{Glauber:densities:63,Sudarshan:P_rep:63,%
Cohen:densities:66}.  Equivalent constructions are widely applied to
problems in time-frequency analysis or
optics~\cite{Wahlberg:Wigner_Gauss:05,Bazarov:Wigner_optics:12}, and
an example of the map from quantum to classical 2FFI Wigner functions
is given in~\cite{Smith:DP:08}.}  It is defined in terms of the
representation of unity in Eq.~(\ref{eq:gen_fn_twofield_CKdecomp}), as
\begin{widetext}
\begin{align}
  w_t \! 
  \left( 
    {\phi}^{\dagger}_{\ddagger} , 
    {\phi}_{\ddagger}
  \right)
& \equiv 
  \frac{1}{{\pi}^D}
  \int_{t + \delta t}^T 
  {\mathcal{D}}^D \! {\phi}^{\dagger}
  {\mathcal{D}}^D \! {\phi} \, 
  e^{
    \left( z - {\phi}^{\dagger}_T \right) {\phi}_T - 
    S 
  } 
  e^{
    \left( 
      {\phi}^{\dagger}_{t + \delta t} - 
      {\phi}^{\dagger}_{\ddagger}
    \right) 
    {\phi}_{\ddagger}
  } 
  e^{
    {\psi}_t 
    \left( 
      \log {\phi}^{\dagger}_{\ddagger}
    \right)
  } 
\label{eq:Wigner_def}
\end{align}
\end{widetext}
Eq.~(\ref{eq:Wigner_def}) implies that, for $w_t \! \left(
{\phi}^{\dagger}_{\ddagger} , {\phi}_{\ddagger} \right)$ at any time,
\begin{equation}
  e^{{\psi}_T \left( \theta \right)} = 
  \int 
  d^D \! {\phi}^{\dagger}_{\ddagger} \, 
  d^D \! {\phi}_{\ddagger} \, 
  w_t \! 
  \left( 
    {\phi}^{\dagger}_{\ddagger} , 
    {\phi}_{\ddagger}
  \right)
\label{eq:genfun_from_Wigner}
\end{equation}
In a saddle-point approximation, one identifies arguments $\left(
{\bar{\phi}}^{\dagger}_{\ddagger} , {\bar{\phi}}_{\ddagger} \right)$
for which, to leading exponential order,
\begin{equation}
  e^{{\psi}_T \left( \theta \right)} \sim
  w_t \! 
  \left( 
    {\bar{\phi}}^{\dagger}_{\ddagger} , 
    {\bar{\phi}}_{\ddagger}
  \right)
\label{eq:genfun_from_Wigner_SP}
\end{equation}
Since Eq.~(\ref{eq:genfun_from_Wigner_SP}) approximates the same
function at any time $t$, its total time derivative along the sequence
of stationary points must vanish,
\begin{align}
  0 
& = 
  \frac{d}{dt}
  w_t \! 
  \left( 
    {\bar{\phi}}^{\dagger}_{\ddagger} , 
    {\bar{\phi}}_{\ddagger}
  \right)
\nonumber \\ 
& = 
  \left( 
    \frac{\partial}{\partial t} + 
    \frac{d{\bar{\phi}}^{\dagger}_{\ddagger}}{dt}
    \frac{\partial}{\partial {\bar{\phi}}^{\dagger}_{\ddagger}} + 
    \frac{d{\bar{\phi}}_{\ddagger}}{dt}
    \frac{\partial}{\partial {\bar{\phi}}_{\ddagger}}
  \right)
  w_t \! 
  \left( 
    {\bar{\phi}}^{\dagger}_{\ddagger} , 
    {\bar{\phi}}_{\ddagger}
  \right)
\label{eq:Wigner_stat}
\end{align}
Moreover, the stationary points should coincide with values along the
stationary trajectories~(\ref{eq:Hamiltonian_var_CS}) of the
functional integral~(\ref{eq:gen_fn_twofield_CKdecomp}), which
satasify
\begin{equation}
  {
    \left. 
      \left( 
  \sum_i 
  \frac{
    \partial {\dot{\phi}}^{\dagger}_i 
  }{
    \partial {\phi}^{\dagger}_i
  } + 
  \sum_i 
  \frac{
    \partial {\dot{\phi}}_i 
  }{
    \partial {\phi}_i
  }
      \right) 
    \right| 
  }_{{\bar{\phi}}^{\dagger} , \bar{\phi}} = 
  0 
\label{eq:CS_Liouville_plain}
\end{equation}
Eq.~(\ref{eq:Wigner_stat}) may thus be recast as the conservation law
for a $2D$-dimensional current $\left( {\dot{\phi}}^{\dagger} w ,
{\dot{\phi}} w \right)$,
\begin{equation}
  0 = 
  \frac{\partial w_t}{\partial t} + 
  \sum_i
  \frac{\partial}{\partial {\phi}^{\dagger}_i} 
  \left( 
    {\dot{\phi}}^{\dagger}_i w_t
  \right) + 
  \sum_i
  \frac{\partial}{\partial {\phi}_i} 
  \left( 
    {\dot{\phi}}_i w_t
  \right)
\label{eq:Wigner_stat_CS_rot}
\end{equation}
which is Liouville's theorem.  

$w_t$ is a density of rays for joint base distributions and likelihood
ratios that is conserved along the Doi-Peliti stationary trajectories.
$\log w_t$ is the leading exponential approximation to the value of
the CGF.  It therefore integrates information along the trajectory
from the final-time imposed value of $z$ and the initial-time
structure of the generating function ${\psi}_0 \! \left(
\log {\phi}^{\dagger}_0
\right)$.

The indirect definition~(\ref{eq:Wigner_def}) of the Wigner function
in terms of the functional integral is convenient to manipulate but
perhaps not very self-explanatory.  App.~\ref{sec:Wigner_stat_pt}
gives a direct construction of the stationary-point approximation in
terms of a density $\rho \! \left( \theta \right)$ over the basis
of coherent states $\left| \phi \right)$ and their Laplace
transforms, and verifies that the sequence of stationary points do
indeed coincide with the equations of
motion~(\ref{eq:Hamiltonian_var_CS}). 

\subsubsection*{Constraints and conserved current flows in reduced
dimensions} 

Often systems of interest will evolve under constraints arising from
conservation laws, such as conserved quantities of the stoichiometry
in chemical reaction
networks~\cite{Polettini:open_CNs_I:14,Krishnamurthy:CRN_moments:17,%
Smith:CRN_moments:17}.  Conserved quantities result in flat directions
in the CGF and zero eigenvalues of the Fisher metric.  Since generally
the constraints will involve multiple species, and because the
action-angle canonical transform~(\ref{eq:AA_std_defs}) is defined in
the species basis, it will not be possible to factor out non-dynamical
combinations.  Then the transport
equation~(\ref{eq:Wigner_stat_CS_rot}) for the current of the Wigner
density will occupy only a sub-manifold of the $2D$-dimensional
Doi-Peliti coordinate space needed to define the system.

A convenient way to handle constraints is to work in the eigenbasis of
the Fisher metric which we will index with subscript $\alpha$, where
an action-angle counterpart to the transport
equation~(\ref{eq:Wigner_stat_CS_rot}) reads
\begin{align}
  0 
& = 
  \frac{\partial w_t}{\partial t} + 
  \sum_{\alpha}
  \frac{\partial}{\partial {\theta}^{\alpha}} 
  \left( 
    {\dot{\theta}}^{\alpha} w_t
  \right) + 
  \sum_{\alpha}
  \frac{\partial}{\partial n_{\alpha}} 
  \left( 
    {\dot{n}}_{\alpha} w_t
  \right)
\label{eq:Wigner_stat_AA_rot}
\end{align}
The picture of the Liouville equation as implying a conserved volume
element 
\begin{equation}
  \frac{d}{dt}
  \left( 
    \prod_{\alpha}
    \delta {\theta}^{\alpha}
    \delta n_{\alpha}
  \right) = 
  0 
\label{eq:cons_vol}
\end{equation}
with the product index $\alpha$ taken only over nonzero eigenvalues
of the Fisher metric, remains nondegenerate and has a direct
interpretation in terms of the product of eigenvectors of the Fisher
inner product in independent dimensions.

\subsection{The Fisher metric and cubic tensor in dual canonical
coordinates} 
\label{sec:Fisher_from_psi}

The leading-exponential equivalence of the Wigner density to the CGF
from Eq.~(\ref{eq:genfun_from_Wigner_SP}) suggests that the
$2D$-dimensional differential of the stationary-point CGF should
likewise obey a symplectic transport law, implying a transport law for
the Fisher metric.  To derive those results we return to the
expression of the differential of the CGF in terms of the generalized
Pythagorean theorem~(\ref{eq:triangle_to_Fisher}), and derive the
Fisher metric from the $\psi$-divergence following
Amari~\cite{Amari:inf_geom:01} Sec.~6.2.

Base distributions corresponding to points along stationary paths
under the action~(\ref{eq:CRN_L_genform}) form exponential families,
because they are in the class of coherent-state distributions
described in App.~\ref{sec:Fisher_reduced_sphere}.  Therefore label
importance distributions~(\ref{eq:CGF_tilt_n0_def}) symmetrically as
$\tilde{\rho} \! \left( \theta , \eta \right)$ with the exponential
coordinates in the two action-angle
transformations~(\ref{eq:AA_std_defs}) and~(\ref{eq:dual_AA}).  To
study their independent variations about a reference value $\left(
{\theta}_R , {\eta}_R \right)$, introduce two distinct exponential
families, labeled
\begin{align}
  \rho 
& \equiv 
  {
    \tilde{\rho} \! \left( \theta , \eta \right)
  }_{\eta = {\eta}_R}
\nonumber \\ 
  {\rho}^{\prime} 
& \equiv 
  {
    \tilde{\rho} \! \left( \theta , \eta \right)
  }_{\theta = {\theta}_R}
\label{eq:two_dists}
\end{align}

The $\psi$-divergence $D_{\psi} \! \left( \theta : \eta \right) = \psi
\! \left( \theta \right) + {\psi}^{\ast} \! \left( n \right) - n \theta$, a
Bregman divergence of the CGF, is related to the Kullback-Leibler
divergence of ${\rho}^{\prime}$ from $\rho$ as
\begin{align}
  D_{\psi} \! \left( \theta : \eta \right) 
& = 
  D_{\rm KL} \! \left( {\rho}^{\prime} \parallel \rho \right)
\nonumber \\
& = 
  \sum_{\rn} 
  {\tilde{\rho}}_{\rn} \! \left( {\theta}_R , \eta \right)
  \log 
  \left(
    \frac{
      {\tilde{\rho}}_{\rn} \! \left( {\theta}_R , \eta \right)
    }{
      {\tilde{\rho}}_{\rn} \! \left( \theta , {\eta}_R \right)
    }
  \right) 
\label{eq:psi_Div}
\end{align}
The mixed second partial derivative of $- D_{\psi}$ gives the same
variance that defines the Fisher metric.  At general $\theta$, $\eta$,
it is labeled
\begin{align}
  g^D_{ij}
& = 
  - \frac{\partial}{\partial {\theta}^i}
  \frac{\partial}{\partial {\eta}^j}
  D_{\psi} \! \left( \theta : \eta \right) 
\nonumber \\ 
& = 
  \sum_{\rn}
  \frac{
    \partial
    {\tilde{\rho}}_{\rn} \! \left( {\theta}_R , \eta \right)
  }{
    \partial {\eta}^j
  }
  \frac{
    \partial
    \log {\tilde{\rho}}_{\rn} \! \left( \theta , {\eta}_R \right)
  }{
    \partial {\theta}^i
  }
\nonumber \\ 
& = 
  \sum_{\rn}
  {\tilde{\rho}}_{\rn} \! \left( {\theta}_R , \eta \right)
  \left(
    {\rn}_j - 
    \frac{\partial \psi}{\partial {\eta}^j}
  \right) 
  \left(
    {\rn}_i - 
    \frac{\partial \psi}{\partial {\theta}^i}
  \right) 
\label{eq:g_D_form}
\end{align}
At $\theta = {\theta}_R$, $\eta = {\eta}_R$, the second line of
Eq.~(\ref{eq:g_D_form}) recovers exactly the differential form of the
Pythagorean theorem of Eq.~(\ref{eq:triangle_to_Fisher}) in dual
exponential coordinates.

Two third-order mixed partials define the connection coefficients for
Amari's dually-flat connections on exponential and mixture
coordinates.  Written in all-contravariant indices,\footnote{Note that
it is the dual connection ${{\Gamma}^{D\ast}}^{kij} \equiv 0$, written
in all-\emph{covariant} indices, which vanishes as the affine
connection on the mixture family.}  these are given by
\begin{align}
  {\Gamma}^D_{kij}
& = 
  \frac{\partial}{\partial {\theta}^k}
  g^D_{ij}
\nonumber \\
& = 
  - \frac{
    {\partial}^2 \psi
  }{
    \partial {\theta}^k
    \partial {\theta}^i
  }
  \sum_{\rn}
  {\tilde{\rho}}_{\rn} \! \left( {\theta}_R , \eta \right)
  \left(
    {\rn}_j - 
    \frac{\partial \psi}{\partial {\eta}^j}
  \right) 
\nonumber \\ 
  {\Gamma}^{D\ast}_{kij}
& = 
  \frac{\partial}{\partial {\eta}^k}
  g^D_{ij}
\nonumber \\
& = 
  \sum_{\rn}
  {\tilde{\rho}}_{\rn} \! \left( {\theta}_R , \eta \right)
  {
    \left(
      {\rn}_j - 
      \frac{\partial \psi}{\partial {\eta}^j}
    \right) 
  }^2
  \left(
    {\rn}_i - 
    \frac{\partial \psi}{\partial {\theta}^i}
  \right) 
\nonumber \\ 
& \mbox{} - 
  \frac{
    {\partial}^2 \psi
  }{
    \partial {\eta}^k
    \partial {\eta}^j
  }
  \sum_{\rn}
  {\tilde{\rho}}_{\rn} \! \left( {\theta}_R , \eta \right)
  \left(
    {\rn}_i - 
    \frac{\partial \psi}{\partial {\theta}^i}
  \right) 
\label{eq:Gammas_D_forms}
\end{align}
Evaluated at $\theta = {\theta}_R$ and $\eta = {\eta}_R$, 
\begin{align}
  g^D_{ij} \! \left( {\theta}_R , {\eta}_R \right) 
& = 
  g_{ij} = 
  {
    \left. 
  \frac{
    {\partial}^2 \psi
  }{
    \partial {\theta}^i
    \partial {\theta}^j
  }
    \right| 
  }_{{\theta}_R, {\eta}_R}
\nonumber \\ 
  {\Gamma}^{D}_{kij} \! \left( {\theta}_R , {\eta}_R \right) 
& = 
  0 
\nonumber \\ 
  {\Gamma}^{D\ast}_{kij} \! \left( {\theta}_R , {\eta}_R \right) 
& = 
  T_{kij} = 
  {
    \left. 
  \frac{
    {\partial}^3 \psi
  }{
    \partial {\theta}^k
    \partial {\theta}^i
    \partial {\theta}^j
  }
    \right| 
  }_{{\theta}_R , {\eta}_r}
\label{eq:D_tens_to_tens}
\end{align}
$g_{ij}$ is the Fisher metric introduced in
Eq.~(\ref{eq:Fisher_element}), and $T_{kij}$ is the cubic tensor, also
called the Amari-Chentsov tensor~\cite{Amari:inf_geom:01}.  Below we
remove the subscript $R$ and write $\theta$ and $\eta$ as the
arguments of these tensors.

\subsection{The dual vector fields induced by base-distribution
initial conditions, and final-time tilts}
\label{sec:dual_vec_fields}

From the construction of $g^D$ in Sec.~\ref{sec:Fisher_from_psi}, we
can see how to use the stationary-path equations of motion to induce
two mappings of vector fields that respect the dual arguments of the
$\psi$-divergence.  Variations in the likelihood act on the $\rho$
argument, while variations in the base distribution act on the
${\rho}^{\prime}$ argument, in Eq.~(\ref{eq:two_dists}).  The
stationary-path equations are then used to define a 1-parameter family
of maps of any basis of dual variations in initial base and final tilt
parameters to intermediate times.  Conservation of the Liouville
volume then translates to a conserved inner product of pairs of vector
fields transported respectively under the two branches of the dual
mapping.  Conservation of the inner product will imply a transport
equation for the Fisher metric corresponding to the
equation~(\ref{eq:Wigner_stat_AA_rot}) for the Wigner function.

Introduce two vector fields corresponding to variations in $\theta$ at
the final time $T$, and to variations in $\eta$ at the initial time
$0$.  The first can be independently imposed through the arguments in
${\Psi}_T \! \left( z \right)$, while the second can be independently
imposed in the initial data.  Fields $\delta {\theta}_T$ and $\delta
{\eta}_0$ are written in components as
\begin{align}
  \delta {\theta}_T 
& \equiv 
  \delta {\theta}_T^i
  \frac{\partial}{\partial {\theta}_T^i}
\nonumber \\ 
  \delta {\eta}_0 
& \equiv 
  \delta {\eta}_0^i
  \frac{\partial}{\partial {\eta}_0^i}
\label{eq:vec_fields_bc}
\end{align}

The stationary-path conditions map the dual initial and final
coordinates to pairs of coordinates at any intermediate time, which we
denote $\theta \! \left( {\theta}_T, {\eta}_0, t \right)$, $\eta \!
\left( {\theta}_T, {\eta}_0, t \right)$.  A one-parameter family of
vector fields is defined by assigning to each such coordinate image
$\left( \theta \! \left( {\theta}_T, {\eta}_0, t \right) , \eta \! 
\left( {\theta}_T, {\eta}_0, t \right) \right)$ the field values
\begin{align}
  \delta \theta \! \left( \theta, \eta , t \right)
& \equiv 
  \delta {\theta}_T^i
  \frac{\partial}{\partial {\theta}_T^i}
  \theta \! \left( {\theta}_T, {\eta}_0, t \right)
\nonumber \\ 
  \delta \eta \! \left( \theta, \eta , t \right) 
& \equiv 
  \delta {\eta}_0^i
  \frac{\partial}{\partial {\eta}_0^i}
  \eta \! \left( {\theta}_T, {\eta}_0, t \right)
\label{eq:delta_vec_flds_def}
\end{align}
Under the change of coordinates from $\left( {\theta}_T, {\eta}_0
\right)$ to $\left( \theta , \eta \right)$ at each time $t$, the
vector fields~(\ref{eq:delta_vec_flds_def}) may be written in terms of
the local coordinate differentials as 
\begin{align}
  \delta \theta \! \left( \theta, \eta , t \right)
& \equiv 
  \delta {\theta}^j \! \left( \theta, \eta , t \right)
  \frac{\partial}{\partial {\theta}^j} 
\nonumber \\ 
  \delta \eta \! \left( \theta, \eta , t \right)
& \equiv 
  \delta {\eta}^j \! \left( \theta, \eta , t \right)
  \frac{\partial}{\partial {\eta}^j} 
\label{eq:vec_fld_comps}
\end{align}
Below we suppress the explicit $\left( \theta , \eta , t \right)$
coordinate and time arguments of $\delta {\theta}^j \! \left( \theta,
\eta , t \right)$ and $\delta \eta \! \left( \theta, \eta , t
\right)$, and indicate the time $t$ in a subscript only where it is
needed to avoid confusion.

The vector fields~(\ref{eq:vec_fld_comps}) have a time dependence that
can be defined through the dependences of $\left( \theta \! \left(
{\theta}_T, {\eta}_0, t \right) , \eta \!  \left( {\theta}_T,
{\eta}_0, t \right) \right)$ on the boundary coordinates and then
transformed to the local coordinate system, becoming 
\begin{align}
  {
    \left( 
      \frac{d}{dt}
      \delta \theta 
    \right)
  }^j
& = 
  \delta {\theta}_T^i
  \frac{\partial}{\partial {\theta}_T^i}
  {\dot{\theta}}^j \! \left( {\theta}_T, {\eta}_0, t \right)
\nonumber \\ 
& = 
  \delta {\theta}_t^i
  \frac{\partial}{\partial {\theta}_t^i}
  {\dot{\theta}}^j \! \left( {\theta}_T, {\eta}_0, t \right)
\nonumber \\ 
& = 
  \delta {\theta}_t^i
  \frac{
    {\partial}^2 \mathcal{L}
  }{
    \partial {\theta}^i
    \partial n_j
  }
\nonumber \\ 
  {
    \left( 
      \frac{d}{dt}
      \delta \eta 
    \right)
  }^j
& = 
  \delta {\eta}_0^i
  \frac{\partial}{\partial {\eta}_0^i}
  {\dot{\eta}}^j \! \left( {\theta}_T, {\eta}_0, t \right)
\nonumber \\ 
& = 
  \delta {\eta}_t^i
  \frac{\partial}{\partial {\eta}_t^i}
  {\dot{\eta}}^j \! \left( {\theta}_T, {\eta}_0, t \right)
\nonumber \\ 
& = 
  - \delta {\eta}_t^i
  \frac{
    {\partial}^2 \mathcal{\tilde{L}}
  }{
    \partial {\eta}^i
    \partial n_j
  }
\label{eq:vec_flds_timedep}
\end{align}
Eq.~(\ref{eq:Hamiltonian_var_AA}) is used to arrive at the third form
of each equation in terms of mixed partials of $\mathcal{L}$ and
$\mathcal{\tilde{L}}$.

The coordinate transformation~(\ref{eq:g_as_coord_change}) from
contravariant exponential coordinates to covariant mixture coordinates
may be used in two ways to write the inner product of vector fields
$\delta \theta$ and $\delta \eta$ in mixed form.  From the definition
of the inner product in terms of $g^D$ in Eq.~(\ref{eq:g_D_form}) and
its equivalence to the Hessian definition of $g$ in
Eq.~(\ref{eq:D_tens_to_tens}),
\begin{align}
  {\left( \delta \theta \right)}^i 
  g_{ij}
  {\left( \delta \eta \right)}^j
& \equiv 
  {\left( \delta \theta \right)}^i 
  {\left( {\delta}_{\eta} n \right)}_i
\nonumber \\ 
& \equiv 
  {\left( {\delta}_{\theta} n \right)}_j 
  {\left( \delta \eta \right)}^j
\label{eq:inner_prod_two_ways}
\end{align}
Although the field variable $n$ is the same in either action-angle
transform~(\ref{eq:AA_std_defs}) or~(\ref{eq:dual_AA}), the two
displacements ${\delta}_{\eta} n $ and ${\delta}_{\theta} n$ are
independent vector fields.

\subsubsection{The conserved inner product of dual vector fields, and
directional transport of the metric}

Eq.~(\ref{eq:vec_flds_timedep}) has a symmetric form but evolves
$\delta \theta$ and $\delta \eta$ respectively using $\mathcal{L}$ and
$\mathcal{\tilde{L}}$, making it not immediately apparent that the
inner product is preserved.  Writing the field $\delta \eta$ in its
dual mixture coordinate as in the first line of
Eq.~(\ref{eq:inner_prod_two_ways}) the time derivative becomes
\begin{align}
  {
    \left( 
      \frac{d}{dt}
      \delta n
    \right)
  }_i
& = 
  \delta n_{0j}
  \frac{\partial}{\partial n_{0j}}
  {\dot{n}}_i \! \left( {\theta}_T, n_0, t \right)
\nonumber \\ 
& = 
  \delta n_{tj}
  \frac{\partial}{\partial n_{tj}}
  {\dot{n}}_i \! \left( {\theta}_T, n_0, t \right)
\nonumber \\ 
& = 
  - \delta n_{tj}
  \frac{
    {\partial}^2 \mathcal{L}
  }{
    \partial n_j
    \partial {\theta}^i
  }
\label{eq:vec_flds_timedep_nucoord}
\end{align}
The condition~(\ref{eq:symplect_cond}) is met and we have 
\begin{align}
  {
    \left( 
      \frac{d}{dt}
      \delta {\theta}
    \right)
  }^j
  {\left( \delta n \right)}_j + 
  {\left( \delta \theta \right)}^i
  {
    \left( 
      \frac{d}{dt}
      \delta n
    \right)
  }_i = 
  0 
\label{eq:inn_prod_cons}
\end{align}

Using Eq.~(\ref{eq:inn_prod_cons}) to evaluate the change in the inner
product written as ${\left( \delta \theta \right)}^i g_{ij} {\left(
\delta \eta \right)}^j$, substituting the
derivatives~(\ref{eq:Hamiltonian_var_AA}) for $\dot{\theta}$ and
$\dot{\eta}$, and grouping terms, we obtain the transport equation for
the metric along stationary paths
\begin{align}
  0 
& = 
  {\dot{\theta}}^k
  \frac{\partial g_{ij}}{\partial {\theta}^k} + 
  \frac{
    {\partial}^2 \mathcal{L}
  }{
    \partial {\theta}^i
    \partial n_k
  }
  g_{kj} + 
  \frac{\partial g_{ij}}{\partial {\eta}^k}
  {\dot{\eta}}^k - 
  g_{ik}  
  \frac{
    {\partial}^2 \mathcal{\tilde{L}}
  }{
    \partial n_k
    \partial {\eta}^j
  }
\nonumber \\ 
& = 
  \left( 
    \frac{
      \partial \mathcal{L}
    }{
      \partial n_k
    }
    \frac{\partial g_{ij}}{\partial {\theta}^k} + 
    \frac{
      {\partial}^2 \mathcal{L}
    }{
      \partial {\theta}^i
      \partial n_k
    }
    g_{kj} 
  \right) - 
  \left(
    \frac{\partial g_{ij}}{\partial {\eta}^k}
    \frac{
      \partial \mathcal{\tilde{L}}
    }{
      \partial n_k
    } + 
    g_{ik}  
    \frac{
      {\partial}^2 \mathcal{\tilde{L}}
    }{
      \partial n_k
      \partial {\eta}^j
    }
  \right) 
\nonumber \\ 
& = 
  \frac{
    \partial 
  }{
    \partial {\theta}^i
  } 
  \left(
    {\dot{\theta}}^k g_{kj}
  \right) + 
  \frac{
    \partial 
  }{
    \partial {\eta}^j
  } 
  \left(
    g_{ik}
    {\dot{\eta}}^k
  \right) 
\label{eq:g_par_trans}
\end{align}
The tensor transport equation from Eq.~(\ref{eq:g_par_trans}) can be
compared to Eq.~(\ref{eq:Wigner_stat_AA_rot}) for the transport of the
Wigner density.

\subsection{Dual connections respecting the symplectic structure of
canonical transformations in the 2-field system}
\label{sec:dual_connections}

The transport relations derived so far make use of the symplectic
structure of maps generated by time-translation along Doi-Peliti
stationary paths, but they are not specifically geometric.  We now
turn to geometric constructions that respect the symplectic structure,
render its maps coordinate invariant under canonical transformations,
and express the special roles of affine transport in some coordinates
such as coherent states through the definition of appropriate
Riemannian connections.

\subsubsection{Conservation of the inner product through the
combined effects of two maps}

The inner product~(\ref{eq:inner_prod_two_ways}) is preserved through
the complementary action of two maps, one generated by the
time-dependence of $\theta$, and the other by the time-dependence of
$\eta$.  By construction, $\delta \theta$ depends on time only through
$\dot{\theta}$, and $\delta \eta$ only through $\dot{\eta}$, while
the metric has no explicit time dependence but changes under both maps
as the location $\left( \theta , \eta \right)$ changes.  Denoting by
${ \left. d / dt \right| }_{\dot{\theta}}$ and ${ \left. d / dt
\right| }_{\dot{\eta}}$ these separate components of change, the time
derivative of the inner product can be partitioned into two canceling
terms:
\begin{align}
  {
    \left(
      {
        \left. 
          \frac{d}{dt}
        \right|
      }_{\dot{\theta}}
      {\delta}_{\theta} n
    \right)
  }_j
  {\left( \delta \eta \right)}^j 
& \equiv 
  \left[
    {
      \left( 
        \frac{d}{dt}
        \delta \theta 
      \right)
    }^i
    g_{ij} + 
    {\left( \delta \theta \right)}^i 
    {\dot{\theta}}^k
    \frac{\partial g_{ij}}{\partial {\theta}^k}
  \right]
  {\left( \delta \eta \right)}^j 
\nonumber \\   
  {\left( \delta \theta \right)}^i 
  {
    \left(
      {
        \left. 
          \frac{d}{dt}
        \right|
      }_{\dot{\eta}}
      {\delta}_{\eta} n
    \right)
  }_i  
& \equiv 
  {\left( \delta \theta \right)}^i 
  \left[ 
    g_{ij}
    {
      \left( 
        \frac{d}{dt}
        \delta \eta 
      \right)
    }^j + 
    {\dot{\eta}}^k
    \frac{\partial g_{ij}}{\partial {\eta}^k}
    {\left( \delta \eta \right)}^j 
  \right] 
\label{eq:two_simple_parts}
\end{align}
Connection coefficients may be added within either ${\left( { \left. d
/ dt \right| }_{\dot{\theta}} \, {\delta}_{\theta} n \right)}_j$ or
${\left( { \left. d / dt \right| }_{\dot{\eta}} \, {\delta}_{\eta} n
\right)}_i$ to make the components of change in the vector field and
metric coordinate-invariant, without altering the duality between
independent variations in the base distribution and in the likelihood
ratio. 

To introduce a connection we first replace the total derivative $d/dt$
with a partial-derivative decomposition expressing the same
transformation as a flow: 
\begin{align}
  {
    \left( 
      \frac{d}{dt}
      \delta \theta 
    \right)
  }^i
& \equiv 
  {
    \left( 
      \frac{\partial}{\partial t}
      \delta \theta 
    \right)
  }^i + 
  {\dot{\theta}}^k
  \frac{\partial}{\partial {\theta}^k}
  {\left( \delta \theta \right)}^i + 
  {\dot{\eta}}^k
  \frac{\partial}{\partial {\eta}^k}
  {\left( \delta \theta \right)}^i
\nonumber \\ 
  {
    \left( 
      \frac{d}{dt}
      \delta \eta 
    \right)
  }^j
& \equiv 
  {
    \left( 
      \frac{\partial}{\partial t}
      \delta \eta 
    \right)
  }^j + 
  {\dot{\theta}}^k
  \frac{\partial}{\partial {\theta}^k}
  {\left( \delta \eta \right)}^j + 
  {\dot{\eta}}^k
  \frac{\partial}{\partial {\eta}^k}
  {\left( \delta \eta \right)}^j
\label{eq:dts_to_flows}
\end{align}
Connection coefficients are defined from the pullbacks ${\left(
\partial / \partial {\theta}^k \right)}^{\prime}$ or ${\left(
\partial / \partial {\eta}^k \right)}^{\prime}$ of infinitesimally
transformed basis vectors in the tangent spaces to the two exponential
families, 
\begin{align}
  \frac{\partial}{\partial {\theta}^k}
  {
    \left( 
      \frac{\partial}{\partial {\theta}^j}
    \right)
  }^{\prime} 
& \equiv 
  {{\Gamma}^{\left( \theta \right)}_{kj}}^i
  \left( 
    \frac{\partial}{\partial {\theta}^i}
  \right)
\nonumber \\ 
  \frac{\partial}{\partial {\theta}^k}
  {
    \left( 
      \frac{\partial}{\partial {\eta}^j}
    \right)
  }^{\prime} 
& \equiv 
  {{\Gamma}^{\left( \eta \right)}_{kj}}^i
  \left( 
    \frac{\partial}{\partial {\eta}^i}
  \right)
\nonumber \\ 
  \frac{\partial}{\partial {\eta}^k}
  {
    \left( 
      \frac{\partial}{\partial {\eta}^j}
    \right)
  }^{\prime}
& \equiv 
  {{\Gamma}^{\left( \theta \right) \ast}_{kj}}^i
  \left( 
    \frac{\partial}{\partial {\eta}^i}
  \right)
\nonumber \\ 
  \frac{\partial}{\partial {\eta}^k}
  {
    \left( 
      \frac{\partial}{\partial {\theta}^j}
    \right)
  }^{\prime}
& \equiv
  {{\Gamma}^{\left( \eta \right) \ast}_{kj}}^i
  \left( 
    \frac{\partial}{\partial {\theta}^i}
  \right)
\label{eq:theta_eta_Gammas}
\end{align}
Superscripts ${\Gamma}^{\left( \theta \right)}$ or ${\Gamma}^{\left(
\eta \right)}$ refer to the subspace of basis vectors $\partial /
\partial \theta$ or $\partial / \partial \eta$ being pulled back, and
the designation $\Gamma$ or ${\Gamma}^{\ast}$ distinguishes the
connection associated with $\theta$ displacement or $\eta$
displacement, respectively.  Because the $\dot{\eta}$ component of
time translation does not act in $\delta \theta$ and vice versa, we
set connection coefficients ${{\Gamma}^{\left( \eta
\right)}_{kj}}^i$ and ${{\Gamma}^{\left( \theta \right) \ast}_{kj}}^i$
to zero.

Covariant derivatives of the vector fields $\delta \theta$ and $\delta
\eta$ in the connections $\Gamma$, ${\Gamma}^{\ast}$ of
Eq.~(\ref{eq:theta_eta_Gammas}) are defined as
\begin{align}
  {
    \left( 
      {\nabla}^{\left( \theta \right)}_k
      \delta \theta 
    \right)
  }^i 
& = 
  \frac{\partial}{\partial {\theta}^k}
  {
    \left( 
      \delta \theta 
    \right)
  }^i + 
  {{\Gamma}^{\left( \theta \right)}_{kj}}^i
  {
    \left( 
      \delta \theta 
    \right)
  }^j
\nonumber \\ 
  {
    \left( 
      {\nabla}^{\left( \eta \right)}_k
      \delta \theta 
    \right)
  }^i
& = 
  \frac{\partial}{\partial {\eta}^k}
  {
    \left( 
      \delta \theta 
    \right)
  }^i 
\nonumber \\ 
  {
    \left( 
      {\nabla}^{\left( \theta \right) \ast}_k
      \delta \eta 
    \right)
  }^j 
& = 
  \frac{\partial}{\partial {\theta}^k}
  {
    \left( 
      \delta \eta 
    \right)
  }^j 
\nonumber \\ 
  {
    \left( 
      {\nabla}^{\left( \eta \right) \ast}_k
      \delta \eta 
    \right)
  }^j 
& = 
  \frac{\partial}{\partial {\eta}^k}
  {
    \left( 
      \delta \eta 
    \right)
  }^j + 
  {{\Gamma}^{\left( \eta \right) \ast}_{ki}}^j
  {
    \left( 
      \delta \eta 
    \right)
  }^i
\label{eq:dual_pair_covars}
\end{align}
The covariant part of the flow decomposition in
Eq.~(\ref{eq:dts_to_flows}) is defined by subtraction of the nonzero
connection coefficients from the total
derivatives~(\ref{eq:vec_flds_timedep}), as  
\begin{align}
& 
  {
    \left( 
      \frac{\partial}{\partial t}
      \delta \theta 
    \right)
  }^j + 
  {\dot{\theta}}^k
  {
    \left( 
      {\nabla}^{\left( \theta \right)}_k
      \delta \theta 
    \right)
  }^j + 
  {\dot{\eta}}^k
  {
    \left( 
      {\nabla}^{\left( \eta \right)}_k
      \delta \theta 
    \right)
  }^j 
\nonumber \\ 
& \equiv 
  {
    \left( 
      \frac{d}{dt}
      \delta \theta 
    \right)
  }^j + 
  {\dot{\theta}}^k
  {{\Gamma}^{\left( \theta \right)}_{ki}}^j
  {
    \left( 
      \delta \theta 
    \right)
  }^i
\nonumber \\ 
& 
  {
    \left( 
      \frac{\partial}{\partial t}
      \delta \eta 
    \right)
  }^j + 
  {\dot{\theta}}^k
  {
    \left( 
      {\nabla}^{\left( \theta \right) \ast}_k
      \delta \eta 
    \right)
  }^j + 
  {\dot{\eta}}^k
  {
    \left( 
      {\nabla}^{\left( \eta \right) \ast}_k
      \delta \eta 
    \right)
  }^j 
\nonumber \\ 
& \equiv 
  {
    \left( 
      \frac{d}{dt}
      \delta \eta 
    \right)
  }^j + 
  {\dot{\eta}}^k
  {{\Gamma}^{\left( \eta \right) \ast}_{ki}}^j
  {
    \left( 
      \delta \eta 
    \right)
  }^i
\label{eq:vec_tot_der_decomp}
\end{align}
Compensating covariant derivatives of the metric are 
\begin{align}
  {\nabla}^{\left( \theta \right)}_k
  g_{ij}
& = 
  \frac{\partial}{\partial {\theta}^k} 
  g_{ij} - 
  {{\Gamma}^{\left( \theta \right)}_{ki}}^l
  g_{lj} 
\nonumber \\ 
  {\nabla}^{\left( \eta \right) \ast}_k
  g_{ij}
& = 
  \frac{\partial}{\partial {\eta}^k} 
  g_{ij} - 
  {{\Gamma}^{\left( \eta \right) \ast}_{kj}}^l
  g_{il}
\label{eq:covar_metric}
\end{align}
Eq.~(\ref{eq:vec_tot_der_decomp}) extracts a coordinate-invariant
component of the time derivative of vector fields $\delta \theta$ and
$\delta \eta$ under canonical transformations, while
Eq.~(\ref{eq:covar_metric}) extracts the corresponding
coordinate-invariant part of the change in the Fisher metric.

\subsubsection{Referencing arbitrary dual connections to dually flat
connections in the exponential family}

The manifold for a Doi-Peliti system with $D$ independent components
has dimension $2D$, with parallel subspaces for the base distribution
and likelihood ratio.  The dual
connections~(\ref{eq:theta_eta_Gammas}) act within these two
independent subspaces, in contrast to the dually-flat connections
${\Gamma}^D$ and ${\Gamma}^{D\ast}$ of Eq.~(\ref{eq:Gammas_D_forms}),
which act within the same $D$-dimensional exponential family.
Although the Fisher metric is a function only of the overall
importance distribution, which aggregates dependence from the base
distribution and likelihood, the symplectic transformations from
translation along stationary paths separate components of variation
from within the two independent subspaces.  The subspace decomposition
cannot be recovered from the importance distribution alone, and thus
no connection defined only from the properties of the Fisher metric is
sufficient to identify the dual symplectic connections for a
Doi-Peliti system.

Nonetheless, we may relate the symplectic dual connections to Amari's
dually flat connections and the Amari-Chentsov tensor through the
relation (see~\cite{Amari:inf_geom:01} Eq.~(6.27))
\begin{equation}
  {\partial}_k g^D_{ij} = 
  {\Gamma}^D_{kij} + 
  {\Gamma}^{D \ast}_{kji} = 0 + T_{kji}
\label{eq:dual_par_g}
\end{equation}
Substituting Eq.~(\ref{eq:dual_par_g}) into
Eq.~(\ref{eq:covar_metric}) gives expressions for the dual covariant
derivatives of the Fisher metric
\begin{align}
  {\nabla}^{\left( \theta \right)}_k
  g_{ij}
& = 
  T_{kji} - 
  {\Gamma}^{\left( \theta \right)}_{kij}
\nonumber \\ 
  {\nabla}^{\left( \eta \right) \ast}_k
  g_{ij}
& = 
  T_{kji} - 
  {\Gamma}^{\left( \eta \right) \ast}_{kji}
\label{eq:covar_metric_minuspar}
\end{align}

\subsubsection{Flat connections for coherent-state coordinates} 

Of particular interest in Doi-Peliti theory will be the canonical
transformations~(\ref{eq:AA_std_defs}) and~(\ref{eq:dual_AA}) between
coherent-state and number-potential (or action-angle) coordinates.  We
note that the forms of the connection coefficients for which affine
transport in fields ${\phi}^{\dagger}$ is flat in the likelihood
subspace, and affine transport in fields $\varphi$ is flat in the
base-distribution subspace, are\footnote{In the basis of the original
species counts, by Eq.~(\ref{eq:AA_std_defs}) and
Eq.~(\ref{eq:dual_AA}), the measures between action-angle and
coherent-state basis vectors are 
\begin{align}
  \frac{\partial}{\partial {\theta}^i}
& = 
  {\phi}^{\dagger i}
  \frac{\partial}{\partial {\phi}^{\dagger i}}
& 
  \frac{\partial}{\partial {\phi}^{\dagger i}}
& = 
  e^{- {\theta}^i}
  \frac{\partial}{\partial {\theta}^i}
\nonumber \\ 
  \frac{\partial}{\partial {\eta}^i}
& = 
  {\varphi}^i
  \frac{\partial}{\partial {\varphi}^i}
& 
  \frac{\partial}{\partial {\varphi}^i}
& = 
  e^{- {\eta}^i}
  \frac{\partial}{\partial \eta}
\label{eq:basis_tilt_fam}
\end{align}
Therefore the connection coefficients~(\ref{eq:dual_Christoffels_CS})
in the species basis are  
\begin{align}
  {{\Gamma}^{\left( \theta \right)}_{ki}}^j
& = 
  {\delta}_{ki}
  {\delta}_i^j
\nonumber \\ 
  {{\Gamma}^{\left( \eta \right) \ast}_{ki}}^j
& = 
  {\delta}_{ki}
  {\delta}_i^j
\label{eq:Gamma_CS_evals}
\end{align}
For various reasons it will, however, not be convenient to work in the
species basis, and the resulting connection coefficients will not
generally be constant.
}
\begin{align}
  {{\Gamma}^{\left( \theta \right)}_{kj}}^i
& = 
  \frac{\partial {\theta}^i}{\partial {\phi}^{\dagger l}}
  \frac{\partial}{\partial {\theta}^k}
  \left(
    \frac{\partial {\phi}^{\dagger l}}{\partial {\theta}^j}
  \right) 
\nonumber \\ 
  {{\Gamma}^{\left( \eta \right)}_{kj}}^i
& = 
  0 
\nonumber \\ 
  {{\Gamma}^{\left( \theta \right) \ast}_{kj}}^i
& = 
  0 
\nonumber \\ 
  {{\Gamma}^{\left( \eta \right) \ast}_{kj}}^i
& = 
  \frac{\partial {\eta}^i}{\partial {\varphi}^l}
  \frac{\partial}{\partial {\eta}^k}
  \left(
    \frac{\partial {\varphi}^l}{\partial {\eta}^j}
  \right) 
\label{eq:dual_Christoffels_CS}
\end{align}

\subsection{On the roles of coherent-state versus number-potential
coordinates in the Doi-Peliti representation}
\label{sec:CS_vs_AA}

The Doi-Peliti solution method is almost always introduced through the
coherent-state
representation~\cite{Mattis:RDQFT:98,Kamenev:DP:02,Baez:QTSM:17}, and
for many applications such as chemical reaction
networks~\cite{Baez:QTRN:13,Baez:QTRN_eq:14,Krishnamurthy:CRN_moments:17,%
Smith:CRN_moments:17} or evolutionary population
processes~\cite{Smith:evo_games:15}, coherent states are also the
``native'' representation in the sense that the Liouville operator is
a finite-order (generally low-order) polynomial in fields.  Moreover,
for the importance-sampling interpretation emphasized in this paper,
the coherent-state representation separates the nominal distribution
and likelihood ratio.

On the other hand, Legendre duality is defined with respect to
potential fields, which are the tilt coordinates $\theta$ in the
exponential family of importance distributions, and it is in these
coordinates, not the coherent-state coordinates, that the Fisher
metric corresponds to the Hessian of the CGF.  Indeed, it is not
generally possible to define a dual coordinate system from the Hessian
of the CGF in coherent-state fields, as we illustrate for the worked
example in Sec.~\ref{sec:why_not_invertible}.  

The use of Riemannian connections neatly expresses the role of each
coordinate system.  The elementary eigenvalues of divergence or
convergence of bases and tilts, and of information susceptibilities,
are often simple in coherent-state coordinates, where they are
eigenvalues of coordinate divergence or convergence.  In the dual
connections~(\ref{eq:dual_Christoffels_CS}), covariant derivatives
retain these elementary eigenvalues, while inheriting from the
exponential family the Fisher geometry that defines
contravariant/covariant coordinate duality.  A concrete example is
given in the next section.

\section{A worked example: the 2-state linear system}
\label{sec:example_two_state}

The foregoing constructions are nicely illustrated in minimal form in
a simple, exactly solvable model.  It is the stochastic process for
$N$ independent random walkers on a network with two states and
bidirectional hopping between them.  The statistical mechanics of
transients, time-dependent generating functionals, and large
deviations for this system has been didactically covered within the
Doi-Peliti framework in~\cite{Smith:LDP_SEA:11}.  Though simple, the
model is nonetheless rich enough to illustrate the complementary roles
of coherent states and number-potential coordinates in Doi-Peliti
theory -- the former as the ``native'' coordinates in which the system
is simple, and the latter as the coordinate system carrying the Fisher
geometry -- and the way this relation is captured by the dual
coherent-state connection~(\ref{eq:dual_Christoffels_CS}) different
from both the Levi-civita connection and the dually-flat
connections~(\ref{eq:D_tens_to_tens}) of Nagaoka and
Amari~\cite{Nagaoka:dual_geom:82,Amari:methods_IG:00}. 

\subsection{Two-argument and one-argument generating functions on
distributions with a conserved quantity}

The two-state model describes (for example) a one-particle chemical
reaction in a well-mixed reactor with the schema 
\begin{equation}
  a 
  \overset{
    {\bm k}_{+}
  }{
    \underset{
      {\bm k}_{-}
    }{
      \rightleftharpoons
    }
  }
  b 
\label{eq:2state_schema}
\end{equation}
The probability per unit time for a reaction event is given by rate
constants ${\bm k}_{+}$ and ${\bm k}_{-}$, and proportional sampling
(the microphysics underlying mass-action rate laws).

A distribution initially in binomial form~(\ref{eq:multinom_form})
will retain that form at all times under the master equation for the
schema~(\ref{eq:2state_schema}), even with time-dependent
coefficients.  Here for simplicity we will take ${\bm k}_{+}$ and
${\bm k}_{-}$ to be fixed.  Therefor the distribution at any time is
specified by descaled mean values ${\nu}_a = \left< {\rn}_a \right> /
N$, ${\nu}_b = \left< {\rn}_b \right> /
N$, with ${\nu}_a + {\nu}_b = 1$.  

Although the system has only one dynamical degree of freedom, it is
instructive to compute both the 2-argument generating function with
independent weights $z_a$ on ${\rn}_a$ and $z_b$ on ${\rn}_b$, and the
1-argument generating function for the difference coordinate $\rn
\equiv \left( {\rn}_b - {\rn}_a \right) / 2$, to illustrate the role
of conservation laws and the geometry of the coherent-state
connection.  The 2-argument CGF~(\ref{eq:CGF_onetime_def}) for the
binomial distribution is
\begin{equation}
  e^{\psi \left( \log z \right)} \equiv 
  \sum_{\rn}
  z_a^{{\rn}_a}
  z_b^{{\rn}_b}
  {\rho}_{{\rn}_a , {\rn}_b} = 
  {
    \left[ 
      z_a {\nu}_a + z_b {\nu}_b
    \right] 
  }^N
\label{eq:psi_binom_twovar}
\end{equation}
Because the total number $N = \left( {\rn}_b + {\rn}_a \right)$ is
fixed, the normalized 1-variable distribution may be written 
\begin{equation}
  {\rho}_{\rn} = 
  {\sqrt{{\nu}_b {\nu}_a}}^N
  \left(
    \begin{array}{c}
      N \\ n 
    \end{array}
  \right)
  {
    \left( \frac{{\nu}_b}{{\nu}_a} \right) 
  }^{\rn}
\label{eq:rho_bin_onearg_form}
\end{equation}
and the terms in the generating function~(\ref{eq:psi_binom_twovar})
regrouped as  
\begin{align}
  e^{\psi \left( \log z \right)} 
& = 
  {\sqrt{z_b z_a}}^N
  \sum_{\rn}
  {\rho}_{\rn} 
  {
    \left( \frac{z_b}{z_a} \right) 
  }^{\rn}
\label{eq:psi_binom_twovar_split}
\end{align}

Introducing rotated coordinates on the exponential family of tilts
\begin{align}
  h 
& \equiv 
  \left( {\theta}_b + {\theta}_a \right) / 2 
\nonumber \\ 
  \theta 
& \equiv 
  \left( {\theta}_b - {\theta}_a \right) 
\label{eq:h_theta_def}
\end{align}
and dividing the two-argument MGF~(\ref{eq:psi_binom_twovar_split}) by
${\sqrt{z_a z_b}}^N$, we obtain an expression for the one-argument MGF
in the difference coordinate $\rn$: 
\begin{align}
  e^{
    \psi \left( \log z \right) - 
    N h
  } 
& = 
  \sum_{\rn}
  {\rho}_{\rn} 
  e^{\theta \rn}
\label{eq:psi_binom_onevar}
\end{align}
In what follows, $\psi \left( \log z \right)$ will always be used to
refer to the 2-argument CGF~(\ref{eq:psi_binom_twovar}), and the
1-argument generating function, when needed, will be written out
explicitly as $\psi \! \left( \log z \right) - N h$, as in
Eq.~(\ref{eq:psi_binom_onevar}).

\subsection{Generator and conserved volume element in coherent-state
coordinates} 

The master equation for the 2-state system is developed
in~\cite{Smith:LDP_SEA:11}, but introduces further notation, and will
not be needed here.  We move directly to the expression for the
Liouville function of Eq.~(\ref{eq:CRN_L_genform}) after conversion to
field variables, which is
\begin{equation}
  \mathcal{L} = 
  {\bm k}_{+} 
  \left( 
    {\phi}^{\dagger}_a - 
    {\phi}^{\dagger}_b 
  \right)
  {\phi}_a + 
  {\bm k}_{-} 
  \left( 
    {\phi}^{\dagger}_b - 
    {\phi}^{\dagger}_a 
  \right)
  {\phi}_b 
\label{eq:Liouville_fields}
\end{equation}
In what follows, math boldface will be reserved for parameters in the
generator such as ${\bm k}_{\pm}$ or functions of these such as the
associated steady states used in Eq.~(\ref{eq:Baish_trans_only}).

Two descalings reduce the problem to parameters which are
dimensionless ratios.  The first defines a time coordinate $\tau$ in
units of the sum of rate parameters, 
\begin{equation}
  \frac{d\tau}{dt} \equiv 
  {\bm k}_{+} + {\bm k}_{-} 
\label{eq:dtau_dt_def}
\end{equation}
The second expresses the equilibrium steady state under
generator~(\ref{eq:Liouville_fields}) in terms of relative hopping
rates,
\begin{align}
  \frac{{\bm k}_{+}}{{\bm k}_{+} + {\bm k}_{-}}
& = 
  \frac{{\bm n}_b}{N} \equiv 
  {\bm \nu}_b
& 
  \frac{{\bm k}_{-}}{{\bm k}_{+} + {\bm k}_{-}}
& = 
  \frac{{\bm n}_a}{N} \equiv 
  {\bm \nu}_a
\label{eq:norms_def}
\end{align}
As for the discrete index $\rn$, define  ${\bm \nu} \equiv \left( {\bm
\nu}_b - {\bm \nu}_a \right) / 2$. 

Conservation of total number $N$ results in a generator $\mathcal{L}$
that is a function only of the difference variable $\left(
{\phi}^{\dagger}_b - {\phi}^{\dagger}_a \right)$.  Therefore it is
natural to rotate the coherent-state fields to components
corresponding to conserved $N$ and dynamical $\rn / N$, and their dual
coordinates in the generating-function argument $z$:
\begin{align}
  {\phi}^{\dagger} 
& \equiv 
  {\phi}^{\dagger}_b -  
  {\phi}^{\dagger}_a 
& 
  \hat{\phi} 
& \equiv 
  \left( {\phi}_b - {\phi}_a \right) / 2N 
\nonumber \\
  {\Phi}^{\dagger} 
& \equiv 
  \left( {\phi}^{\dagger}_b + {\phi}^{\dagger}_a \right) / 2 . 
& 
  \hat{\Phi} 
& \equiv 
  \left( {\phi}_b + {\phi}_a \right) / N 
\label{eq:fields_rotate_basis}
\end{align}

In rotated fields~(\ref{eq:fields_rotate_basis}) the
action~(\ref{eq:CRN_L_genform}) becomes
\begin{eqnarray}
  S 
& = & 
  N \int d\tau 
  \left[ 
    - {\partial}_{\tau}
    {\Phi}^{\dagger}
    \hat{\Phi} - 
    {\partial}_{\tau}
    {\phi}^{\dagger}
    \hat{\phi} + 
    {\phi}^{\dagger}
    \left( 
      \hat{\phi} - 
      {\bm \nu} \hat{\Phi}
    \right)
  \right]
\nonumber \\
& \equiv & 
  N \int d\tau 
  \left( 
    - {\partial}_{\tau}
    {\Phi}^{\dagger}
    \hat{\Phi} - 
    {\partial}_{\tau}
    {\phi}^{\dagger}
    \hat{\phi} + 
    \hat{\mathcal{L}}
  \right) 
\label{eq:action_twofields_descaled}
\end{eqnarray}
A descaled Liouville function has been introduced as $N \left( {\bm
k}_{+} + {\bm k}_{-} \right) \hat{\mathcal{L}} \equiv \mathcal{L}$.
Absence of the field ${\Phi}^{\dagger}$ from $\hat{\mathcal{L}}$
implies constancy of the expectation for $\hat{\Phi}$.\footnote{It
implies constancy of a tower of higher-order correlation functions
expressing exact conservation of the underlying variable $N$, though
we do not develop the 2FFI representation of correlation functions in
this paper.}

\subsubsection{Splitting the symplectic structure between
coherent-state conjugate field pairs}

Although $\hat{\Phi}$ obeys certain time-translation invariances in
correlation functions, its value even along stationary paths will not
generally be 1.  Therefore the coherent-state variables cannot
directly be interpreted as mean values of number variables in the
nominal distribution or mean weights in its dual likelihood ratio.
To express the functions that are these expectation values, we recall
the mean number components in the importance distribution, which are
bilinear quantities in ${\phi}^{\dagger}$ and $\phi$, and then
introduce a pair of dual number coordinates that, while not linear
functions of the coherent-state fields, are functions respectively of
${\phi}^{\dagger}$ or of $\phi$ extracted by making use of the
steady-state measure under the instantaneous value ${\bm
\nu}$ in the generator~(\ref{eq:Liouville_fields}).  (Along stationary
paths, where some components of ${\phi}^{\dagger}$ or $\phi$ are
invariant, these dual number fields will become linear functions of
the remaining dynamical components of ${\phi}^{\dagger}$ or $\phi$, as
we show below.)

The two components of the normalized number field in action-angle
coordinates~(\ref{eq:AA_std_defs}) are given by 
\begin{align}
  \frac{1}{N}
  {\phi}^{\dagger}_b 
  {\phi}_b 
& \equiv 
  {\nu}_b \equiv 
  \left( \frac{1}{2} + \nu \right)
& 
  \frac{1}{N}
  {\phi}^{\dagger}_a 
  {\phi}_a 
& \equiv 
  {\nu}_a \equiv 
  \left( \frac{1}{2} - \nu \right)
\label{eq:number_fields_def}
\end{align}

Recall that the instantaneous steady state under the generating
process is the scale variable for the dualizing canonical
transform~(\ref{eq:Baish_trans_only}).  To see how this reference
steady state is used to separate the two conjugate variables (base and
tilt) in the symplectic transformations, it is helpful to recast
Eq.~(\ref{eq:number_fields_def}) as

\begin{align}
  \frac{1}{2}
  \frac{
    \left( 
      {\phi}^{\dagger}_b 
      {\phi}_b - 
      {\phi}^{\dagger}_a 
      {\phi}_a 
    \right) 
  }{
    \left( 
      {\phi}^{\dagger}_b 
      {\phi}_b + 
      {\phi}^{\dagger}_a 
      {\phi}_a 
    \right) 
  }
& \equiv 
  \frac{
  \left( 
    {\nu}_b - {\nu}_a
  \right) 
  }{
    2 
  } \equiv 
  \nu
\label{eq:nu_as_ratio}
\end{align}

The action of the tilt alone can be isolated, without regard to the
underlying nominal distribution, by referencing the action of the
${\phi}^{\dagger}$ fields to the steady state rather than to $\phi$,
defining an offset $\underline{\nu}$ as 
\begin{align}
  \frac{1}{2}
  \frac{
    \left( 
      {\phi}^{\dagger}_b {\bm \nu}_b - 
      {\phi}^{\dagger}_a {\bm \nu}_a
    \right) 
  }{
    \left( 
      {\phi}^{\dagger}_b {\bm \nu}_b + 
      {\phi}^{\dagger}_a {\bm \nu}_a
    \right) 
  }
& \equiv 
  \frac{
  \left( 
    {\underline{\nu}}_b - {\underline{\nu}}_a
  \right) 
  }{
    2 
  } \equiv 
  \underline{\nu}
\label{eq:und_nu_def}
\end{align}

Likewise, the mean value $\bar{\nu}$ of $\rn / N$ in the base
(nominal) distribution is isolated by referencing the value of $\phi$
to the uniform measure $1$ instead of the dynamic measure
${\phi}^{\dagger}$, as
\begin{align}
  \frac{1}{2}
  \frac{
    \left( 
      {\phi}_b - {\phi}_a
    \right) 
  }{
    \left( 
      {\phi}_b + {\phi}_a
    \right) 
  }
& \equiv 
  \frac{
  \left( 
    {\bar{\nu}}_b - {\bar{\nu}}_a
  \right) 
  }{
    2 
  } \equiv 
  \bar{\nu}
\label{eq:bar_nu_def}
\end{align}

\subsubsection{Stationary-path solutions and Liouville volume element}
\label{sec:examp_stat_path}

Solutions to the stationary-path equations of
motion~(\ref{eq:Hamiltonian_var_CS}) for the Liouville
function~(\ref{eq:Liouville_fields}) are evaluated in
App.~\ref{sec:CS_statpath_solns}.  

Stationary-path approximations to the time-dependent density $\rho$
would be binomial distributions even if the exact $\rho$ were not (the
stationary point is always a pure coherent state), so the CGF at any
time has the form~(\ref{eq:psi_binom_twovar}), with fields $z$
replaced by the stationary-path values of ${\phi}^{\dagger}$ and the
mean values $\nu$ from Eq.~(\ref{eq:rho_bin_onearg_form}) replaced by
corresponding components of $\phi$.

In particular, the initial-time generating function ${\psi}_0 \left(
\log {\phi}^{\dagger}_{a0} , \log {\phi}^{\dagger}_{b0} \right)$
appearing in Eq.~(\ref{eq:gen_fn_twofields_nosource}) carries the
mean value ${\bar{\nu}}_0$ in the starting density ${\rho}_{0}$,
imposed as an initial-data parameter.  It is through this function
that the final-time tilt data in the form of the parameter
${\underline{\nu}}_T$, propagated forward to the stationary-path
values of ${\phi}^{\dagger}_{a0}$ and ${\phi}^{\dagger}_{b0}$,
determines the stationary path values for the fields $\phi$ of the
base distribution, establishing the potential for information coupling
between initial properties of the base distribution and final-time
queries in the generating function ${\psi}_T$.  

${\psi}_0$ is evaluated in Eq.~(\ref{eq:gamma_0_form}), and the value
is shown to depend only on an overlap parameter between initial and
final data which we denote 
\begin{equation}
  \Lambda \equiv 
  \frac{
    \left( {\bar{\nu}}_0 - {\bm \nu} \right) 
    \left( {\underline{\nu}}_T - {\bm \nu} \right) 
  }{
    \left( \frac{1}{4} - {\bm \nu}^2 \right) 
  } 
\label{eq:Lambda_def}
\end{equation}

\begin{figure}[ht]
\begin{center} 
  \includegraphics[scale=0.6]{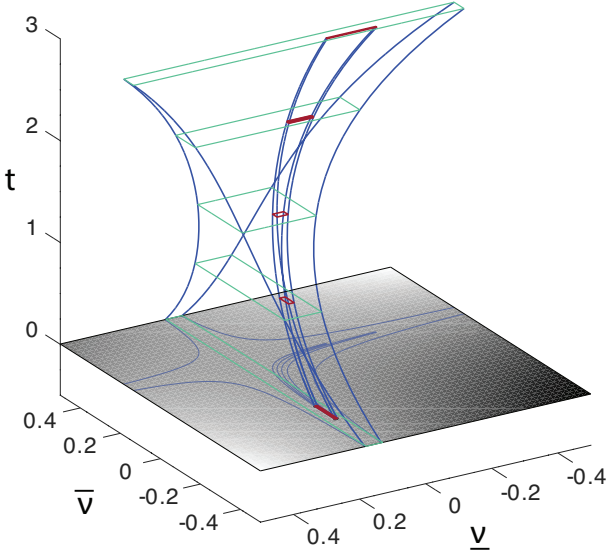} 
  \caption{
  Four trajectories (heavy blue contours), plotted in coordinates
  $\left( \underline{\nu} , \bar{\nu} \right)$, that bound a region
  specified by ${\underline{\nu}}_T = -1/6 \pm 0.075$, ${\bar{\nu}}_0
  = -1/4 \pm 0.05$.  The steady state under the generating process
  sets ${\bm \nu} = 1/6$.  A time interval $T = 3$ between the input
  distribution and the final-time generating function is shown.  Small
  rectangles (heavy dark red) show the area $\delta \underline{\nu} \,
  \delta \bar{\nu}$ at five equally spaced times from start to end.
  The outer four trajectories (thin blue) show the possible range of
  joint images of $-\sfrac{1}{2} \le {\bar{\nu}}_0 \le \sfrac{1}{2}$
  and $-\sfrac{1}{2} \le {\underline{\nu}}_T \le \sfrac{1}{2}$.  Large
  rectangles (thin green) show the constriction of the possible range
  $\propto e^{-T}$.  Projections of the total range and the inner
  trajectories are shown in thin lines on the base plane.  Shading of
  the base plane is a grayscale plot of ${\hat{\Phi}}_0^2$, which is
  constant along trajectories but variable over the $\left(
  \underline{\nu} , \bar{\nu} \right)$ coordinate range.  Min and max
  of ${\hat{\Phi}}_0^2$ are respectively $0.83$ and $1.1$.
  \label{fig:L_volumes} 
  } 
\end{center}
\end{figure}

The stationary-path values of the displacement
coordinates~(\ref{eq:und_nu_def}) and~(\ref{eq:bar_nu_def}) are shown
in Equations~(\ref{eq:true_mean_tau}) and~(\ref{eq:null_mean_tau}) to
follow simple exponential laws
\begin{align}
  \bar{\nu} - {\bm \nu} 
& = 
  \left( {\bar{\nu}}_0 - {\bm \nu} \right) 
  e^{-\tau}
\nonumber \\ 
  \underline{\nu} - {\bm \nu} 
& = 
  \left( {\underline{\nu}}_T - {\bm \nu} \right) 
  e^{\tau - T}
\label{eq:und_bar_nus_sol}
\end{align}
Thus under independent variations of ${\bar{\nu}}_0$ and
${\underline{\nu}}_T$ as described in Sec.~\ref{sec:dual_vec_fields},
the trajectories of the coherent state fields ${\phi}^{\dagger}$ and
$\phi$ trace out an invariant volume, illustrated in
Fig.~\ref{fig:L_volumes}.  

\subsubsection{Invariant cumulant-generating function and the
incompressible phase-space density}

The stationary value of ${\hat{\Phi}}_0$, obtained from the gradient
of ${\psi}_0$ with respect to the components ${\phi}^{\dagger}_{0a}$
and ${\phi}^{\dagger}_{0b}$, is computed in
Eq.~(\ref{eq:phis_bothsol_ofz}).  It differs from unity -- the reason
constructions~(\ref{eq:und_nu_def}) and~(\ref{eq:bar_nu_def}) were
needed -- and it is equal to the stationary value of $\hat{\Phi}$ at
all times as a consequence of conservation of total number $N$.  The
value depends only on $\Lambda$ and $T$ in the combination 
\begin{align}
  {\hat{\Phi}}_0 = 
  \frac{
    1 
  }{
    1 + \Lambda e^{-T}
  }
\label{eq:hatPhi_0_value}
\end{align}
Moreover, as a consequence of the conserved Liouville volume element
from Eq.~(\ref{eq:und_bar_nus_sol}), the stationary-point evaluation
of the CGF at all times takes the same form as
Eq.~(\ref{eq:psi_binom_twovar}) and evaluates to the constant
\begin{align}
  \frac{\psi}{N}
& = 
  \log 
  \left[ 
    \frac{{\underline{\nu}}_a}{{\bm \nu}_a} 
    {\bar{\nu}}_a + 
    \frac{{\underline{\nu}}_b}{{\bm \nu}_b}
    {\bar{\nu}}_b
  \right] 
\nonumber \\
& = 
  - \log {\hat{\Phi}}_0
\label{eq:gen_fn_eval}
\end{align}

${\hat{\Phi}}_0$ in Eq.~(\ref{eq:hatPhi_0_value}) is the basis for all
information densities in this simple linear system.  Through the
stationary-point relation~(\ref{eq:genfun_from_Wigner_SP}) between the
Wigner function and the CGF, $-\log {\hat{\Phi}}_0$ is the
incompressible phase-space density convected along stationary
trajectories by Eq.~(\ref{eq:Wigner_stat_CS_rot}).  As shown below, it
is also the geometrically invariant part of sole nonzero eigenvalue of
the Fisher metric.

\subsection{Fisher metric}

The Fisher metric~(\ref{eq:Fisher_as_innerprod}) for the 2-state
system evaluates, along the stationary path at any time, to  
\begin{align}
  \frac{g}{N} 
& = 
  \frac{\partial \nu}{\partial \theta}
  \left[
    \begin{array}{r}
      1 \\ -1 
    \end{array}
  \right] \! \! 
  \begin{array}{c}
    \left[
      \begin{array}{cc}
        1 & -1 
      \end{array}
    \right] \\
    \phantom{[]}
  \end{array}
\label{eq:g_from_nat}
\end{align}
The nonzero eigenvalue comes from the single-argument generating
function in Eq.~(\ref{eq:psi_binom_onevar}) for the difference
coordinate $\rn$, and the zero eigenvalue comes from the linear CGF $h
N$ for the conserved quantity $N$.

The term $\partial \nu / \partial \theta$ in Eq.~(\ref{eq:g_from_nat})
may be converted, after some algebra, to the form 
\begin{align}
  \frac{\partial \nu}{\partial \theta}
& = 
  \frac{
    \left(
      \frac{1}{4} - 
      {\underline{\nu}}^2
    \right) 
    \left(
      \frac{1}{4} - 
      {\bar{\nu}}^2
    \right) 
  }{
    \left(
      \frac{1}{4} - 
      {\bm \nu}^2
    \right) 
  }
  {\hat{\Phi}}_0^2
\label{eq:g_coeff_as_measures}
\end{align}
The measure terms $\left( \sfrac{1}{4} - {\underline{\nu}}^2 \right)$
and $\left( \sfrac{1}{4} - {\bar{\nu}}^2 \right)$ appearing in
Eq.~(\ref{eq:g_coeff_as_measures}) follow the
divisions~(\ref{eq:und_nu_def}) and~(\ref{eq:bar_nu_def}) into
independent dimensions of base and tilt variation, and we will show
that their effects are canceled in an appropriate covariant
derivative.  The remaining dependence of the eigenvalue on the initial
and final data is all carried in ${\hat{\Phi}}_0^2$.

\subsection{Dual coordinates for base and tilt, and the additive
exponential family}

To relate the Fisher metric in Eq.~(\ref{eq:g_from_nat}) to the
construction of Sec.~\ref{sec:Fisher_from_psi} from the
$\psi$-divergence and to dually-symplectic parallel transport, we
first express the base and tilt displacements~(\ref{eq:und_nu_def})
and~(\ref{eq:bar_nu_def}) in terms of the coordinates in their
respective exponential families.

Introduce reference values for the fields $\theta$ and $h$ defined in
Eq.~(\ref{eq:h_theta_def}), corresponding to the steady-state measure
under the parameters of the generating process, denoted
\begin{align}
  {\bm \theta} 
& \equiv 
  \log 
  \left( 
    \frac{
      \frac{1}{2} + {\bm \nu} 
    }{
      \frac{1}{2} - {\bm \nu} 
    }
  \right)
\nonumber \\ 
  {\bm h} 
& \equiv 
  \frac{1}{2}
  \log 
  \left( 
    \frac{1}{4} - {\bm \nu}^2 
  \right) = 
  - \log 
  \left[ 
    2 
    \ch 
    \frac{{\bm \theta}}{2}
  \right]
\label{eq:bar_EFs_def}
\end{align}

It is clear, in the 2-argument generating
function~(\ref{eq:psi_binom_twovar}), that one component of variation
in $z$ couples only to the conserved quantity $N$ and is not needed.
It is sufficient therefore to vary along an affine coordinate in $z$
that couples to the dynamical argument $\rn$, and the natural choice
is to fix the component of $z$ corresponding to the component of
${\phi}^{\dagger}$ that is invariant under the stationary-path
equations of motion, given in Eq.~(\ref{eq:phi_dagg_ofz}).  The
resulting contour for $z$ at final time $T$ becomes 
\begin{align}
  z_b {\bm \nu}_b + 
  z_a {\bm \nu}_a 
& = 
  {\Phi}^{\dagger}_T + 
  {\bm \nu} 
  {\phi}^{\dagger}_T = 
  1 
\nonumber \\ 
  z_b - z_a 
& = 
  {\phi}^{\dagger}_T = 
  \frac{
    {\underline{\nu}}_T - {\bm \nu}
  }{
    \frac{1}{4} - {\bm \nu}^2
  }
\label{eq:z_coords}
\end{align}
The quantity in the first line of Eq.~(\ref{eq:z_coords}) is preserved
at all times by Eq.~(\ref{eq:phi_dagg_ofz}), and the quantity in the
second line obeys the exponential law of Eq.~(\ref{eq:null_mean_tau})
repeated as~(\ref{eq:und_bar_nus_sol}).

By the definition~(\ref{eq:und_nu_def}), the
contour~(\ref{eq:z_coords}) which is affine in coherent-state fields
${\phi}^{\dagger}$ is written in the coordinates on the exponential
family of tilts as
\begin{align}
  \underline{\nu}
& \equiv  
  \frac{1}{2}
  \Th
  \left( 
    \frac{
      \theta + {\bm \theta}
    }{
      2 
    }
  \right) 
\nonumber \\ 
  0 
& \equiv 
  h + {\bm h} + 
  \log 
  \left[ 
    2 
    \ch 
    \left(
      \frac{
        \theta + {\bm \theta}
      }{
        2 
      }
    \right)
  \right]
\nonumber \\ 
& = 
  h - 
  \frac{1}{2}
  \log 
  \left( 
    \frac{1}{4} - {\underline{\nu}}^2 
  \right) + 
  \frac{1}{2}
  \log 
  \left( 
    \frac{1}{4} - {\bm \nu}^2 
  \right) 
\label{eq:CS_in_EF_defs}
\end{align}
Likewise, in the dual exponential representation~(\ref{eq:dual_AA}) of
the family of base distributions, the
definition~(\ref{eq:bar_nu_def}) giving the mean number offset in the
nominal distribution is expressed
\begin{align}
  \bar{\nu} 
& = 
  \frac{1}{2}
  \Th
  \left( 
    \frac{
      \eta + {\bm \theta}
    }{
      2 
    }
  \right) 
\label{eq:theta_tau_def}
\end{align}
in which $\eta \equiv \left( {\eta}_b - {\eta}_a \right)$ in the dual
action-angle system~(\ref{eq:dual_AA}), analogously to $\theta$ in
Eq.~(\ref{eq:h_theta_def}).

Because the two exponential coordinates (base and tilt) are additive,
the mean of samples in the importance distribution can likewise be
written
\begin{align}
  \nu 
& = 
  \frac{1}{2}
  \Th
  \left( 
    \frac{
      \theta + \eta + {\bm \theta}
    }{
      2 
    }
  \right) 
\label{eq:full_LDF_theta_eta}
\end{align}
It follows that the eigenvalue~(\ref{eq:g_coeff_as_measures}) in the
Fisher metric also has the simple expression
\begin{equation}
  \frac{
    \partial \nu
  }{
    \partial \theta
  } =
  \frac{
    \partial \nu
  }{
    \partial \eta
  } =
  \frac{1}{4} -
  {\nu}^2
\label{eq:dnu_deta_Fisher}
\end{equation}
exhibiting the equivalence of the $\psi$-divergence
expression~(\ref{eq:g_D_form}) and the
Hessian~(\ref{eq:D_tens_to_tens}) for this quantity.

\subsection{Why coherent-state fields do not generally produce
invertible coordinate transformations} 
\label{sec:why_not_invertible}

The Hessian matrix is not a tensor under coordinate transform, so it
is clear that the Hessian of $\psi$ with respect to the argument $z$
equivalent to the coherent-state response field ${\phi}^{\dagger}$
will not be the Fisher metric.  However, since coherent states are in
many ways a native basis for Doi-Peliti theory, as noted in
Sec.~\ref{sec:CS_vs_AA}, we may ask whether some other coordinate
duality can be defined from the coherent-state Hessian of $\psi$.  In
fact such a duality cannot generally be defined, and it is instructive
to see where it fails, to better understand why the affine
connection~(\ref{eq:dual_Christoffels_CS}) and not the Fisher geometry
captures the special role of coherent states.

A divergence under the Hessian of $\psi$ in coherent-state variables,
which we will denote $\Delta \delta s^2$ for reasons to become clear
in a moment, if converted from the coordinates $\underline{\nu}$ to
coordinates $\theta$ along the $z$-affine
contour~(\ref{eq:CS_in_EF_defs}), evaluates as
\begin{align}
  \frac{1}{N}
  \Delta ds^2 
& \equiv 
  {
    \left( \delta \theta \right) 
  }^2
  {
    \left( 
      \frac{1}{4} - 
      {\underline{\nu}}^2
    \right) 
  }^2
  \frac{
    {\partial}^2
  }{
    \partial {\underline{\nu}}^2
  }
  \left(
    \frac{
      \psi
    }{
      N 
    }
  \right) 
\nonumber \\
& \equiv 
  - {
    \left( \delta \theta \right) 
  }^2
  {
    \left( 
      \nu - 
      \underline{\nu}
    \right) 
  }^2
\label{eq:diff_ds}
\end{align}
Unlike the Fisher metric, Eq.~(\ref{eq:diff_ds}) is
negative-semidefinite, and degenerates if $\nu = \underline{\nu}$,
which is shown in Eq.~(\ref{eq:SP_nu_diffs_compact}) to hold for all
$z$ if ${\bar{\nu}}_0 = {\bm \nu}$.  At degenerate solutions, we
cannot use the Hessian of $\psi$ to define a base-field variation
$\delta \phi$ as a dual coordinate for a variation produced by a field
$\delta {\phi}^{\dagger}$, as we \emph{could} use the Hessian in the
exponential family to produce a variation $\delta n$ as a dual
coordinate to a variation $\delta \theta$.

The source of the degeneration has a nice description in terms of
intrinsic and extrinsic curvatures, and advection, in the natural
geometry on the exponential family. The geometric distance
element~(\ref{eq:Fisher_element}), with $\theta$ and $h$ varied
independently, is
\begin{align}
  \frac{1}{N}
  \delta s^2 
& = 
  {\left( \delta \theta \right)}^2
  \frac{{\partial}^2}{\partial {\theta}^2}
  \left( 
    \frac{\psi}{N} - h
  \right) + 
  {\left( \delta h \right)}^2
  \frac{{\partial}^2}{\partial h^2}
  h
\nonumber \\ 
& = 
  {\left( \delta \theta \right)}^2
  \frac{\partial \nu}{\partial \theta} + 
  {\left( \delta h \right)}^2
  0 
\label{eq:ds2_full_coord}
\end{align}

The $z$-affine contour~(\ref{eq:CS_in_EF_defs}) specifies a function
$h \! \left( \theta \right)$ with extrinsic curvature in the affine
coordinate manifold of the exponential family, along which the
distance element is
\begin{align}
  \frac{1}{N}
  \delta s^2_{\rm CS-ext} 
& = 
  {\left( \delta \underline{\nu} \right)}^2
  \frac{
    d^2
  }{
    d {\underline{\nu}}^2
  }
  \left( -h \! \left( \theta \right) \right) 
\nonumber \\ 
& = 
  {\left( \delta \theta \right)}^2
  \left(
    \frac{1}{4} + 
    {\underline{\nu}}^2
  \right) 
\label{eq:ds2_ref_CS}
\end{align}
The second coherent-state coordinate derivative of $\psi$ along the
contour~(\ref{eq:CS_in_EF_defs}) can be decomposed as
\begin{align}
  \frac{1}{N}
  \Delta \delta s^2 
& = 
  {\left( \delta \underline{\nu} \right)}^2
  \frac{
    d^2
  }{
    d {\underline{\nu}}^2
  }
  \left(
    \frac{
      \psi
    }{
      N 
    }
  \right) 
\nonumber \\ 
& = 
  {\left( \delta \underline{\nu} \right)}^2
  \left[ 
    \frac{
      d^2
    }{
      d {\underline{\nu}}^2
    }
    \left(
      \frac{
        \psi
      }{
        N 
      } - 
      h 
    \right) + 
    \frac{
      d^2
    }{
      d {\underline{\nu}}^2
    }
    h 
  \right] 
\nonumber \\ 
& = 
  {\left( \delta \theta \right)}^2
  \left( 
    \frac{
      d^2
    }{
      d {\theta}^2
    } + 
    2 \underline{\nu}
    \frac{
      d
    }{
      d \theta
    } 
  \right) 
  \left(
    \frac{
      \psi
    }{
      N 
    } - 
    h 
  \right) + 
  {\left( \delta \underline{\nu} \right)}^2
  \frac{
    d^2
  }{
    d {\underline{\nu}}^2
  }
  h 
\nonumber \\ 
& = 
  {\left( \delta \theta \right)}^2
  \left[
    \frac{\partial \nu}{\partial \theta} + 
    2 \underline{\nu} \nu
  \right] + 
  {\left( \delta \underline{\nu} \right)}^2
  \frac{
    d^2
  }{
    d {\underline{\nu}}^2
  }
  h 
\nonumber \\ 
& = 
  \frac{1}{N}
  \left( 
    \delta s^2 - 
    \delta s^2_{\rm CS-ext} 
  \right) + 
  {\left( \delta \theta \right)}^2
  2 \underline{\nu} \nu
\label{eq:Delt_ds2_decom}
\end{align}
With some algebra, the expression~(\ref{eq:Delt_ds2_decom}) is shown
to equal that in Eq.~(\ref{eq:diff_ds}).  The extrinsic curvature of
the embedded contour $h \! \left( \theta \right)$ and the convected
quantity $- 2 \underline{\nu} \nu$ cancel against the intrinsic Fisher
curvature, rendering the duality invisible to the fields
${\phi}^{\dagger}$ at degenerate points.

\subsection{Flat transport in the coherent-state connection}

The correct way to capture the simplifying role of coherent-state
coordinates for simple models such as the 2-state system is with the
dual connections of Sec.~\ref{sec:dual_connections}.

We first recognize, from the forms~(\ref{eq:Lambda_def})
or~(\ref{eq:conserved_Lambda}) of $\Lambda$, a completely-descaled
coordinate system for the dynamical parts of the coherent-state
fields, define by 
\begin{align}
  v 
& \equiv 
  \frac{
    \left( \underline{\nu} - {\bm \nu} \right)
  }{
    \sqrt{
      \frac{1}{4} - 
      {\bm \nu}^2
    } 
  }
& 
  u 
& \equiv 
  \frac{
    \left( \bar{\nu} - {\bm \nu} \right)
  }{
    \sqrt{
      \frac{1}{4} - 
      {\bm \nu}^2
    } 
  }
\label{eq:lightcone}
\end{align}
The eigenvalue of the Fisher metric in
Eq.~(\ref{eq:g_coeff_as_measures}) then reduces to
\begin{equation}
  \frac{
    {\partial}^2
  }{
    \partial {\theta}^2
  } \! 
  \left( 
    \frac{{\psi}}{N} 
  \right) = 
  \frac{
    \left( dv / d\theta \right)
    \left( du / d\eta \right)
  }{
    {
      \left[ 1 + u v \right]
    }^2
  }
\label{eq:metric_uv_basis}
\end{equation}
The role of the factors $\left( \sfrac{1}{4} - {\underline{\nu}}^2
\right) = \sqrt{\sfrac{1}{4} - {\bm \nu}^2} \left( \partial v /
\partial \theta \right)$ and $\left( \sfrac{1}{4} - {\bar{\nu}}^2
\right) = \sqrt{\sfrac{1}{4} - {\bm \nu}^2} \left( \partial u /
\partial \eta \right)$ in Eq.~(\ref{eq:g_coeff_as_measures}) as
measure terms is now explicit, and they can be absorbed by a change of
variables to $u$ and $v$. By Eq.~(\ref{eq:conserved_Lambda}) and the
definitions~(\ref{eq:lightcone}) and~(\ref{eq:hatPhi_0_value}),
$\left[ 1 + u v \right] = \left[ 1 + \Lambda e^{-T} \right] =
1/{\hat{\Phi}}_0$, so the Fisher inner
product~(\ref{eq:triangle_to_Fisher}) may be written
\begin{align}
  \delta \theta \, 
  {\delta}_{\eta} n
  \left< 
    \frac{\partial}{\partial \theta} , 
    \frac{\partial}{\partial n}
  \right> = 
  \delta \underline{\nu} \, 
  \delta \bar{\nu}
  \left<
    \frac{\partial}{\partial \underline{\nu}} , 
    \frac{\partial}{\partial \bar{\nu}}
  \right>
& = 
  \delta u \, \delta v \, 
  {\hat{\Phi}}_0^2 
\label{eq:Fisher_lightcone}
\end{align}

\subsubsection{Connection coefficients and absorption of measure terms}

In this linear model, time evolution of ${\phi}^{\dagger}$ and $\phi$
has no cross-dependence once the initial values have been fixed
through the gradients of ${\psi}_0$ as explained in
Sec.~\ref{sec:examp_stat_path}.  Thus  ${ \left( {\nabla}^{\left( \eta
\right)}_k \delta \theta \right) }^j = 0$ and ${\left(
{\nabla}^{\left( \theta \right) \ast}_k \delta \eta \right)}^j = 0$. 

App.~\ref{sec:examp_CS_conn_coeff} computes connection coefficients
and covariant derivatives for the vector fields corresponding to
Eq.~(\ref{eq:vec_tot_der_decomp}), and for the metric tensor
corresponding to Eq.~(\ref{eq:covar_metric}).
Eq.~(\ref{eq:model_dual_covars_vecs}) in the appendix gives the
covariant part of the time derivatives of $\delta \theta$ and $\delta
\eta$ as 
\begin{align}
  \left( 
    \frac{\partial}{\partial \tau}
    \delta \theta 
  \right) + 
  \dot{\theta}
  \left( 
    {\nabla}_{\theta}
    \delta \theta 
  \right)
& = 
  \delta \theta 
\nonumber \\ 
  \left( 
    \frac{\partial}{\partial t}
    \delta \eta 
  \right) + 
  \dot{\eta}
  \left( 
    {\nabla}^{\ast}_{\eta}
    \delta \eta 
  \right)
& = 
  - \delta \eta 
\label{eq:model_dual_covars_vecs_short}
\end{align}
capturing the simple exponential scaling~(\ref{eq:und_bar_nus_sol}) of
the coherent-state fields in the exponential-family coordinates.

The covariant part of the change in the Fisher metric, from 
Eq.~(\ref{eq:covar_metric}) is computed in
Eq.~(\ref{eq:model_dual_covars_g}) to be 
\begin{align}
  \dot{\theta}
  {\nabla}_{\theta} g 
& = 
  \left( 
    \dot{v}
    \frac{\partial}{\partial v}
    \log 
    {\hat{\Phi}}_0^2
  \right) 
  g 
\nonumber \\ 
  \dot{\eta}
  {\nabla}^{\ast}_{\eta} g 
& = 
  \left(
    \dot{u}
    \frac{\partial}{\partial u}
    \log 
    {\hat{\Phi}}_0^2
  \right) 
  g 
\label{eq:model_dual_covars_g_short}
\end{align}
Only the dependence in the Fisher eigenvalue ${\hat{\Phi}}_0^2$ from
Eq.~(\ref{eq:Fisher_lightcone}) appears.  

The two lines of Eq.~(\ref{eq:model_dual_covars_g_short}) (which are
equal and opposite) scale as $\sim e^{-T}$, and have an interpretation
similar to that of a Le Chatelier principle.  The term $\Lambda
e^{-T}$ in Eq.~(\ref{eq:hatPhi_0_value}) for ${\hat{\Phi}}_0$ is a
susceptibility of the initial stationary value ${\phi}_0$ to the
perturbation by the tilt variable ${\phi}^{\dagger}_T = z$, attenuated
exponentially from time $T$ to time $0$.  The role of this
attenuation, which takes ${\hat{\Phi}}_0 \rightarrow 1$ as $T
\rightarrow \infty$, becomes clearer as a constraint on the total
extractable information when we consider in
Sec.~\ref{sec:LD_rats_estimators} the range of all initial
distributions ${\rho}_0$ and all tilts $z$.

\subsubsection{Duality of dynamics and inference in Doi-Peliti theory}

The natural separation of the coordinate transformation of the inner
product of vector fields $\delta \theta$ and $\delta \eta$ generated
by time translation is not between exponential and mixture
coordinates, as in the dually-flat connections of
Amari~\cite{Amari:inf_geom:01}, but rather between the symplectically
dual contributions from changes in $\theta$ and in $\eta$.  The two
contributions group as 
\begin{align}
  0 
& = 
  \frac{d}{dt}
  \left( 
    \delta \theta \, 
    g \, 
    \delta \eta 
  \right)
\nonumber \\ 
& = 
    \left( 
      \frac{\partial}{\partial t}
      \delta \theta 
    \right)
  {\delta}_{\eta} \nu + 
  \dot{\theta}
  {\nabla}^{\left( \theta \right)}_{\theta} \! 
  \left( 
    \delta \theta \, 
    g 
  \right) 
  \delta \eta 
\nonumber \\ 
& \mbox{} + 
  {\delta}_{\theta} \nu
  \left( 
    \frac{\partial}{\partial t}
    \delta \eta 
  \right) + 
  \delta \theta \, 
  \dot{\eta}
  {\nabla}^{\left( \eta \right) \ast}_{\eta} \! 
  \left( 
    g \, 
    \delta \eta
  \right) 
\label{eq:covar_decomp_innprod}
\end{align}
The two rows of Eq.~(\ref{eq:covar_decomp_innprod}) add covariant
contributions from Eq.~(\ref{eq:model_dual_covars_vecs_short}) and
Eq.~(\ref{eq:model_dual_covars_g_short}) in the combinations
\begin{align}
    \left( 
      \frac{\partial}{\partial t}
      \delta \theta 
    \right)
  {\delta}_{\eta} \nu + 
  \dot{\theta}
  {\nabla}^{\left( \theta \right)}_{\theta} \! 
  \left( 
    \delta \theta \, 
    g 
  \right) 
  \delta \eta
& = 
  \left( 
    \delta \theta \, 
    {\delta}_{\eta} \nu
  \right) 
  \left( 
    1 + 
    \dot{v}
    \frac{\partial}{\partial v}
    \log 
    {\hat{\Phi}}_0^2
  \right)   
\nonumber \\ 
  {\delta}_{\theta} \nu
  \left( 
    \frac{\partial}{\partial t}
    \delta \eta 
  \right) + 
  \delta \theta \, 
  \dot{\eta}
  {\nabla}^{\left( \eta \right) \ast}_{\eta} \! 
  \left( 
    g \, 
    \delta \eta
  \right) 
& = 
  \left( 
    {\delta}_{\theta} \nu \, 
    \delta \eta 
  \right) 
  \left( 
    - 1 + 
    \dot{u}
    \frac{\partial}{\partial u}
    \log 
    {\hat{\Phi}}_0^2
  \right)   
\label{eq:model_symp_par_timeder}
\end{align}
Eq.~(\ref{eq:model_symp_par_timeder}) captures in the clearest way
possible the symplectic balance of distribution dynamics (through
$\eta$) and inference (through $\theta$) in Doi-Peliti theory, through
both the direct effects of the exponential growth and decay
eigenvalues $\left( \pm 1 \right)$ and the Le Chatelier-like
susceptibility of the density ${\hat{\Phi}}_0$.

\subsection{The Fisher information density and large-deviation ratios
as sample estimators}
\label{sec:LD_rats_estimators}

The interpretation of the vector inner product as a convected density
of information can be illustrated by using ratios of large-deviation
probabilities to define a sample estimator for differences in the tilt
coordinate $\eta$ between two base distributions.

Suppose that we sample from a binomial nominal distribution at a
parameter $\eta$ that is to be estimated.  Recall from
Eq.~(\ref{eq:IS_est_bound_short}) that the probability for the value
$\rn$ of a sample to exceed a threshold $n$ is given in terms of the
large-deviation function by
\begin{equation}
  P \! \left( \rn \ge n \mid \eta \right) \sim
  e^{
    - {\psi}^{\ast} \left(n ; \eta \right)
  }
\label{eq:P_esc_LD}
\end{equation}

In a 1-dimensional system,\footnote{In one dimension, the conditional
probability is a ratio because the only way to escape beyond $n_B$ is
to have also exceeded $n_A < n_B$.  In higher dimensions, a similar
construction of the conditional can be made, but escapes must be
computed along the local least-action trajectories under the
action~(\ref{eq:CRN_L_genform}), and conditions computed for
thresholds that lie in sequence along those trajectories.  The leading
exponential approximations to such probabilities are the standard
first-passage constructions of Freidlin-Wentzel
theory~\cite{Freidlin:RPDS:98}.} for two threshold values $n_B > n_A$,
the conditional probability for $\rn$ to surpass $n_B$ given that it
has surpassed $n_A$ is the ratio
\begin{align}
  P \! \left( n_B \mid n_A ; \eta \right)
& \equiv 
  \frac{
    P \! \left( \rn \ge n_B \mid \eta \right) 
  }{
    P \! \left( \rn \ge n_A \mid \eta \right) 
  }
\nonumber \\ 
& \sim 
  e^{
    - \left[ 
      {\psi}^{\ast} \left(n_B ; \eta \right) - 
      {\psi}^{\ast} \left(n_A ; \eta \right)
    \right] 
  }
\label{eq:P_LD_cond_def}
\end{align}
The ratio~(\ref{eq:P_LD_cond_def}) can be estimated from samples of
the indicator function $h_{\rn}$ for thresholds $n$ as described in
Sec.~\ref{sec:indicator_LDF}.

App.~\ref{sec:LDF_estimator_alg} shows that if two such conditional
probabilities are compared from distributions at unknown parameters
${\eta}_2$ and ${\eta}_1$, the log ratio is related to the
large-deviation thresholds and the $\eta$ values as
\begin{align}
  \log 
  \left( 
    \frac{
      P \! \left( n_B \mid n_A ; {\eta}_2 \right) 
    }{
      P \! \left( n_B \mid n_A ; {\eta}_1 \right) 
    }
  \right) 
& \sim 
  \int_{{\eta}_1}^{{\eta}_2} \! \! 
  \int_{n_A}^{n_B}
  d_{\theta} n \, d\eta 
\nonumber \\ 
& = 
  \left( n_B - n_A \right)
  \left( {\eta}_2 - {\eta}_1 \right)
\label{eq:LDF_int_result}
\end{align}
where $d_{\theta} n \, d\eta$ is one of the two forms of the
(differential) inner product appearing in
Eq~(\ref{eq:inner_prod_two_ways}).

Thus 
\begin{align}
  \frac{
    \log 
    \left( 
      \frac{
        P \! \left( n_B \mid n_A ; {\eta}_2 \right) 
      }{
        P \! \left( n_B \mid n_A ; {\eta}_1 \right) 
      }
    \right) 
  }{
    \left( n_B - n_A \right)
  } \sim 
  \left( {\eta}_2 - {\eta}_1 \right)
\label{eq:sample_est}
\end{align}
is a sample estimator for the difference of exponential parameters in
the two underlying distributions. 

The quantity~(\ref{eq:LDF_int_result}) may be computed at any time,
for instance the final time $T$ when the thresholds $n_B$ and $n_A$
are imposed as experimental conditions, and ${\eta}_2$ and ${\eta}_1$
characterize evolved nominal distributions at time $T$ from any pair
of initial conditions at some earlier time $t = 0$.  If we use the
stationary-path conditions to propagate values of $\theta$ and $\eta$
through time, and define $V \! \left( \tau \right)$ to be the area
inside the image of the rectangle in Eq.~(\ref{eq:LDF_int_result})
along these stationary trajectories, time-invariance of the inner
product, and the Liouville conservation of volume elements in dual
coordinates, implies that
\begin{equation}
  \frac{d}{d\tau}
  \int_{V}
  d_{\theta} n \, d\eta = 
  0 
\label{eq:V_info_const}
\end{equation}

Note that, with a coordinate transform to coherent-state variables and
a corresponding redefinition of the boundary of $V$, the
relation~(\ref{eq:V_info_const}) could be recast using
Eq.~(\ref{eq:Fisher_lightcone}) as
\begin{equation}
  \frac{d}{d\tau}
  \int_{V}
  dv \, du \, 
  {\hat{\Phi}}_0^2 = 
  0 
\label{eq:V_info_const_uv}
\end{equation}
which is the conserved integral graphed in Fig~\ref{fig:L_volumes}.

In Eq.~(\ref{eq:V_info_const_uv}) ${\hat{\Phi}}_0^2$, the
2-dimensional differential of the scaled CGF $\psi / N = - \log
{\hat{\Phi}}_0$, appears explicitly as the density of overlap of $dv$
with $du$ that, like $\psi$ itself, is constant along stationary
paths.  ${\hat{\Phi}}_0^2$ is not independent of the position $\left(
v , u \right)$ within the volume $V$, but because the volume element
moves along with the conserved density, the integral measures a fixed
quantity of Fisher information as it is transported through different
domains of base and tilt.

Although the limits of integration for $\int dv \, du$ in
Eq.~(\ref{eq:V_info_const_uv}) are bounded, the limits on $\left(
{\eta}_2 - {\eta}_1 \right)$ in Eq.~(\ref{eq:LDF_int_result}) are not,
so formally the range of the sample estimator~(\ref{eq:sample_est})
remains unbounded over any duration $T$.  However, for any
\emph{fixed} values of ${\left( n_B - n_A \right)}_{t = T}$ and
starting uncertainty ${\left( {\eta}_2 - {\eta}_1 \right)}_{t = 0}$,
the total information obtainable from large-deviations sampling about
differences in the initial conditions is finite and decreases as
$e^{-T}$.  In Fig~\ref{fig:L_volumes} this limit is seen in the way
any fixed ranges are squeezed exponentially at the ``waist'' as $T
\rightarrow \infty$.  The contraction of boundaries, rather than
the asymptotic behavior of the eigenvalue in the Fisher metric,
measures the loss of information between initial distributions and
final observations with increasing separation between the two.

\section{Conclusions: the duality of dynamics and inference for
irreversible and reversible processes}
\label{sec:conclusions}

The three-part structure of the Fisher metric, dual Riemannian
connections, and symplectic parallel transport of the Wigner density,
vector fields, and the metric tensor, elegantly expresses the
transport properties along 2FFI stationary paths in terms of geometric
invariants.  It resolves a feature of 2-field constructions that at
first seems paradoxical: if memory of initial conditions is
continuously lost to dissipation, what concept of time-reversal is
implied by invertibility of the map along stationary rays?  The answer
from the perspective of importance sampling is that, even if samples
are finite, their expectations are computed in continuous-valued
distributions, and deformations of measure through the Radon-Nikodym
derivative can locally compensate for concentration of measure in the
nominal distribution by expanding sensitivity of likelihood ratios.
Locally in sampling space, then, time is immaterial as it is in
Hamiltonian mechanics; the mappings along stationary trajectories make
it possible to interpret sampling protocols from different times in an
evolving distribution simply as coordinate transformations of a fixed
sampling protocol on the original distribution.  On the other hand,
for any \emph{fixed} ranges of parameter variation in the initial
conditions, and fixed large-deviation thresholds compared at late
time, the integrated Liouville density contracts monotonically with
the separation between the two times, reflecting the absolute loss of
information that can be recovered.

We have wanted to establish a concrete interpretation of time-duality
in 2FFI theories as a duality of dynamics and inference, to provide an
alternative to the interpretation in terms of \emph{physical} reversal
of paths that is the starting point in most of the literature on
fluctuation theorems in stochastic thermodynamics.  Microscopic
reversibility can always be added later to any class of 2FFI
constructions as a restriction on the scope of phenomena under study,
and both stronger conclusions and additional interpretations will then
follow from the added constraints.  Where the existence of a duality
in the mathematics itself does not depend on any such additional
assumptions, taking the inference interpretation to reflect the core
concepts, directly expressing Kolmogorov's forward/backward adjoint
duality, frames the special case of microscopic reversibility as one
in which the system's own dynamics contains an image of certain
sampling protocols over itself.

Even if one only cares about microscopically reversible processes,
making explicit the step of self-modeling, and having a concrete
interpretation of conserved densities such as the Fisher information
constructed here, provides a bridge between trajectory reversal in
low-level mechanics and operations for sample estimation of the kind
that are used by control systems.  Linking limitations from path
probability in a system's autonomous dynamics to concepts of
information capacity in control
loops~\cite{Ashby:cybernetics:56,Ashby:req_var:58,Conant:regulator:70}
promises a way to study the limits on spontaneous emergence of
dynamical hierarchy, which has been a desired application for
stochastic
thermodynamics~\cite{England:statphys_selfrepl:13,Perunov:adaptation:15}.
These are intended topics for future work.

\subsection*{Acknowledgments}

The author thanks Supriya Krishnamurthy for ongoing collaboration and
the Stockholm University Physics Department for hospitality while much
of this work was done, and Nathaniel Virgo for helpful discussion.
The work was supported in part by NASA Astrobiology CAN-7 award
NNA17BB05A through the College of Science, Georgia Institute of
Technology, and by the Japanese Ministry of Education, Culture,
Sports, Science, and Technology (MEXT) through the World Premiere
International program.

\appendix 

\section{Fisher spherical embeddings}
\label{sec:Fisher_spheres}

\subsection{The embedding for general distributions on finite state
spaces} 
\label{sec:Fisher_general_sphere}

Eq.~(\ref{eq:Fisher_element}) in the text can be written in the form 
\begin{equation}
  \delta s^2 = 
  4 
  \delta {\theta}^i
  \delta {\theta}^j
  \sum_{\rn}
  \frac{
    \partial 
    \sqrt{{\tilde{\rho}}_{\rn}^{\left( \theta \right)}}
  }{
    \partial {\theta}^i 
  }
  \frac{
    \partial 
    \sqrt{{\tilde{\rho}}_{\rn}^{\left( \theta \right)}}
  }{
    \partial {\theta}^j 
  }
\label{eq:Fisher_element_sqrt}
\end{equation}

Let $\left| \left\{ \rn \right\} \right|$ be the cardinality of the
set of states on which ${\rho}_{\rn}$ is defined (for example, in
chemistry, only a sub-lattice of all integer-valued vectors in the
positive orthant may ever be accessible as counts, given a system's
stoichiometry and conserved quantities).  Suppose $\left| \left\{ \rn
\right\} \right|$ is finite in order illustrate the Fisher
embedding geometry for distributions over finite state spaces.  All
possible base distributions $\rho$ fall within the simplex of
dimension $\left| \left\{ \rn \right\} \right| - 1$.

Now let $\left\{ {\alpha}_1, {\alpha}_2,  \ldots {\alpha}_{\left|
\left\{ \rn \right\} \right| - 1} \right\}$ be angles associated with
independent rotation axes in ${\mathbb{R}}^{\left| \left\{ \rn
\right\} \right|}$.  Any distribution can be embedded in
${\mathbb{R}}^{\left| \left\{ \rn \right\} \right|}$ by arranging the
states $\rn$ in an (arbitrary) order ${\rn}_1 , {\rn}_2 , \ldots
{\rn}_{\left| \left\{ \rn \right\} \right|}$, and writing
\begin{align}
  p_{{\rn}_1}
& \equiv 
  \cos^2 {\alpha}_1
\nonumber \\ 
  p_{{\rn}_2}
& \equiv 
  \sin^2 {\alpha}_1
  \cos^2 {\alpha}_2
\nonumber \\ 
  p_{{\rn}_3}
& \equiv 
  \sin^2 {\alpha}_1
  \sin^2 {\alpha}_2
  \cos^2 {\alpha}_3
\nonumber \\ 
& \: \: \vdots 
\nonumber \\ 
  p_{{\rn}_{\left| \left\{ \rn \right\} \right| - 1}}
& \equiv 
  \sin^2 {\alpha}_1
  \sin^2 {\alpha}_2 \cdots 
  \sin^2 {\alpha}_{\left| \left\{ \rn \right\} \right| - 2}
  \cos^2 {\alpha}_{\left| \left\{ \rn \right\} \right| - 1}
\nonumber \\ 
  p_{{\rn}_{\left| \left\{ \rn \right\} \right|}}
& \equiv 
  \sin^2 {\alpha}_1
  \sin^2 {\alpha}_2 \cdots 
  \sin^2 {\alpha}_{\left| \left\{ \rn \right\} \right| - 2}
  \sin^2 {\alpha}_{\left| \left\{ \rn \right\} \right| - 1}
\label{eq:Fisher_embed_probs}
\end{align}
A recursive calculation gives the line
element~(\ref{eq:Fisher_element_sqrt}) in terms of the angle
coordinates on the radius-2 sphere as
\begin{align}
  \delta s^2 
& = 
  4 \left[ 
  \vphantom{
    \delta {\alpha}_{\left| \left\{ \rn \right\} \right|}^2 
  }
  \delta {\alpha}_1^2 + 
  \sin^2 {\alpha}_1 \, 
  \delta {\alpha}_2^2 + 
  \ldots 
  \right. 
\nonumber \\
& \mbox{} + 
  \left. 
  \left( 
    \sin^2 {\alpha}_1 \ldots 
    \sin^2 {\alpha}_{\left| \left\{ \rn \right\} \right| - 2}
  \right) 
  \delta {\alpha}_{\left| \left\{ \rn \right\} \right| - 1}^2 
  \right] 
\label{eq:Fisher_element_angles}
\end{align}

\subsection{Embeddings in reduced dimension for exponential families on
the multinomial} 
\label{sec:Fisher_reduced_sphere}

The Poisson~(\ref{eq:Poisson_form}) and
multinomial~(\ref{eq:multinom_form}) distributions are both in a class
recognized by Anderson, Craciun, and Kurtz
(ACK)~\cite{Anderson:product_dist:10} in connection with uniqueness of
stationary solutions for chemical reaction networks.  All factorial
moments are powers their first moments, causing the CGF for many
particles to scale as a multiple of a single-particle CGF.  It is not
then surprising that the expression~(\ref{eq:Fisher_element}) for the
Fisher metric in terms of a distribution ${\tilde{\rho}}_{\rn}$ with
possibly indefinitely many independent terms, for the ACK
distributions projects to a function of the same form in terms of
expected numbers $n_i$ over the $D$ independent species.

To see how this works for a distribution ${\rho}^{\left( n_0
\right)}_{\rn}$ with multinomial form~(\ref{eq:multinom_form}),
express the expected number fractions as
\begin{equation}
  \frac{n_{0i}}{N} = 
  \frac{
    e^{{\eta}_i}
  }{
    \sum_{j=1}^D e^{{\eta}_j}
  } 
\label{eq:multnom_exp_n0}
\end{equation}
Then the mean in the distribution ${\tilde{\rho}}_{\rn}^{\left( \theta
; n_0 \right)}$ is  
\begin{equation}
  \frac{n_i \! \left( \theta \right)}{N} = 
  \frac{
    e^{{\eta}_i + {\theta}_i}
  }{
    \sum_{j=1}^D e^{{\eta}_j + {\theta}_j}
  } \equiv 
  {\nu}_i
\label{eq:multnom_exp_n}
\end{equation}
and the CGF $\psi \! \left(
\theta ; n_0 \right)$ evaluates (up to a constant offset) to
\begin{equation}
  \psi \! \left( \theta ; n_0 \right) = 
  N \log 
  \left[ 
    \sum_{j = 1}^D e^{{\eta}_j + {\theta}_j}
  \right] 
\label{eq:gen_fun_multinom}
\end{equation}

The Hessian giving the Fisher metric is 
\begin{equation}
  g_{ij} \! \left( \theta \right) = 
  N 
  \left\{
    {\nu}_i {\delta}_{ij} - 
    {\nu}_i 
    {\nu}_j 
  \right\}
\label{eq:multnom_Fisher}
\end{equation}
where ${\nu}_i$ is the function of $\eta + \theta$ in
Eq.~(\ref{eq:multnom_exp_n}).  Inverting
Eq.~(\ref{eq:multnom_Fisher}), and projecting onto the $\sum_i
{\theta}_i = 0$ to fix the undetermined component of $\theta$, gives
the inverse
\begin{equation}
  g^{ij} \! \left( n \right) = 
  \frac{1}{N} 
  \left\{
    \frac{1}{{\nu}_i} {\delta}_{ij} - 
    \frac{1}{D}
    \left[ 
      \frac{1}{{\nu}_i} + 
      \frac{1}{{\nu}_j} - 
      \frac{1}{D}
      \sum_k 
      \frac{1}{{\nu}_k} 
    \right] 
  \right\}
\label{eq:multnom_Fisher_inv}
\end{equation}
One can check that both $g \! \left( \theta \right)$ and $g^{-1} \!
\left( n \right)$ sum to zero on either index, and the product
\begin{equation}
  g^{ij} \! \left( n \right)  
  g_{jk} \! \left( \theta \right) = 
  {\delta}^i_k - 
  \frac{1}{D}
\label{eq:inverse_verif}
\end{equation}
is the identity on the subspace $\sum_{j=1}^D {\theta}^j = 0$ or
$\sum_{j=1}^D n_j = N$.

If a shift of the tilted distribution in the exponential
family is indexed with coordinate $\delta n$, with $\sum_{j=1}^D
\delta n_j = 0$, the Fisher distance element from
Eq.~(\ref{eq:Fisher_element_ns}) becomes 
\begin{equation}
  \delta s^2 = 
  \sum_{j=1}^D
  \frac{\delta n_j^2}{n_j}
\label{eq:Fisher_element_red}
\end{equation}
which is the same function of $n$ as the function of $\tilde{\rho}$ in
the third line of Eq.~(\ref{eq:Fisher_element}).  

\section{Sample means and variances in the large-deviation
approximation to threshold indicator expectations}
\label{sec:app_IS_LDF_alg}

The expectation of the tilted indicator function from
Eq.~(\ref{eq:IS_no_bias}) may be written in a series of inequalities
culminating in the expression for the CGF, as 
\begin{align}
  {\left< {\tilde{h}}^{\theta} \right>}_{\left( \theta ; n_0 \right)} 
& = 
  \sum_{\rn > \bar{n}}
  e^{
    - \theta \rn 
  }
  e^{
    \theta \rn 
  }
  {\rho}^{\left( n_0 \right)}_{\rn}
\nonumber \\ 
& \le 
  e^{
    - \theta \bar{n}
  }
  \sum_{\rn > \bar{n}}
  e^{
    \theta \rn 
  }
  {\rho}^{\left( n_0 \right)}_{\rn}
\nonumber \\ 
& \le 
  e^{
    - \theta \bar{n}
  }
  \sum_{\rn}
  e^{
    \theta \rn 
  }
  {\rho}^{\left( n_0 \right)}_{\rn}
\nonumber \\ 
& = 
  e^{
    \psi \left( \theta ; n_0 \right) - \theta \bar{n} 
  }
\label{eq:IS_est_bound}
\end{align}
providing Eq.~(\ref{eq:IS_est_bound_short}) in the text.

The variance of the same sample estimator has a corresponding bound 
\begin{align}
  {
    \left< 
      { \left( {\tilde{h}}^{\theta} \right) }^2
    \right>
  }_{\left( \theta ; n_0 \right)} - 
  {\left< {\tilde{h}}^{\theta} \right>}_{\left( \theta ; n_0 \right)}^2 
& = 
  {
    \left< 
      { \left( {\tilde{h}}^{\theta} \right) }^2
    \right>
  }_{\left( \theta ; n_0 \right)} - 
  {\left< h \right>}_{\left( n_0 \right)}^2 
\nonumber \\ 
& = 
  {
    \left< 
      e^{
        \psi \left( \theta ; n_0 \right) - \theta \rn 
      }
      {\tilde{h}}^{\theta} 
    \right>
  }_{\left( \theta ; n_0 \right)} - 
  {\left< h \right>}_{\left( n_0 \right)}^2 
\nonumber \\ 
& \le  
  e^{
    \psi \left( \theta ; n_0 \right) - \theta \bar{n} 
  }
  {
    \left< 
      {\tilde{h}}^{\theta} 
    \right>
  }_{\left( \theta ; n_0 \right)} - 
  {\left< h \right>}_{\left( n_0 \right)}^2 
\nonumber \\ 
& = 
  e^{
    \psi \left( \theta ; n_0 \right) - \theta \bar{n} 
  }
  {\left< h \right>}_{\left( n_0 \right)} - 
  {\left< h \right>}_{\left( n_0 \right)}^2 
\label{eq:IS_var_def}
\end{align}
giving Eq.~(\ref{eq:IS_var_def_short}) in the text.

To estimate the tightness of the bounds, begin by observing that in
the large-deviation scaling regime~(\ref{eq:LD_scaling}), with all
cumulants generated as derivatives of $\psi \sim N$, the expansion of
central moments in terms of cumulants bounds the scaling of the $k$th
central moment as 
\begin{equation}
  \left< 
    {
      \left( \rn - \left< \rn \right> \right)
    }^k 
  \right> / 
  {\left< \rn \right>}^k
  \le 
  \mathcal{O} \! 
  \left( 
    N^{ 
      - \left\lceil k/2 \right\rceil 
    }
  \right) 
\label{eq:cent_mom_scaling}
\end{equation}

The log ratio we wish to bound is the Bregman divergence
\begin{equation}
  \log 
  \left[
    \frac{
      \sum_{\rn > \bar{n}}
      e^{- \theta \left( \rn - \bar{n} \right)}
      e^{\theta \rn} {\rho}_{\rn}
    }{
      \sum_{\rn}
      e^{\theta \rn} {\rho}_{\rn}
    }
  \right]   = 
  \theta \bar{n} - 
  \psi \! \left( \theta , n_0 \right) + 
  \log {\left< h^{\left( \bar{n} \right)} \right>}_{\left( n_0 \right)}
\label{eq:log_ratio_Bregman}
\end{equation}
The maximum of Eq.~(\ref{eq:log_ratio_Bregman}) occurs at $\theta \! 
\left( \bar{n} \right)$ from Eq.~(\ref{eq:psi_Legend_to_phi}), and the
width of the transition for the log ratio to change by $\mathcal{O} \!
\left( N^0 \right)$ is given by
\begin{equation}
  \delta \theta \approx 
  {
    \left( 
  {
    \left. 
      \frac{{\partial}^2 \psi}{{\partial \theta}^2}
    \right| 
  }_{
    \theta \left( \bar{n} \right) 
  } 
    \right) 
  }^{-1} \sim 
  N^{-1/2}
\label{eq:trans_width}
\end{equation}

To estimate its maximum value we write
Eq.~(\ref{eq:log_ratio_Bregman}) as the sum of log ratios of the two
inequalities in Eq.~(\ref{eq:IS_est_bound}), and observe that they
have boundary values
\begin{align}
  {
    \left. 
      \log 
      \left[
        \frac{
          \sum_{\rn > \bar{n}}
          e^{- \theta \left( \rn - \bar{n} \right)}
          e^{\theta \rn} {\rho}_{\rn}
        }{
          \sum_{\rn > \bar{n}}
          e^{\theta \rn} {\rho}_{\rn}
        }
      \right] 
    \right| 
  }_{
    \theta = 0 
  }
& = 
  0 
\nonumber \\ 
  {
    \left. 
      \log 
      \left[
        \frac{
          \sum_{\rn > \bar{n}}
          e^{\theta \rn} {\rho}_{\rn}
        }{
          \sum_{\rn}
          e^{\theta \rn} {\rho}_{\rn}
        }
      \right] 
    \right| 
  }_{
    \theta \rightarrow \infty
  }
& = 
  0 
\label{eq:log_rat_limits}
\end{align}
The values of these ratios at intermediate $\theta$ are then obtained
by integrating the derivatives
\begin{align}
  \frac{\partial}{\partial \theta}
  \log 
  \left[
    \frac{
      \sum_{\rn > \bar{n}}
      e^{- \theta \left( \rn - \bar{n} \right)}
      e^{\theta \rn} {\rho}_{\rn}
    }{
      \sum_{\rn > \bar{n}}
      e^{\theta \rn} {\rho}_{\rn}
    }
  \right] 
& = 
  - \frac{
    \sum_{\rn > \bar{n}}
    \left(
      \rn - \bar{n}
    \right) 
    e^{\theta \rn} {\rho}_{\rn}
  }{
    \sum_{\rn > \bar{n}}
    e^{\theta \rn} {\rho}_{\rn}
  }
\nonumber \\ 
  \frac{\partial}{\partial \theta}
  \log 
  \left[
    \frac{
      \sum_{\rn > \bar{n}}
      e^{\theta \rn} {\rho}_{\rn}
    }{
      \sum_{\rn}
      e^{\theta \rn} {\rho}_{\rn}
    }
  \right] 
& = 
  \frac{
    \sum_{\rn > \bar{n}}
    \left(
      \rn - n \! \left( \theta \right)
    \right) 
    e^{\theta \rn} {\rho}_{\rn}
  }{
    \sum_{\rn > \bar{n}}
    e^{\theta \rn} {\rho}_{\rn}
  }
\label{eq:change_log_ratio}
\end{align}
Both log derivatives in Eq.~(\ref{eq:change_log_ratio}) are monotone
if $\psi$ is convex, and their values sum to $\bar{n} - n \! \left(
\theta \right)$, the derivative of Eq~(\ref{eq:log_ratio_Bregman}), 
where $n \! \left( \theta \right) \equiv {\left< \rn
\right>}_{{\tilde{\rho}}^{\left( \theta , n_0 \right)}}$.  

Next, observe the leading-order scaling of the expectation ${\left<
{\left( \rn - \bar{n} \right)}^2 \right>}_{\rn > \bar{n}} \approx
{\left< \left( \rn - \bar{n} \right) \right>}^2_{\rn > \bar{n}}$ on
the half-line (as for any second moment), and likewise for $\left( \rn
- n \! \left( \theta \right) \right)$.  

At $\theta \! \left( \bar{n} \right)$, where $n \! \left( \theta
\right) = \bar{n}$, the two derivatives~(\ref{eq:change_log_ratio})
are equal and opposite, and the boundary of the half-line $\rn >
\bar{n}$ is also the symmetry point of ${\left( \rn - \bar{n}
\right)}^2$.  Because the skewness and higher-order central moments
grow more slowly than ${\bar{n}}^k$ by
Eq.~(\ref{eq:cent_mom_scaling}), ${\left< {\left( \rn - \bar{n}
\right)}^2 \right>}_{\rn > \bar{n}} \approx {\left< {\left( \rn -
\bar{n} \right)}^2 \right>}_{\rn} \times \left\{ 1 + \mathcal{O} \!
\left( 1 / N \right) \right\}$, and given the
large-deviations scaling of central
moments~(\ref{eq:cent_mom_scaling}), the derivatives in
Eq.~(\ref{eq:change_log_ratio}) scale as 
\begin{equation}
  {
    \left. 
  \frac{
    \sum_{\rn > \bar{n}}
    \left(
      \rn - \bar{n} 
    \right) 
    e^{\theta \rn} {\rho}_{\rn}
  }{
    \sum_{\rn > \bar{n}}
    e^{\theta \rn} {\rho}_{\rn}
  }
    \right| 
  }_{
    n \left( \theta \right) = \bar{n}
  } \sim 
  \sqrt{
    \left<
      {
        \left( \rn - \bar{n} \right)  
      }^2
    \right> 
  } \sim 
  N^{1/2}
\label{eq:change_log_atlim}
\end{equation}
Over the range $\pm \delta \theta$ from Eq.~(\ref{eq:trans_width}),
where the total log ratio changes by $\mathcal{O} \! \left( N^0
\right)$, the integral of the first
derivative~(\ref{eq:change_log_atlim}) saturates the lower limit in
the first line of Eq.~(\ref{eq:log_rat_limits}), and the upper limit
in the second line of Eq.~(\ref{eq:log_rat_limits}), to within $\le
\mathcal{O} \! \left( N^{1/2} \right)$, implying that the log-ratios
themselves at the midpoint scale as $\le \mathcal{O} \! \left( N^{1/2}
\right)$.   Hence also the total log ratio that is their sum scales as
$\log \left[ {\left< h^{\left( \bar{n} \right)} \right>}_{\left( n_0
\right)} / e^{- {\psi}^{\ast} \left(\bar{n} \right)} \right] \le
\mathcal{O} \! \left( N^{1/2} \right)$, the result used in the text.

\section{Review of Doi Hilbert space and Peliti functional integral
constructions}
\label{sec:DP_review}

\subsection{Doi operator algebra and inner product}
\label{sec:DP_review_Doi_part}

The main constructs in the Doi operator
formulation~\cite{Doi:SecQuant:76,Doi:RDQFT:76} of moment-generating
functions as formal power series are as follows:

The identification~(\ref{eq:a_adag_defs}) of $z$ and $\partial /
\partial z$ with raising and lowering operators $a^{\dagger}$ and $a$
allows the commutator algebra 
\begin{align}
  \left[ 
    a_i , a^{\dagger}_j 
  \right] = 
  {\delta}_{ij} . 
\label{eq:comm_relns}
\end{align}
to stand for the commutator algebra between components of $\partial /
\partial z$ and factors of $z$, applied by function composition acting
to the right on MGFs.

Monomials $z^{\rn}$ from Eq.~(\ref{eq:product_conventions}) are basis
elements in a linear space of MGFs, built up by multiplication on the
number 1.  A bracket notation for states and an inner product are
introduced by the pair of denotations 
\begin{align}
  1 
& \rightarrow 
  \left| 0 \right)
& 
  \int d^D \! z \, 
  {\delta}^D \! \left( z \right) 
& \rightarrow 
  \left( 0 \right|
\label{eq:null_states}
\end{align}
Each monomial $z^{\rn}$ is denoted as a \textit{number state}
\begin{align}
  \prod_{i = 1}^D
  z_i^{{\rn}_i} \times 
  1 
& \rightarrow 
  \prod_{i = 1}^D
  {
    a_i^{\dagger} 
  }^{{\rn}_i} 
  \left| 0 \right) \equiv 
  \left| \rn \right) .
\label{eq:number_states}
\end{align}
The number states are eigenstates of the set of \textit{number
operators} ${\hat{n}}_i \equiv a^{\dagger}_i a^i$ (no Einstein sum):
\begin{align}
   {\hat{n}}_i
  \left| \rn \right) =
  {\rn}_i
  \left| \rn \right) .
\label{eq:num_states_eigs}
\end{align}

Dual to each number state is a conjugate projection operator 
\begin{align}
  \left( {\rm m} \right| 
& \equiv 
  \left( 0 \right| 
  \prod_{i = 1}^D
  \frac{
    {
      a_i^{\dagger} 
    }^{{\rm m}_i} 
  }{
    {\rm m}_i ! 
  }
\leftarrow 
  \int d^D \! z \, 
  {\delta}^D \! \left( z \right) 
  \prod_{i = 1}^D
  \frac{
    {
      \left( \partial / \partial z_i \right) 
    }^{
      {\rm m}_i
    }
  }{
    {\rm m}_i ! 
  }
\label{eq:number_projectors}
\end{align}
From the commutation relations of variables and their derivatives it
follows that the number states and projectors have overlap
\begin{equation}
  \left( 
    {\rm m} \, 
  \right| \! 
  \left. 
    \rn 
  \right) = 
  {\delta}^D_{{\rm m} \rn}
\label{eq:number_overlap}
\end{equation}
the Kronecker $\delta$ symbol on $D$ indices.  The number states and
projectors are complete, and a sum of Eq.~(\ref{eq:number_overlap}) on
${\rm m}$ is the \textit{Glauber norm}
\begin{align}
  \left( 0 \right|
  e^{\sum_i a_i}
  \left| \rn \right) = 1 , \quad
  \forall \rn , 
\label{eq:Glauber_inn_prod}
\end{align}
which defines the asymmetric inner product on the Hilbert space of
generating functions.  

Replacing the uniform measure $\sum_i a_i \equiv 1^T a$ in
Eq.~(\ref{eq:Glauber_inn_prod}) with the scalar product $z a$ gives
the map~(\ref{eq:Glauber_to_FG}) to $\Psi \! \left( z \right)$ in the
main text.  $\Psi \! \left( 1 \right) \equiv 1$; the Glauber norm of
the Laplace transform of any normalized distribution is the trace of
the probability distribution $\sum_{\rn} {\rho}_{\rn}$.

\subsection{Coherent states and Peliti functional integral}
\label{sec:Peliti_CS_alg}

The uniform measure in the 2D-dimensional integral for the
representation of unity~(\ref{eq:field_int_ident_short}) in the main
text is known as the Haar measure.  Using the
definition~(\ref{eq:coh_st_eig_a}) for coherent states
and~(\ref{eq:dual_phi_states}) for their dual projectors, and
expanding the exponential functions as sums, 

\begin{align}
& \lefteqn{
  \int 
  \frac{
    d^D \! {\phi}^{\dagger}
    d^D \! \phi
  }{
    {\pi}^D
  }
  \left| \phi \right) 
  \left( \phi \right| 
}
\nonumber \\ 
& = 
  \int 
  \frac{
    d^D \! {\phi}^{\dagger}
    d^D \! \phi
  }{
    {\pi}^D
  }
  \prod_{i = 1}^D
  e^{- {\phi}^{\dagger}_i {\phi}_i}
  \sum_{{\rn}_i}
  \sum_{{\rm m}_i}
  \frac{
    {\phi}_i^{{\rn}_i}
    {{\phi}^{\dagger}_i}^{{\rm m}_i}
  }{
    {\rn}_i ! 
  }
  \left| \rn \right) 
  \left( {\rm m} \right| 
\nonumber \\ 
& = 
  \sum_{\rn}
  \left| \rn \right) 
  \left( \rn \right| = 
  I 
\label{eq:field_int_ident}
\end{align}
The phase component in each integral $d{\phi}^{\dagger}_i \,
d{\phi}_i$ vanishes unless ${\rn}_i = {\rm m}_i$, and the remaining
modulus component produces a Gamma-function canceling the ${\rn}_i !$.
Thus the Haar measure on coherent states is equivalent to the uniform
measure on states $\rn$ of the classical probability
distribution.\footnote{Aaronson~\cite{Aaronson:QC_Democritus:13}
(p.~123) has raised this equivalence as one of the reasons only the
complex L2 norm of quantum mechanics results in a correspondence
principle with the classical laws of probability.  It is interesting
that the representation of probability components ${\rho}_{\rn}$ as
squared amplitudes (though only real-valued) also underlies the
natural spherical embedding of App.~(\ref{sec:Fisher_spheres}) for the
Fisher metric.}

Mapping backward through the correspondences between analytic
functions and Doi state vectors from
App.~\ref{sec:DP_review_Doi_part}, an evaluation of the integral in
terms of Dirac $\delta$-functions\footnote{There is a notational
subtlety in writing the complex area integral $\int d {\phi}^{\dagger}
d \phi \equiv \int_0^{\infty} d\left| \phi \right| \int_0^{2\pi}
\left| \phi \right| d\arg \phi$ with respect to $\delta$-functions
evaluated as complex contour integrals.  For example, in $D = 1$, the
integral kernel in Eq.~(\ref{eq:unity_as_shift}) is written in the two
notations as
\begin{align*}
&
\lefteqn{
  \int 
  \frac{
    d {\phi}^{\dagger}
    d \phi 
  }{
    \pi
  } 
  e^{z^{\prime} \phi}
  e^{
    {\phi}^{\dagger}
    \left( \partial / \partial z - \phi \right) 
  } = 
  e^{z^{\prime} \partial / \partial z}
  \int 
  \frac{
    d {\phi}^{\dagger}
    d \phi 
  }{
    \pi
  } 
  e^{
    \left( {\phi}^{\dagger} - z^{\prime} \right)
    \left( \partial / \partial z - \phi \right) 
  }
} 
\nonumber \\ 
& = 
  \int_0^{\infty}
  d {\left| \phi \right|}^2
  e^{- {\left| \phi \right|}^2}
  \sum_{n , m = 0}^{\infty}
  \frac{
    {\left( z^{\prime} \left| \phi \right| \right)}^n 
  }{
    n!
  }
  \frac{
    {\left( \left| \phi \right| \partial / \partial z \right)}^m
  }{
    m!
  }
  \int_0^{2 \pi}
  \frac{
    d \arg \phi
  }{
    2 \pi 
  }
  e^{
    i \left( n - m \right) \arg \phi
  }
\nonumber \\
& = 
  \sum_{n = 0}^{\infty}
  \left[ 
    \int_0^{\infty}
    d {\left| \phi \right|}^2
    e^{- {\left| \phi \right|}^2}
    \frac{
      {\left| \phi \right|}^{2n} 
    }{
      n!
    }
  \right] 
  \frac{
    {\left( z^{\prime} \partial / \partial z \right)}^n
  }{
    n!
  } = 
  \sum_{n = 0}^{\infty}
  \frac{
    {\left( z^{\prime} \partial / \partial z \right)}^n
  }{
    n!
  }
\end{align*}
The measure $\int d {\phi}^{\dagger} d \phi$ over a single complex
variable must therefore be used in evaluating $\delta$-functions as 
\begin{align*}
  \int 
  \frac{
    d {\phi}^{\dagger}
    d \phi 
  }{
    \pi
  } 
  e^{
    \left( {\phi}^{\dagger} - z^{\prime} \right)
    \left( \partial / \partial z - \phi \right) 
  }
& \equiv 
  \int 
  d {\phi}^{\dagger}
  \delta \! \left( {\phi}^{\dagger} - z^{\prime} \right)
\end{align*}
where a factor of $2 \pi$ would be required if ${\phi}^{\dagger}$ and
$\phi$ were distinct complex variables integrated over independent
contours.  This use of the measure will be needed to understand the
normalization of the Wigner function in later sections.}  shows that
the effect of the representation of unity is the map
\begin{align}
& 
\lefteqn{
  \int 
  \frac{
    d^D {\phi}^{\dagger}
    d^D \phi 
  }{
    {\pi}^D
  }
  \left| \phi \right) 
  \left( {\phi}^{\dagger} \right| 
} 
\nonumber \\ 
& \leftarrow  
  \int 
  \frac{
    d^D {\phi}^{\dagger}
    d^D \phi 
  }{
    {\pi}^D
  }
  e^{z^{\prime} \phi} 1 
  \int d^D \! z \, {\delta}^D \! \left( z \right)
  e^{
    {\phi}^{\dagger} 
    \left( \partial / \partial z - \phi \right) 
  }
\nonumber \\ 
& = 
  \int d^D \! z \, {\delta}^D \! \left( z \right)
  e^{z^{\prime} \partial / \partial z}  
\nonumber \\ 
& = 
  z \mapsto z^{\prime}
\label{eq:unity_as_shift}
\end{align}
To keep track of the scoping rules for application of complex
functions and derivatives would require introducing a distinct set of
variables $z_{k \delta t}$ for each interval in the
quadrature~(\ref{eq:Liouville_quadrature}).  In the Doi operator
algebra this scoping is handled by the bracket inner product, and the
map~(\ref{eq:unity_as_shift}) becomes simply the identity map on
$a^{\dagger}$ and $a$.

The integration measure that results from inserting a copy of the
representation of unity~(\ref{eq:field_int_ident_short}) between each
interval of time evolution in Eq.~(\ref{eq:Liouville_quadrature}) is
called a skeletonized measure.  Its limit as the interval length
$\delta t \rightarrow 0$ 
\begin{equation}
  \int_0^T 
  {\mathcal{D}}^D \! {\phi}^{\dagger} 
  {\mathcal{D}}^D \! {\phi} \equiv 
  \lim_{\delta t \rightarrow 0}
  \prod_{k = 1}^{T / \delta t}
  \int 
  \frac{
    d^D \! {\phi}^{\dagger}_{k \delta t}
    d^D \! {\phi}_{k \delta t}
  }{
    {\pi}^D
  }
\label{eq:funct_int_measure_def}
\end{equation}
defines the functional integration measure used in
Eq.~(\ref{eq:gen_fn_twofields_nosource}) and elsewhere.

\section{Stationary-point approximations to the Wigner function}
\label{sec:Wigner_stat_pt}

The functional integral provides the most direct route to the current
conservation law~(\ref{eq:Wigner_stat}) for the Wigner function.  It
is possible, with somewhat more work, to derive the same relations
directly from stationary variation of the generating function, and in
the process to gain some more intuition for what the Wigner function
quantifies.

\subsubsection*{The Wigner function in terms of an explicit density
over coherent-state parameters}

Begin by writing any state vector as the integral of a density in the
coherent-state basis:
\begin{equation}
  \left| \Psi \right) \equiv 
  \int d^D \! \phi \, 
  \left| \phi \right) 
  \rho \! \left( \phi \right)
\label{eq:state_in_Peliti_basis}
\end{equation}
The generalized Glauber norm~(\ref{eq:Glauber_to_FG}) returns the
analytic representation of the MFG:
\begin{align}
  \Psi \! \left( z \right) 
& = 
  \left( 0 \right|
  e^{z a}
  \left| \Psi \right) 
\nonumber \\ 
& = 
  \int d^D \! \phi \, 
  \left( 0 \right|
  e^{z a}
  \left| \phi \right) 
  \rho \! \left( \phi \right)
\nonumber \\ 
& = 
  \int d^D \! \phi \, 
  e^{\left( z - 1 \right) \phi}
  \rho \! \left( \phi \right)
\label{eq:Glauber_to_FG_CSdens}
\end{align}

Now evaluate the integral in Eq.~(\ref{eq:Wigner_def}) at time $t
\rightarrow T$, where the stationary value of the field ${\phi}^{\dagger}$
will coincide with the imposed argument $z$, 
\begin{align}
  w_T \! 
  \left( 
    {\phi}^{\dagger}_{\ddagger} , 
    {\phi}_{\ddagger}
  \right) 
& = 
  \frac{1}{{\pi}^D}
  \int d^D \! \phi \, 
  e^{
    \left(
      {\phi}^{\dagger}_{\ddagger} - z 
    \right) 
    \left( 
      \phi - {\phi}_{\ddagger}
    \right)
  }
  e^{\left( z - 1 \right) \phi}
  \rho \! \left( \phi \right)
\label{eq:Wigner_at_T}
\end{align}
It follows then that 
\begin{equation}
  \int 
  d^D \! {\phi}^{\dagger}_{\ddagger} \, 
  w_T \! 
  \left( 
    {\phi}^{\dagger}_{\ddagger} , 
    {\phi}_{\ddagger}
  \right) = 
  e^{\left( z - 1 \right) {\phi}_{\ddagger}} 
  \rho \! \left( {\phi}_{\ddagger} \right)
\label{eq:Wigner_int_dag}
\end{equation}
A second integral over the ${\phi}_{\ddagger}$ fields yields the two
equivalent expressions~(\ref{eq:genfun_from_Wigner})
and~(\ref{eq:Glauber_to_FG_CSdens}).

\subsubsection*{Stationary-point approximations}

The stationary value $\bar{\phi} \! \left( z \right)$ of the tilted
density $e^{\left( z-1 \right) \phi} \rho \! \left( \phi \right)$ is
given by
\begin{equation}
  {
    \left. 
      \frac{\partial \log \rho}{\partial \phi}
    \right| 
  }_{\bar{\phi} \left( z \right)} = 
  1 - z 
\label{eq:SP_cond}
\end{equation}
The two stationarity conditions on the arguments of $w_T$ follow from
Eq.~(\ref{eq:Wigner_at_T}) as 
\begin{align}
  {
    \left. 
      \frac{\partial}{\partial {\phi}_{\ddagger}}
      w_T \! 
      \left( 
        {\phi}^{\dagger}_{\ddagger} , 
        {\phi}_{\ddagger}
      \right) 
    \right| 
  }_{{\phi}^{\dagger}_{\ddagger} = z}
& = 
  \left( 
    z - {\phi}^{\dagger}_{\ddagger} 
  \right)
  w_T \! 
  \left( 
    z, 
    {\phi}_{\ddagger}
  \right) 
\nonumber \\ 
  {
    \left. 
      \frac{\partial}{\partial {\phi}^{\dagger}_{\ddagger}}
      w_T \! 
      \left( 
        {\phi}^{\dagger}_{\ddagger} , 
        {\phi}_{\ddagger}
      \right) 
    \right| 
  }_{{\phi}^{\dagger}_{\ddagger} = z}
& = 
  \int d^D \! \phi \, 
  \left( 
    \phi - {\phi}_{\ddagger}
  \right)
  e^{\left( z - 1 \right) \phi}
  \rho \! \left( \phi \right) 
\nonumber \\ 
& \sim
  \left( 
    \bar{\phi} - {\phi}_{\ddagger}
  \right)
  w_T \! 
  \left( 
    z, 
    {\phi}_{\ddagger}
  \right) 
\label{eq:Wigner_SP_cond_z}
\end{align}
where the stationary-point approximation~(\ref{eq:SP_cond}) to the
mean gives the leading exponential approximation in the second
expression. 

The Wigner function in Eq.~(\ref{eq:Wigner_at_T}), at argument $z$,
exactly equals integral~(\ref{eq:Glauber_to_FG_CSdens}) 
\begin{equation}
  w_T \! 
  \left( 
    z, 
    {\phi}_{\ddagger}
  \right) = 
  \Psi \! \left( z \right) \sim 
  {
    \left. 
      e^{\left( z - 1 \right) \phi} 
      \rho \! \left( \phi \right)
    \right| 
  }_{\bar{\phi} \left( z \right)}
\label{eq:Wigner_late_stat}
\end{equation}
independent of the value of ${\phi}_{\ddagger}$.  While the first line
of Eq.~(\ref{eq:Wigner_SP_cond_z}) shows that $z$ is a
stationary-point argument for ${\phi}^{\dagger}_{\ddagger}$, the
second line shows that only when ${\phi}_{\ddagger} = \bar{\phi} \!
\left( z \right)$ is the other argument also a stationary value.

\subsubsection*{Time dependence along a stationary path}

Suppose now that from such a compatible pair $\left( z , \bar{\phi} \!
\left( z \right) \right)$, we wish to extend $z$ and $\bar{\phi}$
along a trajectory that preserves stationarity.  The total time
derivative of $w_T$ with respect to its final-time argument is given
by 
\begin{align}
& 
\lefteqn{
  \frac{d}{dT}
  w_T \! 
  \left( 
    {\phi}^{\dagger}_{\ddagger} , 
    {\phi}_{\ddagger}
  \right) 
} 
\nonumber \\
& = 
  \left\{ 
    \frac{dz}{dT}
    {\phi}_{\ddagger} + 
    \frac{d {\phi}^{\dagger}_{\ddagger} }{dT}
    \frac{ \partial }{ \partial {\phi}^{\dagger}_{\ddagger} } + 
    \frac{d {\phi}_{\ddagger} }{dT}
    \frac{ \partial }{ \partial {\phi}_{\ddagger}} 
  \right\} 
  w_T \! 
  \left( 
    {\phi}^{\dagger}_{\ddagger} , 
    {\phi}_{\ddagger}
  \right) 
\nonumber \\ 
& \mbox{} - 
  \frac{1}{{\pi}^D}
  \int d^D \! \phi \, 
  e^{
    \left(
      {\phi}^{\dagger}_{\ddagger} - z 
    \right) 
    \left( 
      \phi - {\phi}_{\ddagger}
    \right)
  }
  \mathcal{L} \! 
  \left( 
    {\phi}^{\dagger}_{\ddagger} , 
    \phi
  \right) 
  e^{\left( z - 1 \right) \phi}
  \rho \! \left( \phi \right)
\label{eq:dw_dt_form}
\end{align}
(Note that Eq.~(\ref{eq:dw_dt_form}) includes only contributions to
the derivative of $w_T$ from quantities defined before time $T$; this
derivative is different from the total derivative $d/dt$ of $w_t$ in
the functional integral~(\ref{eq:Wigner_def}), which also includes
effects of the functional integral after time $t$.)

Ensuring that $z + dt \left( dz / dt \right)$ is a stationary value if
$z$ is one requires that 
\begin{align}
  0 
& = 
  {
    \left. 
      \frac{\partial}{\partial {\phi}_{\ddagger}}
      \frac{d}{dt}
      w_T \! 
      \left( 
        {\phi}^{\dagger}_{\ddagger} , 
        {\phi}_{\ddagger}
      \right) 
    \right| 
  }_{{\phi}^{\dagger}_{\ddagger} = z} 
\nonumber \\ 
& \sim 
  \left\{ 
    \frac{dz}{dt} - 
    {
      \left. 
        \frac{\partial}{\partial \phi}
        \mathcal{L} \! 
        \left( 
          z , \phi       
        \right) 
      \right| 
    }_{\bar{\phi} \left( z \right)}
  \right\}
  w_T \! 
  \left( 
    z , 
    {\phi}_{\ddagger}
  \right) 
\label{eq:vanishing_linear}
\end{align}
The term $\partial \mathcal{L} / \partial \phi$ is obtained by an
integration by parts over $d^D \! \phi$, and evaluated in the
stationary-point approximation.  All other terms from
Eq.~(\ref{eq:dw_dt_form}) vanish at ${\phi}^{\dagger}_{\ddagger} = z$.
Thus preservation of the stationary-argument condition for
${\phi}_{\ddagger}$ gives the stationary-path equation for $dz / dt$.

To identify the time-dependence of the stationary argument
${\phi}_{\ddagger}$, we work directly from the stationary-point
condition~(\ref{eq:SP_cond}).  The total time derivative of that
equation is 
\begin{align}
  - {
    \left. 
      \frac{\partial \mathcal{L}}{\partial \bar{\phi}}
    \right| 
  }_z
& = 
  \frac{d}{dt}
  \left(
    {
      \left. 
        \frac{\partial \log \rho}{\partial \phi}
      \right| 
    }_{\bar{\phi} \left( z \right)}
  \right) 
\nonumber \\ 
& = 
  \frac{d \bar{\phi}}{dt}
  \left(
    {
      \left. 
        \frac{{\partial}^2 \log \rho}{\partial {\phi}^2}
      \right| 
    }_{\bar{\phi}}
  \right) + 
  {
    \left. 
      \frac{\partial}{\partial \bar{\phi}}
    \right| 
  }_t
  \left(
    {
      \left. 
        \frac{
          \partial
          \log \rho \! \left( \phi \right)
        }{
          \partial t
        }
      \right| 
    }_{\bar{\phi} \left( z \right)}
  \right) 
\nonumber \\ 
& = 
  - \frac{d \bar{\phi}}{dt}
  {
    \left. 
      \frac{\partial z}{\partial \bar{\phi}}
    \right| 
  }_t - 
  {
    \left. 
      \frac{\partial}{\partial \bar{\phi}}
    \right| 
  }_t
  {
    \left. 
      \mathcal{L} \! 
      \left( 
        z , \bar{\phi}
      \right) 
    \right| 
  }_{z \left( \bar{\phi} \right)}
\nonumber \\ 
& = 
  - {
    \left. 
      \frac{\partial z}{\partial \bar{\phi}}
    \right| 
  }_t 
  \left(
    \frac{d \bar{\phi}}{dt} + 
    {
      \left. 
        \frac{
          \partial \mathcal{L} 
        }{
          \partial z 
        } 
      \right| 
    }_{\bar{\phi}}
  \right) - 
  {
    \left. 
      \frac{
        \partial \mathcal{L} 
      }{
        \partial \bar{\phi}
      } 
    \right| 
  }_z 
\label{eq:phi_SP_with time}
\end{align}
In passing from the second to the third line of
Eq.~(\ref{eq:phi_SP_with time}), to obtain an explicit expression for
${\left. \partial \log \rho \! \left( \phi \right) / \partial t
\right| }_{\bar{\phi}}$ in terms of $\mathcal{L} \! \left( z ,
\bar{\phi} \right)$, we evaluate $z$ as an inverse function of
$\bar{\phi}$ from Eq.~(\ref{eq:SP_cond}).  This functional dependence
contributes the term $\partial \mathcal{L} / \partial z$ in the final
line, from which we obtain the stationary-path equation for the
trajectory of $\bar{\phi} \! \left( z \right)$ along which
${\phi}_{\ddagger}$ is to be evaluated:
\begin{align}
  \frac{d\bar{\phi}}{dt}
& = 
  - \frac{\partial}{\partial z}
  \mathcal{L} \! 
  \left( z , \bar{\phi} \right)
\label{eq:vanishing_linear_other}
\end{align}
Eq.~(\ref{eq:vanishing_linear}) and
Eq.~(\ref{eq:vanishing_linear_other}) imply the 2FFI counterpart to
conservation of energy in Hamiltonian mechanics: $d \mathcal{L} / dt =
0$ along the stationary path if $\partial \mathcal{L} / \partial t = 0$.

\subsubsection*{The stochastic effective action in stationary-path
evaluations} 

Note from Eq.~(\ref{eq:dw_dt_form}) that along the contour identified
to preserve stationarity of $w_T$, 
\begin{align}
  \frac{d}{dT}
  \left( 
    {
      \left. 
        w_T \! 
        \left(
          z  , \bar{\phi} \! \left( z \right)
        \right)
      \right|
    }_{z \left( T \right)}
  \right) 
& = 
  \frac{dz}{dT}
  \bar{\phi} - 
  \mathcal{L} \! 
  \left( 
    z , \bar{\phi} 
  \right) 
  w_T \! 
  \left( 
    z , \bar{\phi} 
  \right) 
\nonumber \\ 
& = 
  \frac{d}{dT}
  \int_T dt 
  \left\{ 
    - \frac{dz}{dt}
    \bar{\phi} + 
    \mathcal{L} \! \left( z , \bar{\phi} \right)
  \right\} 
\nonumber \\ 
& \equiv 
  \frac{d}{dT}
  {\bar{S}}_T 
\label{eq:dw_dt_onpath}
\end{align}
Therefore the extension of the Wigner function to times $t > T$ must
include the stationary-path contribution from the action, which was
present for $t < T$ in the functional integral
definition~(\ref{eq:Wigner_def}).  $w_T$ thus extended satisfies 
\begin{equation}
  \frac{d}{dt}
  \left(
    e^{- {\bar{S}}_T}
    w_T \! 
    \left( 
      z , \bar{\phi} 
    \right) 
  \right) \sim 
  0 
\label{eq:SP_extension_w_T}
\end{equation}
recovering Eq.~(\ref{eq:Wigner_stat}).  

We have termed the stationary-path evaluation of $S$ the
\textit{stochastic effective action}~\cite{Smith:LDP_SEA:11}.  It is
the functional Legendre transform of the large-deviation functional
for trajectories in Doi-Peliti integrals.  The
approximation~(\ref{eq:Wigner_late_stat}), with ${\phi}_{\ddagger}$
set equal to $\bar{\phi} \! \left( z \right)$ given $\rho \! \left(
\phi \right)$, together with the contribution from $S_T$ in
Eq.~(\ref{eq:SP_extension_w_T}), provides the desired interpretation
of the Wigner function in terms of densities in the statistical model
provided by coherent states, and their exponential tilts by likelihood
functions.

\section{Stationary-path solutions for the two-state system} 

The stationary-path equations and both initial and final values for
fields are obtained from vanishing of all terms in the variational
derivative of the exponential argument in
Eq.~(\ref{eq:gen_fn_twofields_nosource}).  We begin with solutions in
coherent-state variables, and then present the forms for the descaled
number coordinates $\nu$, $\underline{\nu}$, and $\bar{\nu}$. 

\subsection{Coherent-state and number-potential solutions}
\label{sec:CS_statpath_solns}

\subsubsection{Stationary-path equations and final-time conditions for
response fields}

The stationary-path equations of motion for the components of the
field ${\phi}^{\dagger}$ from Eq.~(\ref{eq:Hamiltonian_var_CS}), in
the rotated basis~(\ref{eq:fields_rotate_basis}), 
evaluate to
\begin{eqnarray}
  {\partial}_{\tau} {\Phi}^{\dagger}
& = & 
  \frac{\partial \hat{\mathcal{L}}}{\partial \hat{\Phi}} = 
  - {\bm \nu} {\phi}^{\dagger}
\nonumber \\
  {\partial}_{\tau} {\phi}^{\dagger}
& = & 
  \frac{\partial \hat{\mathcal{L}}}{\partial \hat{\phi}} = 
  {\phi}^{\dagger}
\label{eq:Hamilton_EOM_phidagg}
\end{eqnarray}
The final-time values ${\phi}^{\dagger}_T$ are given by variation of
${\hat{\phi}}_T$, as 
\begin{align}
  {\phi}^{\dagger}_a 
& = 
  z_a 
& 
  {\phi}^{\dagger}_b 
& = 
  z_b 
\label{eq:T_bdry_phidagg}
\end{align}

Fixing the magnitude of the combination ${\phi}^{\dagger}_b {\bm
\nu}_b + {\phi}^{\dagger}_a {\bm \nu}_a$ in Eq.~(\ref{eq:und_nu_def})
requires varying $z$ along the contour
\begin{align}
  z_a 
& \equiv 
  \frac{{\underline{\nu}}_{Ta}}{{\bm \nu}_a}
& 
  z_b 
& \equiv 
  \frac{{\underline{\nu}}_{Tb}}{{\bm \nu}_b}
\label{eq:z_parm_denote}
\end{align}
giving Eq.~(\ref{eq:z_coords}) in the main text.  The remaining
time-dependent solutions, with time argument denoted explicitly here
by subscript $\tau$, are given by 
\begin{align}
  {\phi}^{\dagger}_{\tau} 
& = 
  \frac{
    \left( {\underline{\nu}}_T - {\bm \nu} \right) 
  }{
    \left( \frac{1}{4} - {\bm \nu}^2 \right) 
  } 
  e^{\tau - T}  
\nonumber \\ 
  {\Phi}^{\dagger}_{\tau} 
& = 
  1 - 
  {\bm \nu} 
  {\phi}^{\dagger}_{\tau} 
\label{eq:phi_dagg_ofz}
\end{align}

Initial data are specified in the generating function ${\psi}_0$,
which when evaluated at the solutions for ${\phi}^{\dagger}_0$ become
\begin{align}
  \frac{1}{N}
  {\psi}_0 
  \left( 
    \log {{\phi}^{\dagger}_a}_0 , 
    \log {{\phi}^{\dagger}_b}_0 
  \right) 
& = 
  \log 
  \left( 
    {{\phi}^{\dagger}_a}_0 {\underline{\nu}}_{0a} + 
    {{\phi}^{\dagger}_b}_0 {\underline{\nu}}_{0b}
  \right) 
\nonumber \\ 
& = 
  \log 
  \left( 
    {\Phi}^{\dagger}_0 + 
    {\bar{\nu}}_0 
    {\phi}^{\dagger}_0 
  \right) 
\nonumber \\ 
& = 
  \log 
  \left( 
    1 + 
    \Lambda e^{-T}
  \right) 
\label{eq:gamma_0_form}
\end{align}
The final line of Eq.~(\ref{eq:gamma_0_form}) is obtained by combining
the two solutions~(\ref{eq:phi_dagg_ofz}) at $\tau = 0$, and
introduces the combination $\Lambda$ defined in
Eq.~(\ref{eq:Lambda_def}).

\subsubsection{Stationary-path equations and initial-time conditions
for observable fields}

The stationary-path equations of motion for the components of the
field $\phi$ from Eq.~(\ref{eq:Hamiltonian_var_CS}) are obtained by
removing a total derivative $d_t \left( {\phi}^{\dagger} \phi
\right)$ from the action~(\ref{eq:CRN_L_genform}) to shift the
derivative onto $\phi$.  In the rotated
basis~(\ref{eq:fields_rotate_basis}), they evaluate to
\begin{eqnarray}
  {\partial}_{\tau} \hat{\Phi} 
& = & 
  - \frac{\partial \hat{\mathcal{L}}}{\partial {\Phi}^{\dagger}} = 
  0 
\nonumber \\
  {\partial}_{\tau} \hat{\phi} 
& = & 
  - \frac{\partial \hat{\mathcal{L}}}{\partial {\phi}^{\dagger}} = 
  - \left( 
    \hat{\phi} - 
    {\bm \nu} \hat{\Phi} 
  \right)
\label{eq:Hamilton_EOM_phi}
\end{eqnarray}
The total derivative cancels the final-time term ${\left(
{\phi}^{\dagger} \phi \right)}_T$ from the exponential in
Eq.~(\ref{eq:gen_fn_twofields_nosource}) and introduces an
initial-time term ${\left( {\phi}^{\dagger} \phi \right)}_0$.
Variation of this term against ${\psi}_0$ with respect to
${\phi}^{\dagger}_0$ gives the initial-value conditions for the
components of ${\hat{\phi}}_0$, as
\begin{align}
  {\hat{\phi}}_{0a}
& = 
  \frac{
    \partial {\psi}_0
  }{
    \partial {\phi}^{\dagger}_{0a}
  }
& 
  {\hat{\phi}}_{0b}
& = 
  \frac{
    \partial {\psi}_0
  }{
    \partial {\phi}^{\dagger}_{0b}
  }
\label{eq:0_bdry_phi}
\end{align} 
Solutions to the equations of motion~(\ref{eq:Hamilton_EOM_phi}) from
these initial conditions are then 
\begin{align}
  {\hat{\Phi}}_t
& = 
  {\hat{\Phi}}_0 = 
  \frac{
    1 
  }{
    1 + \Lambda e^{-T}
  }
\nonumber \\ 
  {\hat{\phi}}_t
& = 
  {\hat{\Phi}}_0 
  \left[ 
    {\bm \nu} + 
    \left( {\bar{\nu}}_0 - {\bm \nu} \right) 
    e^{- \tau}
  \right] 
\label{eq:phis_bothsol_ofz}
\end{align}

The two displacements defined in equations~(\ref{eq:und_nu_def})
and~(\ref{eq:bar_nu_def}), characterizing respectively the mean in the
nominal distribution and the likelihood ratio applied to the
stationary measure, evaluate to 
\begin{equation}
  {\bar{\nu}}_{\tau} = 
  {\bm \nu} + 
  \left( {\bar{\nu}}_0 - {\bm \nu} \right) 
  e^{-\tau}
\label{eq:true_mean_tau}
\end{equation}
and 
\begin{equation}
  {\underline{\nu}}_{\tau} = 
  {\bm \nu} + 
  \left( {\underline{\nu}}_T - {\bm \nu} \right) 
  e^{\tau - T}
\label{eq:null_mean_tau}
\end{equation}
These results are reproduced (dropping the explicit subscripts $\tau$)
as Eq.~(\ref{eq:und_bar_nus_sol}) in the text.  It follows from
Eq.~(\ref{eq:true_mean_tau}) and Eq.~(\ref{eq:null_mean_tau}) that the
combination
\begin{align}
  \frac{
    \left( {\underline{\nu}}_{\tau} - {\bm \nu} \right) 
    \left( {\bar{\nu}}_{\tau} - {\bm \nu} \right) 
  }{
    \left(
      \frac{1}{4} - 
      {\bm \nu}^2
    \right) 
  } = 
  \Lambda e^{-T} 
\label{eq:conserved_Lambda}
\end{align}
is invariant at its initial value.  

The CGF for a binomial distribution at any time retains the
form~(\ref{eq:psi_binom_twovar}), with ${\underline{\nu}}_a / {\bm
\nu}_a$ and ${\underline{\nu}}_b / {\bm \nu}_b$ replacing $z_a$ and
$z_b$, and ${\bar{\nu}}_a$ and ${\bar{\nu}}_b$ replacing ${\nu}_a$,
and ${\nu}_b$.  From Eq.~(\ref{eq:gamma_0_form}) and the invariant
form~(\ref{eq:conserved_Lambda}), it follows that 
\begin{equation}
  \frac{{\psi}_{\tau}}{N} = 
  \log 
  \left[ 
    1 + \Lambda e^{-T} 
  \right] = 
  -\log 
  {\hat{\Phi}}_0
\label{eq:hatPhi_at_tau}
\end{equation}
giving Eq.~(\ref{eq:gen_fn_eval}) in the text.

Finally, the non-linear mean of sample
values~(\ref{eq:number_fields_def}) in the importance distribution can
be shown to evaluate to
\begin{align}
  {\nu}_{\tau}
& = 
  {\underline{\nu}}_{\tau} + 
  \frac{
    \left(
      \frac{1}{4} - 
      {\underline{\nu}}_{\tau}^2 
    \right) 
  }{
    \left(
      \frac{1}{4} - 
      {\bm \nu}^2
    \right) 
  }
  \left[ 
    \left( {\bar{\nu}}_{\tau} - {\bm \nu} \right) 
    {\hat{\Phi}}_0
  \right]
\label{eq:SP_nu_diffs_compact}
\end{align}
from which the form~(\ref{eq:g_coeff_as_measures}) for $\partial \nu /
\partial \theta$ can be derived.  The ratio of measures $\left(
\sfrac{1}{4} - {\underline{\nu}}_{\tau}^2 \right) / \left(
\sfrac{1}{4} - {\bm \nu}^2 \right)$ in
Eq.~(\ref{eq:SP_nu_diffs_compact}), by which the importance
distribution responds to variations in the initial data, is the
familiar scaling of response functions~\cite{Smith:LDP_SEA:11} in the
Fluctuation-Dissipation Theorem, because expressions of the form $N
\left( \sfrac{1}{4} - {\nu}^2 \right) = N {\nu}_a {\nu}_b$ are the
variance of fluctuations in the binomial.

\subsection{Fisher spherical embedding}

In one dimension, the Pythagorean
theorem~(\ref{eq:triangle_to_Fisher}) for K-L divergences loses the
interpretation of a direction cosine between vector fields, but still
reflects scale changes between coherent-state or exponential families
and the geometric coordinate.

The mean-value Fisher-sphere construction of
Sec.~\ref{sec:Fisher_reduced_sphere}, for one variable, is the
embedding on a circle:
\begin{align}
  \cos^2 \alpha 
& \equiv
  \frac{1}{2} + \nu
&  
  \sin^2 \alpha
& \equiv
  \frac{1}{2} - \nu
\label{eq:Fisher_circle}
\end{align}
The coordinate differential is 
\begin{equation}
  2 \frac{d \alpha}{d \theta} = 
  \sqrt{
    \frac{1}{4} - {\nu}^2
  }
\label{eq:dalpha_deta}
\end{equation}
and the geometric distance element is then 
\begin{equation}
  4 
  {
    \left( \delta \alpha \right) 
  }^2 = 
  \left(
    \frac{1}{4} - {\nu}^2
  \right) 
  {
    \left( \delta \theta \right) 
  }^2 = 
  \frac{\partial \nu}{\partial \theta} 
  {
    \left( \delta \theta \right) 
  }^2 
\label{eq:Fisher_convert}
\end{equation}

The ${\hat{\Phi}}_0^2$ term in Eq.~(\ref{eq:Fisher_lightcone}) becomes
\begin{align}
  {\hat{\Phi}}_0^2 = 
  \frac{
    1 
  }{
    {
      \left[ 1 + u v \right]
    }^2
  } 
& = 
  \frac{
    \left(
      \frac{1}{4} - 
      {\nu}^2
    \right) 
  }{
    \left(
      \frac{1}{4} - 
      {\underline{\nu}}^2
    \right) 
    \left(
      \frac{1}{4} - 
      {\bar{\nu}}^2
    \right) 
  }
\nonumber \\
& = 
  4 \frac{
    \left( \partial \alpha / \partial \theta \right) 
    \left( \partial \alpha / \partial \eta \right) 
  }{
    \left( \partial v / \partial \theta \right) 
    \left( \partial u / \partial \eta \right) 
  }
\nonumber \\
& = 
  4 \frac{\partial \alpha}{\partial v}
  \frac{\partial \alpha}{\partial u}
\label{eq:Fisher_in_alpha_uv}
\end{align}
the invariant Fisher information in coherent-state coordinates.  The
equivalence between the two forms~(\ref{eq:g_D_form})
and~(\ref{eq:D_tens_to_tens}) for the Fisher metric is again recovered
as
\begin{equation}
  \frac{
    {\partial}^2
  }{
    \partial {\theta}^2
  } 
  \left( 
    \frac{{\psi}}{N} 
  \right) = 
  4 \frac{\partial \alpha}{\partial \theta}
  \frac{\partial \alpha}{\partial \eta}
\label{eq:Fisher_circle_alpha}
\end{equation}
showing the variation of the embedding coordinate of the importance
distribution with the tilt multiplied by its variation with the base.

\subsection{Evaluation of the Amari-Chentsov tensor}

From the form~(\ref{eq:lightcone}) of the Fisher metric in coordinates
$\left( v , u \right)$, the Amari-Chentsov tensor on all contravariant
indices can be computed: 
\begin{align}
  \frac{T}{N} 
& \equiv 
  \frac{
    {\partial}^3
  }{
    \partial {\theta}^3
  } \! 
  \left( 
    \frac{{\psi}}{N} 
  \right) 
\nonumber \\
& = 
  - 2 
  \frac{
    \left( dv / d\theta \right)
    \left( du / d\eta \right)
  }{
    {
      \left[ 1 + u v \right]
    }^3
  }
  \left[
    {\bm \nu}
    \left( 1 - u v \right) + 
    \sqrt{
      \frac{1}{4} - {\bm \nu}^2
    }
    \left( u + v \right)
  \right]
\label{eq:Amari_Chentsov}
\end{align}
$T$ is symmetric under $u \leftrightarrow v$, but its magnitude is not
conserved along the stationary-path trajectories.  The measure changes
$\left( dv / d\theta \right)$ and $\left( du / d\eta \right)$ are the
same as those in the Fisher metric, but in addition to these the term
$u + v$ is not invariant.

\subsection{Connection coefficients in the coherent-state connection}
\label{sec:examp_CS_conn_coeff}

Among the nonzero connection coefficients for the dual coherent-state
connections~(\ref{eq:dual_Christoffels_CS}), the only independent
components are for the rotated variables $\theta$ of
Eq.~(\ref{eq:h_theta_def}) and $\eta$ of Eq.~(\ref{eq:theta_tau_def}).
They are
\begin{align}
  {{\Gamma}^{\left( \theta \right)}_{\theta \theta}}^{\theta} 
& = 
  \frac{\partial}{\partial \theta}
  \log 
  \left( 
    \frac{\partial \underline{\nu}}{ \partial \theta}
  \right)
\nonumber \\
  {{\Gamma}^{\left( \eta \right) \ast}_{\eta \eta}}^{\eta}
& = 
  \frac{\partial}{\partial \eta}
  \log 
  \left( 
    \frac{\partial \bar{\nu}}{ \partial \eta}
  \right)
\label{eq:model_dual_Christoffels}
\end{align}
The covariant components of the time derivative of vector fields
$\delta \theta$ and $\delta \eta$ defined in
Eq.~(\ref{eq:vec_tot_der_decomp}), for transport respectively along
$\dot{\theta}$ and $\dot{\eta}$, evaluate to 
\begin{align}
  \left( 
    \frac{\partial}{\partial \tau}
    \delta \theta 
  \right) + 
  \dot{\theta}
  \left( 
    {\nabla}_{\theta}
    \delta \theta 
  \right)
& = 
  \left( 
    \frac{d}{d\tau}
    \delta \theta 
  \right) + 
  \delta \theta \, 
  \dot{\theta}
  \frac{\partial}{\partial \theta}
  \log 
  \left( 
    \frac{\partial \underline{\nu}}{ \partial \theta}
  \right)
\nonumber \\ 
& = 
  \delta \theta 
  \left[ 
    \frac{
      {\partial}^2 \mathcal{\hat{L}}
    }{
      \partial \theta
      \partial \nu
    } + 
    \frac{d}{d\tau}
    \log 
    \left( \frac{1}{4} - {\underline{\nu}}^2 \right)
  \right] 
\nonumber \\ 
& = 
  \delta \theta 
\nonumber \\ 
  \left( 
    \frac{\partial}{\partial t}
    \delta \eta 
  \right) + 
  \dot{\eta}
  \left( 
    {\nabla}^{\ast}_{\eta}
    \delta \eta 
  \right)
& = 
  \left( 
    \frac{d}{d\tau}
    \delta \eta 
  \right) + 
  \delta \eta \, 
  \dot{\eta}
  \frac{\partial}{\partial \eta} 
  \log 
  \left( 
    \frac{\partial \bar{\nu}}{ \partial \eta}
  \right)
\nonumber \\ 
& = 
  - \delta \eta 
  \left[ 
    \frac{
      {\partial}^2 \mathcal{\tilde{\hat{L}}}
    }{
      \partial \nu
      \partial \eta
    } - 
    \frac{d}{d\tau}
    \log 
    \left( \frac{1}{4} - {\bar{\nu}}^2 \right)
  \right] 
\nonumber \\ 
& = 
  - \delta \eta 
\label{eq:model_dual_covars_vecs}
\end{align}

The corrections from the coherent-state connection
coefficients~(\ref{eq:model_dual_Christoffels}) may be understood
immediately by using the first line of Eq.~(\ref{eq:CS_in_EF_defs})
and Eq.~(\ref{eq:theta_tau_def}) together with the time
dependencies~(\ref{eq:und_bar_nus_sol}), to write
\begin{align}
  \delta {\theta}_{\tau} 
& = 
  \delta {\theta}_T
  \frac{
    \left( 
      \frac{1}{4} - 
      {\underline{\nu}}_T^2 
    \right)
  }{
    \left( 
      \frac{1}{4} - 
      {\underline{\nu}}_{\tau}^2 
    \right)
  }
  e^{\tau - T}
\nonumber \\ 
  \delta {\eta}_{\tau} 
& = 
  \delta {\eta}_0
  \frac{
    \left( 
      \frac{1}{4} - 
      {\bar{\nu}}_0^2 
    \right)
  }{
    \left( 
      \frac{1}{4} - 
      {\bar{\nu}}_{\tau}^2 
    \right)
  }
  e^{-\tau}
\label{eq:delta_theta_eta_tau}
\end{align}
The geometric invariants~(\ref{eq:model_dual_covars_vecs}) capture the
exponential growth or decay from Eq.~(\ref{eq:delta_theta_eta_tau}),
while connection terms remove measure factors $\left( \sfrac{1}{4} -
{\underline{\nu}}_{\tau}^2 \right)$, $\left( \sfrac{1}{4} -
{\bar{\nu}}_{\tau}^2 \right)$ for exponential relative to
coherent-state coordinates. 

The Fisher metric $g$, unlike $\delta \theta$ and $\delta \eta$, has
no intrinsic time dependence and changes only due to change in the net
binomial parameter of the importance distribution.  Its covariant
derivatives~(\ref{eq:covar_metric}), with connection
coefficients~(\ref{eq:model_dual_Christoffels}), then produce the two
independent components of variation from $\dot{\theta}$ and
$\dot{\eta}$ of
\begin{align}
  \dot{\theta}
  {\nabla}_{\theta} g 
& = 
  \dot{\theta}
  \left[
    \frac{\partial g}{\partial \theta} - 
    g 
    \frac{\partial}{\partial \theta}
    \log 
    \left( 
      \frac{\partial \underline{\nu}}{ \partial \theta}
    \right)
  \right] 
\nonumber \\
& = 
  \dot{\theta}
  \frac{\partial g}{\partial \theta} - 
  g 
  \frac{d}{d\tau}
  \log 
  \left( 
    \frac{1}{4} - {\underline{\nu}}_{\tau}^2
  \right) 
\nonumber \\
& = 
  \left( 
    \dot{v}
    \frac{\partial}{\partial v}
    \log 
    {\hat{\Phi}}_0^2
  \right) 
  g 
\nonumber \\ 
  -\dot{\eta}
  {\nabla}^{\ast}_{\eta} g 
& = 
  - \dot{\eta}
  \left[
    \frac{\partial g}{\partial \eta} - 
    g 
    \frac{\partial}{\partial \eta}
    \log 
    \left( 
      \frac{\partial \bar{\nu}}{ \partial \eta}
    \right)
  \right] 
\nonumber \\
& = 
  -\dot{\eta}
  \frac{\partial g}{\partial \eta} + 
  g 
  \frac{d}{d\tau}
  \log 
  \left( 
    \frac{1}{4} - {\bar{\nu}}_{\tau}^2
  \right) 
\nonumber \\ 
& = 
  - \left(
    \dot{u}
    \frac{\partial}{\partial u}
    \log 
    {\hat{\Phi}}_0^2
  \right) 
  g 
\label{eq:model_dual_covars_g}
\end{align}
These are reproduced as Eq.~(\ref{eq:model_dual_covars_g_short}) in
the text.

\subsection{Two-dimensional divergences of the large-deviation
function and their integrals}
\label{sec:LDF_estimator_alg}

In the exponential family of tilted distributions with tilting
parameter $\theta$, over a base distribution with exponential
parameter $\eta$, the large-deviation function of two arguments is
constructed as
\begin{equation}
  {\psi}^{\ast} \! \left( n ; \eta \right) = 
  \theta \! \left( n ; \eta \right) n - 
  \psi \! \left( \theta \! \left( n ; \eta \right) ; \eta \right)
\label{eq:LDF_n_eta}
\end{equation}
Its variation with $\eta$ at fixed $n$ is given by 
\begin{equation}
  {
    \left. 
      \frac{\partial {\psi}^{\ast}}{\partial \eta}
    \right| 
  }_{n} = 
  {
    \left. 
      \frac{\partial \theta}{\partial \eta}
    \right| 
  }_{n} 
  \left( 
    n - 
    {
      \left. 
        \frac{\partial \psi}{\partial \theta}
      \right| 
    }_{\eta}
  \right) - 
  {
    \left. 
      \frac{\partial \psi}{\partial \eta}
    \right| 
  }_{\theta}
\label{eq:LDF_grad_eta}
\end{equation}
By definition of $\theta \! \left( n ; \eta \right)$ as the inverse
function of $n \! \left( \theta ; \eta \right) \equiv \partial \psi \!
\left( \theta ; \eta \right) / \partial \theta$, the term $\left( n -
\partial \psi / \partial \theta \right) \equiv 0$ at all $n$.  Therefore
the second derivative 
\begin{align}
  \frac{
    {\partial}^2 {\psi}^{\ast}
  }{
    \partial n \, 
    \partial \eta
  } 
& = 
  - {
    \left. 
      \frac{\partial \theta}{\partial n}
    \right| 
  }_{\eta} 
  \frac{
    {\partial}^2 \psi
  }{
    \partial \theta \, 
    \partial \eta
  } 
\nonumber \\ 
& = 
  - \frac{
    {\partial}^2 {\psi}^{\ast}
  }{
    \partial n^2 
  } 
  \frac{
    {\partial}^2 \psi
  }{
    \partial \theta \, 
    \partial \eta
  } 
\nonumber \\ 
& = 
  - {
    \left( 
      \frac{
        {\partial}^2 \psi
      }{
        \partial {\theta}^2 
      } 
    \right) 
  }^{-1}
  \frac{
    {\partial}^2 \psi
  }{
    \partial \theta \, 
    \partial \eta
  } 
\nonumber \\ 
& = 
  -1 
\label{eq:LDF_mixed_partial}
\end{align}
The third line of Eq.~(\ref{eq:LDF_mixed_partial}) uses additivity of
$\theta$ and $\eta$ to cancel the two factors, which equal
respectively $g^{-1}$ and $g$.

The log-ratio of the conditional large-deviation probabilities
in Eq.~(\ref{eq:P_LD_cond_def}) may be written as the integral
\begin{align}
  \log 
  \left( 
  \frac{
    P \! \left( n_B \mid n_A ; {\eta}_2 \right) 
  }{
    P \! \left( n_B \mid n_A ; {\eta}_1 \right) 
  }
  \right) 
& = 
  \int_{{\eta}_1}^{{\eta}_2}
  d\eta 
  \frac{\partial}{\partial \eta}
  \log P \! \left( n_B \mid n_A ; \eta \right) 
\nonumber \\ 
& = 
  - \int_{{\eta}_1}^{{\eta}_2}
  d\eta 
  \frac{\partial}{\partial \eta}
  \left[
    {\psi}^{\ast} \! \left( n_B ; \eta \right) - 
    {\psi}^{\ast} \! \left( n_A ; \eta \right) 
  \right] 
\nonumber \\
& = 
  - \int_{{\eta}_1}^{{\eta}_2} \! \! 
  \int_{n_A}^{n_B}
  d\eta \, dn 
  \frac{
    {\partial}^2 {\psi}^{\ast}
  }{
    \partial n \, 
    \partial \eta
  } 
\nonumber \\ 
& = 
  \int_{{\eta}_1}^{{\eta}_2} \! \! 
  \int_{n_A}^{n_B}
  d\eta \, dn 
  {
    \left. 
      \frac{\partial \theta}{\partial n}
    \right| 
  }_{\eta} 
  \frac{
    {\partial}^2 \psi
  }{
    \partial \theta \, 
    \partial \eta
  } 
\nonumber \\ 
& = 
  \int_{{\eta}_1}^{{\eta}_2} \! \! 
  \int_{{\theta}_A}^{{\theta}_B}
  d\eta \, d\theta 
  \frac{
    {\partial}^2 \psi
  }{
    \partial \theta \, 
    \partial \eta
  } 
\nonumber \\ 
& = 
  \int_{{\eta}_1}^{{\eta}_2} \! \! 
  \int_{n_A}^{n_B}
  d\eta \, d_{\theta} n 
\nonumber \\ 
& = 
  \left( n_B - n_A \right)
  \left( {\eta}_2 - {\eta}_1 \right)
\label{eq:LDF_int_eval}
\end{align}
giving Eq.~(\ref{eq:LDF_int_result}) in the text.  The conversion from
the fifth to the sixth line in Eq.~(\ref{eq:LDF_int_eval}) makes use
of the two alternative ways of expressing the inner product in
contravariant/covariant coordinates given in
Eq~(\ref{eq:inner_prod_two_ways}).

\vfill 
\eject


\end{document}